\documentclass[prd,aps,twocolumn,preprintnumbers, showpacs, nofootinbib,superscriptaddress,notitlepage]{revtex4-1}
\usepackage{mathrsfs}
\usepackage{amsfonts}
\usepackage{amsmath}
\usepackage{slashed}
\usepackage{array}
\usepackage{verbatim}
\usepackage{epsfig}
\usepackage{graphicx}
\usepackage{color}
\usepackage[dvipsnames]{xcolor}

\newcommand{\beq}{\begin{eqnarray}}
\newcommand{\eeq}{\end{eqnarray}}

\definecolor{red}{rgb}{1,0,0}

\def\be{\begin{equation}}
\def\ee{\end{equation}}
\def\bea{\begin{eqnarray}}
\def\eea{\end{eqnarray}}

\begin{document}

\title{Impact of gauge fixing precision
on the continuum limit \\ of non-local quark-bilinear lattice operators}

\collaboration{\bf{Lattice Parton Collaboration ($\rm {\bf LPC}$)}}

\author{
\includegraphics[scale=0.05]{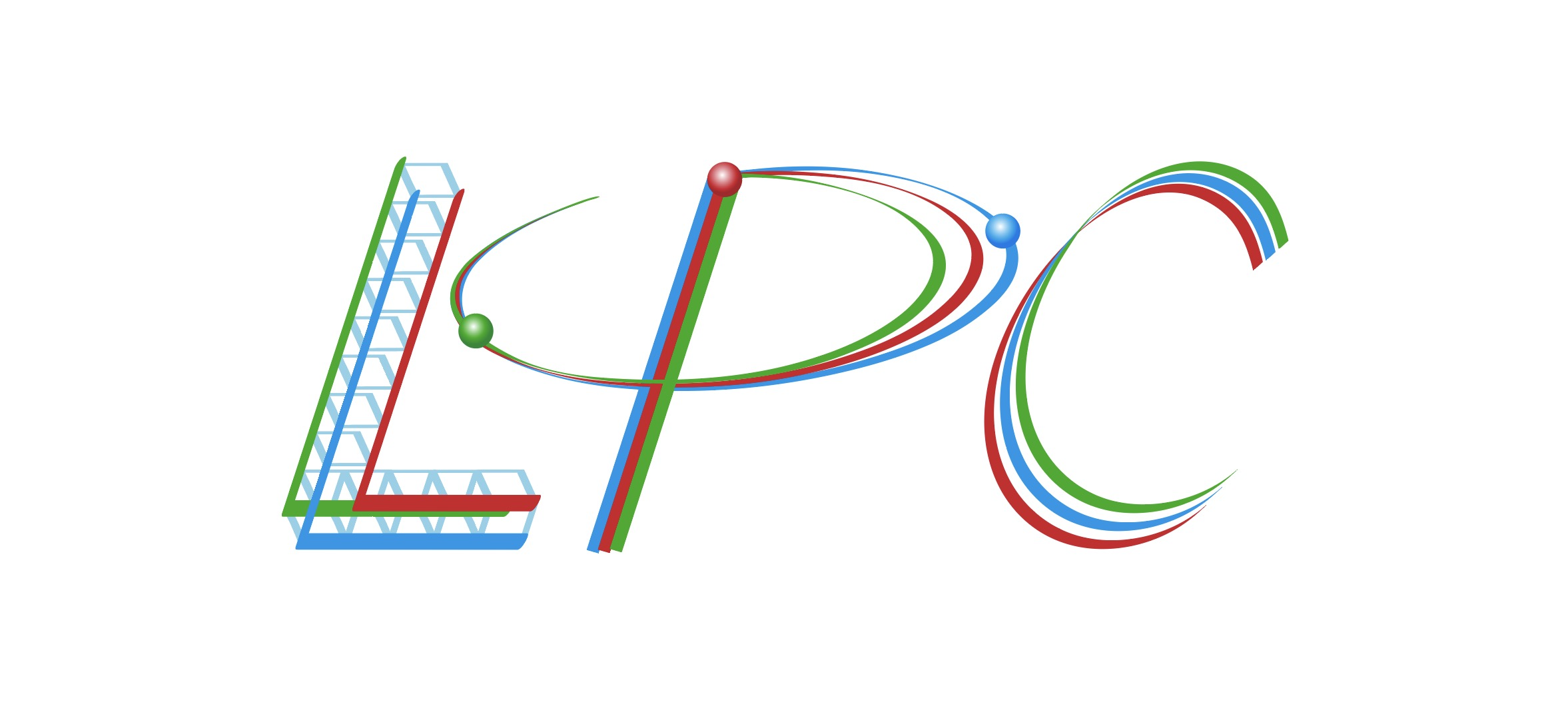}\\
Kuan Zhang}
\email{Corresponding author: zhangkuan@itp.ac.cn}
\affiliation{University of Chinese Academy of Sciences, School of Physical Sciences, Beijing 100049, China}
\affiliation{CAS Key Laboratory of Theoretical Physics, Institute of Theoretical Physics, Chinese Academy of Sciences, Beijing 100190, China}

\author{Yi-Kai Huo}
\affiliation{Physics Department, Columbia University, New York, NY 10027, USA}

\author{Xiangdong Ji}
\affiliation{Department of Physics, University of Maryland, College Park, MD 20742, USA}

\author{Andreas Sch\"afer}
\affiliation{Institut f\"{u}r Theoretische Physik, Universit\"{a}t Regensburg, D-93040 Regensburg, Germany}

\author{Chun-Jiang Shi}
\affiliation{Institute of High Energy Physics, Chinese Academy of Sciences, Beijing 100049, China}

\author{Peng Sun}
\affiliation{Institute of Modern Physics, Chinese Academy of Sciences, Lanzhou, 730000, China}

\author{Wei Wang}
\affiliation{INPAC, Shanghai Key Laboratory for Particle Physics and Cosmology, Key Laboratory for Particle Astrophysics and Cosmology (MOE), School of Physics and Astronomy, Shanghai Jiao Tong University, Shanghai 200240, China}
\affiliation{Southern Center for Nuclear-Science Theory (SCNT), Institute of Modern Physics, Chinese Academy of Sciences, Huizhou 516000, Guangdong Province, China}

\author{Yi-Bo Yang}
\email{Corresponding author: ybyang@itp.ac.cn}
\affiliation{University of Chinese Academy of Sciences, School of Physical Sciences, Beijing 100049, China}
\affiliation{CAS Key Laboratory of Theoretical Physics, Institute of Theoretical Physics, Chinese Academy of Sciences, Beijing 100190, China}
\affiliation{School of Fundamental Physics and Mathematical Sciences, Hangzhou Institute for Advanced Study, UCAS, Hangzhou 310024, China}
\affiliation{International Centre for Theoretical Physics Asia-Pacific, Beijing/Hangzhou, China}

\author{Jian-Hui Zhang}
\affiliation{School of Science and Engineering, The Chinese University of Hong Kong, Shenzhen 518172, China}

\date{\today}

\begin{abstract}
We analyze the gauge fixing precision dependence of some non-local quark-blinear lattice operators interesting in computing parton physics for several measurements, using 5 lattice spacings ranging from 0.032 fm to 0.121 fm. Our results show that gauge dependent non-local measurements are significantly more sensitive to the precision of gauge fixing than anticipated. The impact of imprecise gauge fixing is significant for fine lattices and long distances. For instance, even with the typically defined precision of Landau gauge fixing of $10^{-8}$, the deviation caused by imprecise gauge fixing can reach 12 percent, when calculating the trace of Wilson lines at 1.2 fm with a lattice spacing of approximately 0.03 fm. Similar behavior has been observed in $\xi$ gauge and Coulomb gauge as well. For both quasi PDFs and quasi TMD-PDFs operators renormalized using the RI/MOM scheme, convergence for different lattice spacings at long distance is only observed when the precision of Landau gauge fixing is sufficiently high. To describe these findings quantitatively, we propose an empirical formula to estimate the required precision. 
\end{abstract}

\maketitle

\section{Introduction}

In gauge field theories, the necessity of gauge fixing arises originally from perturbation theory. If we just focus on non-perturbative lattice gauge theory and gauge invariant quantities, we do not really need gauge fixing. Also, any gauge dependent quantity will vanish if averaged over a sufficient number of configurations. 
However, in certain non-perturbative renormalization schemes~\cite{Martinelli:1994ty}, it is unavoidable to use gauge-dependent parton matrix elements to renormalize composite operators in the Landau or other gauges, and then match the results to the $\overline{\text{MS}}$ scheme perturbatively. 
Furthermore, non-perturbative lattice gauge fixing is also necessary to extract information from gauge dependent correlators~\cite{Mandula:1999nj}.

In continuum gauge theory, the gauge potentials $A_{\mu}$ are the fundamental degrees of freedom, and Wilson lines are defined through $A_{\mu}$ as 
\begin{align}\label{eq:Line}
U(x,y)\equiv{\cal P}{\rm exp}\left[ ig_0\int_{y}^{x} \textrm{d}z^{\mu}A^m_{\mu}(z)T^m\right]
.
\end{align}
where $T^m$ are the $\mathrm{SU}(3)$ generators in the fundamental representation, index $m$ is summed over and $g_0$ is the bare coupling constant. 

On the other hand, Wilson link $U_{\mu}(x) = U(x,x+a\hat{\mu})$ are treated as the fundamental degrees of freedom in lattice gauge theory~\cite{Wilson:1974sk}, where $a$ is { the} lattice spacing, and $\hat{\mu}$ is the unit vector in $\mu$ direction. The gauge potential $A_{\mu}$ can be extracted by short Wilson lines up to discretization effects,
\begin{align}\label{eq:A}
A_{\mu}(x+\frac{a}{2}\hat{\mu}) 
&\equiv \bigg[\frac{U_{\mu}(x)-U_{\mu}^{\dagger}(x)}{2ig_0a}\bigg]_\text{Traceless}.
\end{align}
where $g_0$ is the bare coupling at lattice spacing $a$.

Based on the gauge transformation $U^{G}_{\mu}(x) = G(x)U_{\mu}(x)G^{\dagger}(x+a\hat{\mu})$, the discretized gauge condition for the Landau and Coulomb gauge {can be defined as} ($l=4$ for Landau gauge and $l=3$ for Coulomb gauge),
\begin{align}\label{eq:delta}
\Delta^G(x) \equiv &\sum^l_{\mu=1}\big[U^G_{\mu}(x)-U^G_{\mu}(x-a\hat{\mu})-(U^G_{\mu})^{\dagger}(x)\nonumber\\
&+(U^G_{\mu})^{\dagger}(x-a\hat{\mu})\big]/(2ig_0)=0
\end{align}
Above condition corresponds to continuum gauge fixing condition $\sum_{\mu=1}^l\partial_{\mu}A^G_{\mu}=0$ with $A^G_{\mu}=GA^G_{\mu}G^{\dagger}+\frac{1}{ig_0}G\partial_{\mu}G^{\dagger}$, up to discretization errors.

The standard way to fix the Landau and Coulomb gauges in lattice QCD~\cite{Cabibbo:1982zn,Mandula:1987rh,Katz:1987ti} is based on the numerical minimization of the functional
\begin{align}\label{eq:minimization}
F_U[G] \equiv -\frac{1}{3lV}\text{Re}\text{ Tr}\sum_x \sum^l_{\mu=1}U_{\mu}^{G}(x)
,
\end{align}
where $V$ is the lattice volume. The extremum of $F_U[G]$ corresponds to  the discretized gauge condition $\Delta^G(x)=0$ up to discretization errors~\cite{Giusti:2001xf}.

Numerical non-perturbative gauge fixing is never perfect. Therefore, it is necessary to 
introduce a definition for the precision of gauge fixing. Since the latter is unphysical, it allows for a wide range of choices. For an appropriate definition of precision, as it approaches zero, the numerical gauge fixing should approach perfect gauge fixing (here we ignore some subtle issues like Gribov copies). And we expect that most of the gauge dependent measurements we are interested in for our lattice simulation will behave monotonically with this precision. Then we require a certain precision and make sure that the impact of imprecise gauge fixing will not cause a significant systematic deviation compared to the statistical uncertainty and other systematic uncertainties. There are two popular criteria to estimate the quality of gauge fixing. One is defined as,
\begin{align}\label{eq:theta}
\theta^G \equiv \frac{1}{3V}\sum_x\text{Tr}[(\Delta^G(x))^{\dagger}(\Delta^G(x))]
.
\end{align}
Obviously, when $\theta^G$=0, the discretized gauge condition $\Delta^G(x)=0$ is met. The other is determined by the change in the functional $F_U$ during the iterative gauge fixing procedure, given by the equation:
\begin{align}\label{eq:deltaF}
\delta^F(n)=F_U[G(n)]-F_U[G(n-1)],
\end{align}
where $G(n)$ represents the gauge rotation at the $n$-th step. The gauge fixing  will stop at the $m$-th step once $\delta^F(m)$ is smaller than the preassigned value $\delta^F$. When $\delta^F$=0, the functional $F_U[G]$ reaches its extremum, and the discretized gauge condition $\Delta^G(x)=0$ is fulfilled.
To the best of our knowledge, $\delta^F$ is the precision typically defined in Landau gauge fixing codes. The magnitude of $\theta^G$ is proportional to that of $\delta^F$ { as shown in the appendix, but
larger by a factor of $\sim$20}. 
In the following, we will just show the preassigned value of $\delta^F$ as the precision of gauge fixing, except for { the $\xi$ gauge}. The relationship between them has also been explored in other works, such as in Ref.~\cite{Giusti:2001xf}. 

For local quark bi-linear operators, $\delta^F\sim 10^{-7}$ is small enough given the other uncertainty which are at the 0.1\% level or larger~\cite{Lytle:2018evc}. In this work, our interest is gauge-fixing precision study for non-local quark bilinear operators. We find that non-local gauge dependent measurements are much more sensitive to the precision of gauge fixing than local ones. This has important implications for quasi-PDF and related matrix elements renormalization in RI/MOM scheme or fixed-gauge
quark bilinear non-local operator calculations. 

This paper is organized as follows. In {Sec.~\ref{sec:line}}, we will present the numerical setup, investigate the impact of imprecise gauge fixing on longer Wilson lines, and propose an empirical formula to describe the impact of imprecise gauge fixing. In the next two sections, we will show this impact
on LaMET~\cite{Ji:2013dva,Ji:2020ect} lattice simulations.  Quasi PDFs and Coulomb gauge quasi PDFs~\cite{Gao:2023lny} are shown in {Sec.~\ref{sec:pdf}} while quasi TMD-PDFs are shown in Sec~\ref{sec:tmd}. The last section will provide a brief summary.

\section{Numerical setup and Wilson lines}\label{sec:line}

{In this calcualtion}, We use the 2+1+1 flavors (degenerate up and down, strange, and charm degrees of freedom) of highly improved staggered quarks (HISQ) and one-loop Symanzik improved~\cite{Hart:2008sq} gauge ensembles from the MILC Collaboration~\cite{Bazavov:2012xda} at five lattice spacings. We tuned the valence light quark mass {using the HYP smeared tadpole improved clover action} to be around that in the sea, and the information about the ensembles and parameters we used are listed in Table~\ref{tab:milc}. 

\begin{table}[htbp]
  \centering
  \begin{tabular}{|c||ccrc|cc|}
\hline
Tag &  $6/g_0^2$ & $L$ & $T$ & $a(\mathrm{fm})$ & $m^{\textrm{w}}_{q}a$ & $c_{\mathrm{sw}}$ \\
\hline
a12m310 &  3.60 & 24 & 64 & 0.1213(9)    & -0.0695 & 1.0509\\
\hline
a09m310 &  3.78 & 32 & 96 & 0.0882(7)  & -0.0514 & 1.0424 \\
\hline
a06m310 &  4.03 & 48 & 144 & 0.0574(5)  & -0.0398 & 1.0349 \\
\hline
a045m310 &  4.20 & 64 & 192 & 0.0425(5)  & -0.0365 & 1.0314 \\
\hline
a03m310 &  4.37 & 96 & 288 & 0.0318(5)  & -0.0333 & 1.0287 \\
\hline
\end{tabular}
  \caption{Setup of the ensembles, including the bare coupling constant $g_0$, lattice size $L^3\times T$ and lattice spacing $a$. $m^\textrm{w}_q$ is the bare quark masses. The pion masses in the sea are $\sim$310 MeV and the valence pion mass are in the range of 280-320 MeV.}
  \label{tab:milc}
\end{table}

The gauge fixing procedure used in this work is proposed in Ref.~\cite{Cabibbo:1982zn} and can be briefly described as follows~\cite{Edwards:2004sx}:

1. Separate the lattice sites into the even ($\mathrm{mod}(\sum_{i=1}^4x_i,2)=0$) and odd ($\mathrm{mod}(\sum_{i=1}^4x_i,2)=1$) sectors;

2. On each site $x$ of the even sector, calculate $K(x)\equiv \sum_{\mu=1}^{l}\big(U_{\mu}(x)+U_{\mu}^{\dagger}(x-a\hat{\mu})\big)$, obtain the projected SU(2) matrix $a(x)$ from a given 2x2 sub-matrix of $K(x)$, and then apply $a^{\dagger}(x)$ on the left hand side of $U$ and $K$. One can apply the over-relaxation algorithm~\cite{Adler:1981sn,Giusti:2001xf} by replacing $a\equiv a_0+\sigma_i a_i$ into
\begin{align}
a'=\mathrm{cos}(\omega \theta)+\sigma_i a_i*\mathrm{sin}(\omega \theta)/\mathrm{sin}(\theta),
\end{align}
where $\theta=\mathrm{cos}^{-1}a_0$, and $1<\omega<2$ to accelerate the convergence.

3. Repeat the step 2 on all the other 2x2 sub-matrices of $K(x)$ and apply them to $U$ and $K$.

4. Repeat the steps 2-3 for the odd sector.

5. Compute $F_U$, and iterate through steps 2-4 until the change in $F_U$, denoted by $\delta^F$, is less than the required gauge fixing precision.

To get a clear image of imprecise gauge fixing, we choose to look at the ratio of the gauge dependent quantity $X$ under two gauge fixing with precisions $\delta^F$ and $\delta^F_0$,
\begin{align}\label{eq:rGx}
r^{\delta^F}(X) \equiv X^{\delta^F}/X^{\delta^F_0}.
\end{align}
where $\delta^F_0$ is chosen to be ``sufficiently small" to ensure that $X^{\delta^F_0}$ remains unchanged within the statistical uncertainty. In this definition, $r^{\delta^F}(X)$ can deviate from 1 when $\delta^F\gg \delta^F_0$.

Obtaining higher precision in numerical gauge fixing necessitates more iteration steps, resulting in increased computational { cost}, especially for larger lattices such as a03m310. Consequently, in practice, it is necessary to select a specific precision that strikes a balance between systematic error and computational cost. In this study, we opt for the gauge rotation $G_0$ to ensure that the gauge fixing can attain the precision $\delta^F = 10^{-12}$ in most cases. Subsequent discussions will demonstrate the reasonableness of this choice.

\subsection{Straight Wilson line under Landau gauge}

We start with the most straightforward gauge dependent { quantities in lattice simulations, the Wilson lines.} They are defined in Eq.~(\ref{eq:Line}), and can also be expressed as,
\begin{align}\label{eq:LatticeLine}
U(x,x+na\hat{\mu}) \equiv \prod_{k=0}^{n-1}U_{\mu}(x+ka\hat{\mu}),n>0.
\end{align}
They are referred to as long Wilson lines {when $n>1$}. They are gauge dependent and transform under a gauge transformation similar to the short line,
\begin{align}\label{eq:gaugeRot2}
U^{G}(x,x+na\hat{\mu}) = G(x)U(x,x+na\hat{\mu})G^{\dagger}(x+na\hat{\mu}).
\end{align}

\begin{figure}[tbph]
\centering
\includegraphics[width=7cm]{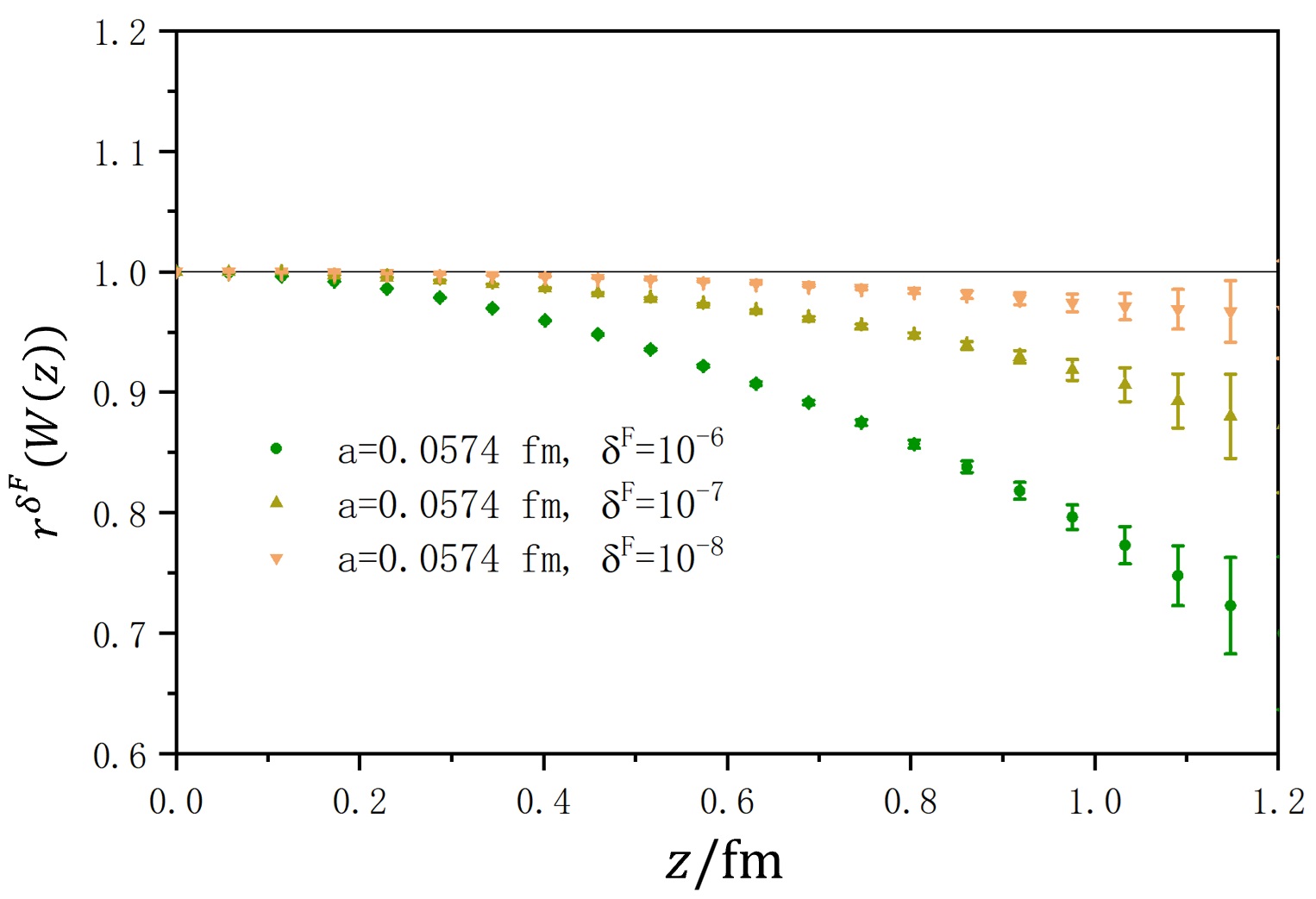}
\includegraphics[width=7cm]{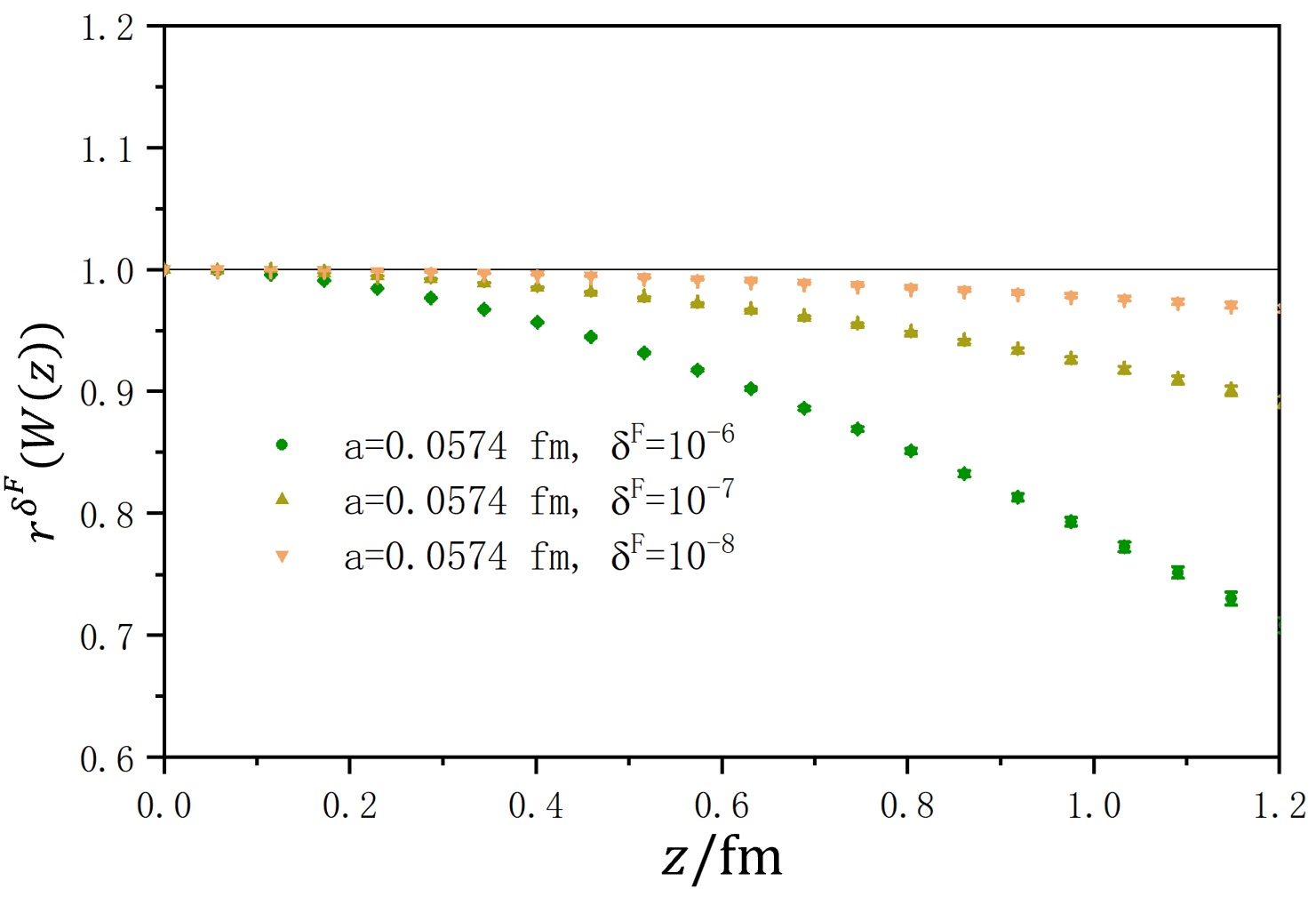}
\includegraphics[width=7cm]{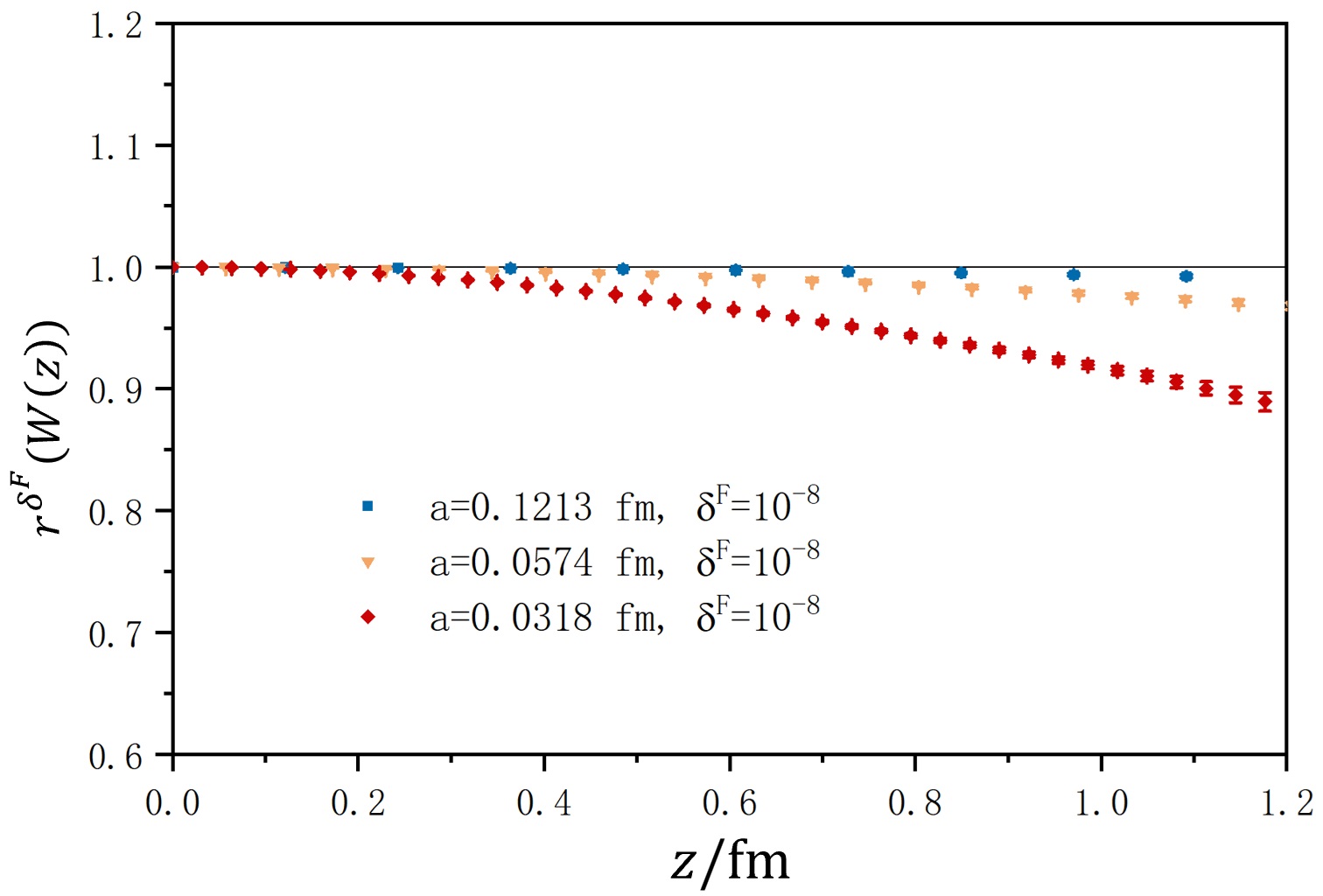}
\caption{The ratio $r^{\delta^F}(W(z))$ defined in Eq.~(\ref{eq:rGx}) of the trace of Wilson lines defined in Eq.~(\ref{eq:TrLine}) for different precisions, with one-step HYP smearing after gauge fixing (upper and middle panels) and without HYP smearing (lower panel) on different ensembles. The result at $\delta^F_0 = 10^{-12}$ is chosen as the denominator. The impact of imprecise gauge fixing is significant at long distance on fine lattice.}
\label{fig:WilsonLine}
\end{figure}

As shown in the upper and middle panels of Fig.~\ref{fig:WilsonLine}, at short distance, the ratio $r^{\delta^F}(W(z))$ of the trace of Wilson line with given length $z$
\begin{align}\label{eq:TrLine}
W(z)\equiv\frac{1}{3}\text{Tr}\left[\sum_x U(x,x+z)\right],
\end{align}
depends only weakly on precision and lattice spacing. But at long distance, the ratio { $r^{\delta^F}(W(z))$  decays} more rapidly for smaller lattice spacings and lower precisions. For example, on a03m310, at $z =$ 1.2~fm, the difference between $\delta^F = 10^{-8}$ and $\delta^F = 10^{-12}$ is 11.6(9) percent. This indicates that even with a high precision of $\delta^F = 10^{-8}$, the Wilson lines remain very sensitive to the precision of gauge fixing, especially for small lattice spacings at long distance. {As for HYP smearing, it enhances the signal but does not result in {additional} systematic { deviations} as shown in the lower panels of Fig.~\ref{fig:WilsonLine} using the a06m310 ensemble at $a\sim 0.06~$fm, in comparison to the case without HYP smearing shown in the middle panel of the same figure.}

To describe and estimate the impact of imprecise gauge fixing, we propose {the following empirical formula 

\begin{align}\label{eq:empirical}
X(\delta^F) = X(0) e^{-c(X)(\delta^F)^{n(X)}},
\end{align}
where $c$ and $n$ are the parameters to be fitted and $X(0)$ corresponds to the observable $X$ using precise enough $G_0(x)$ (for example, $\delta^F=10^{-15}$).

\begin{figure}[tbph]
\centering
\includegraphics[width=7cm]{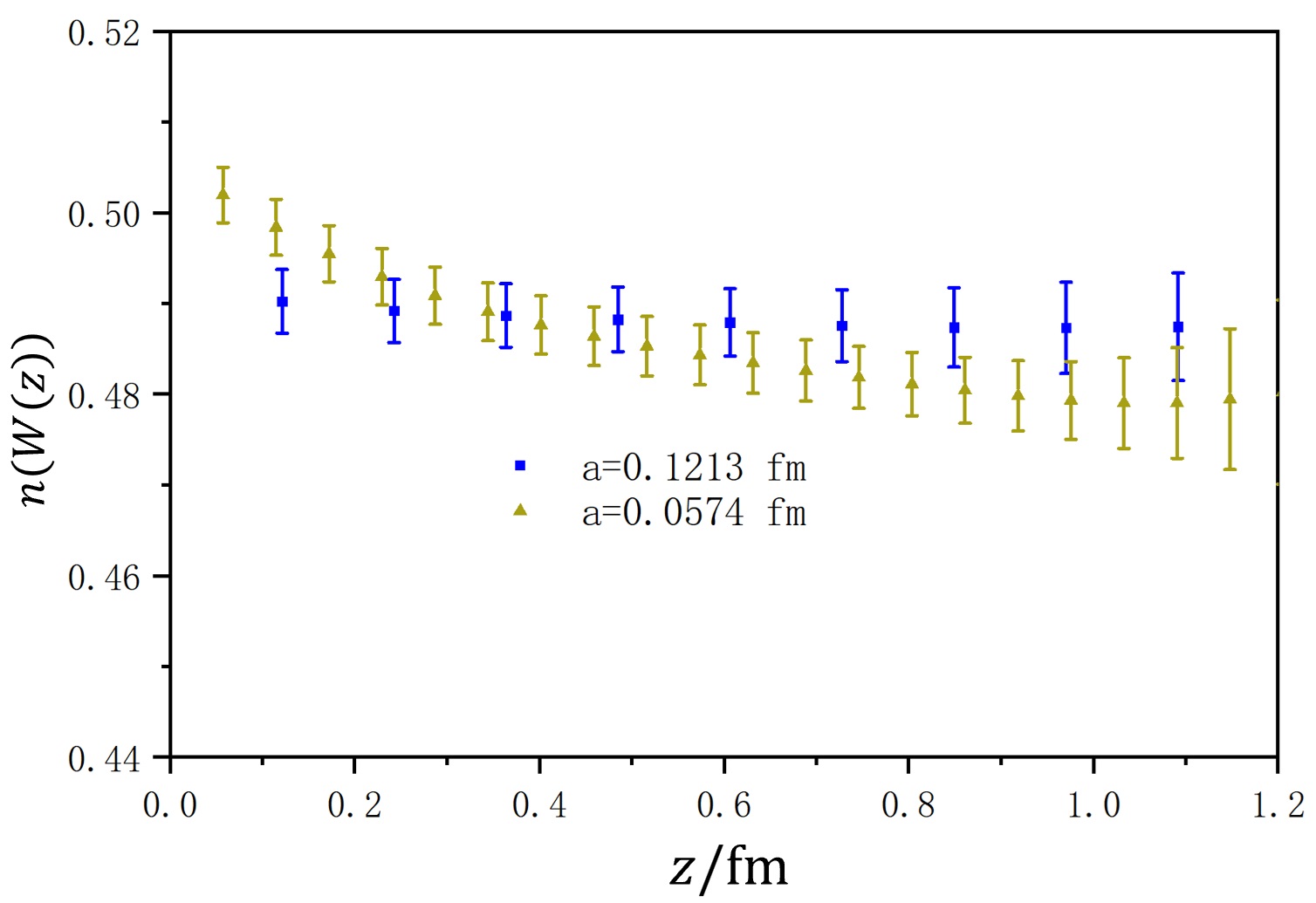}
\includegraphics[width=7cm]{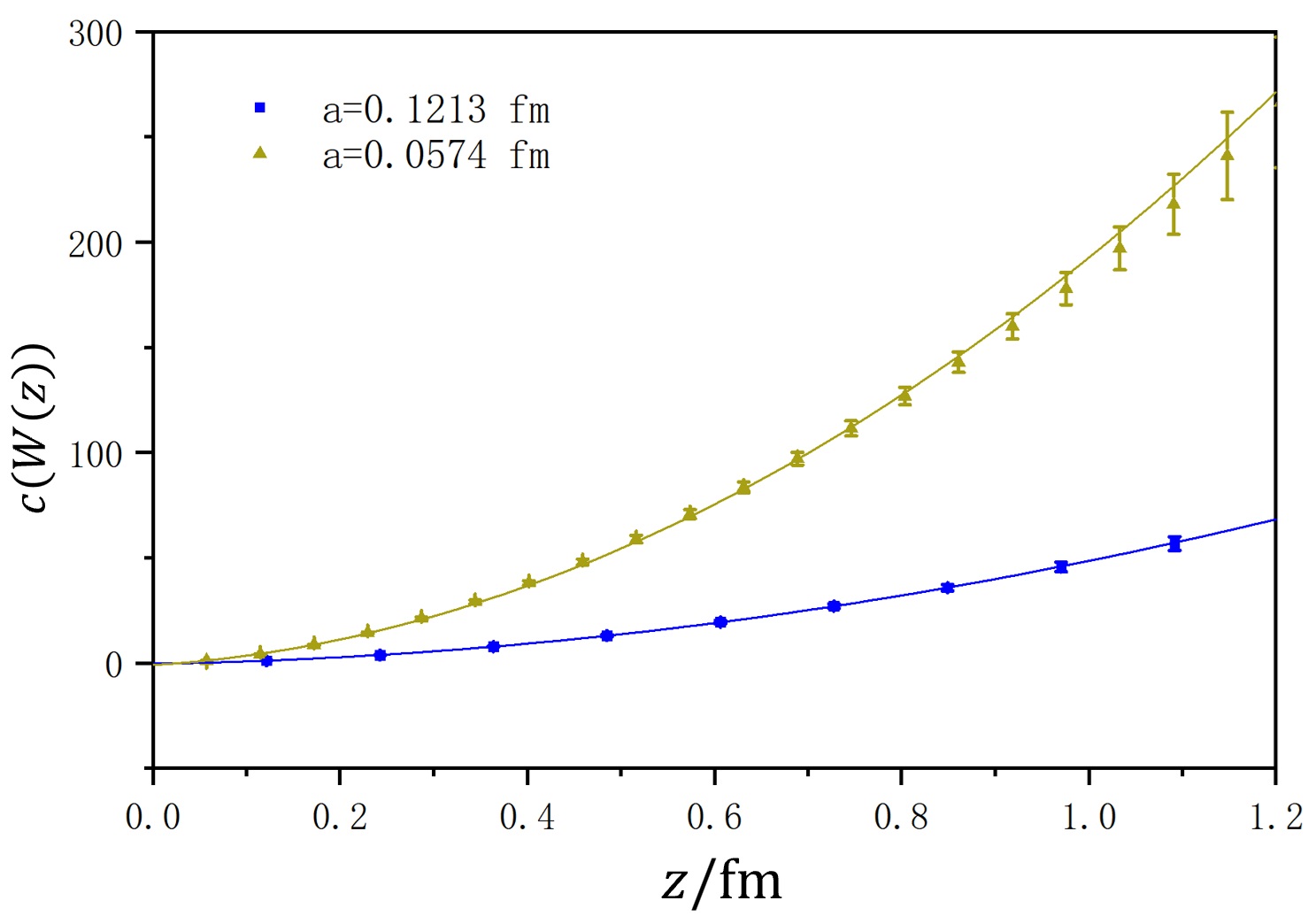}
\includegraphics[width=7cm]{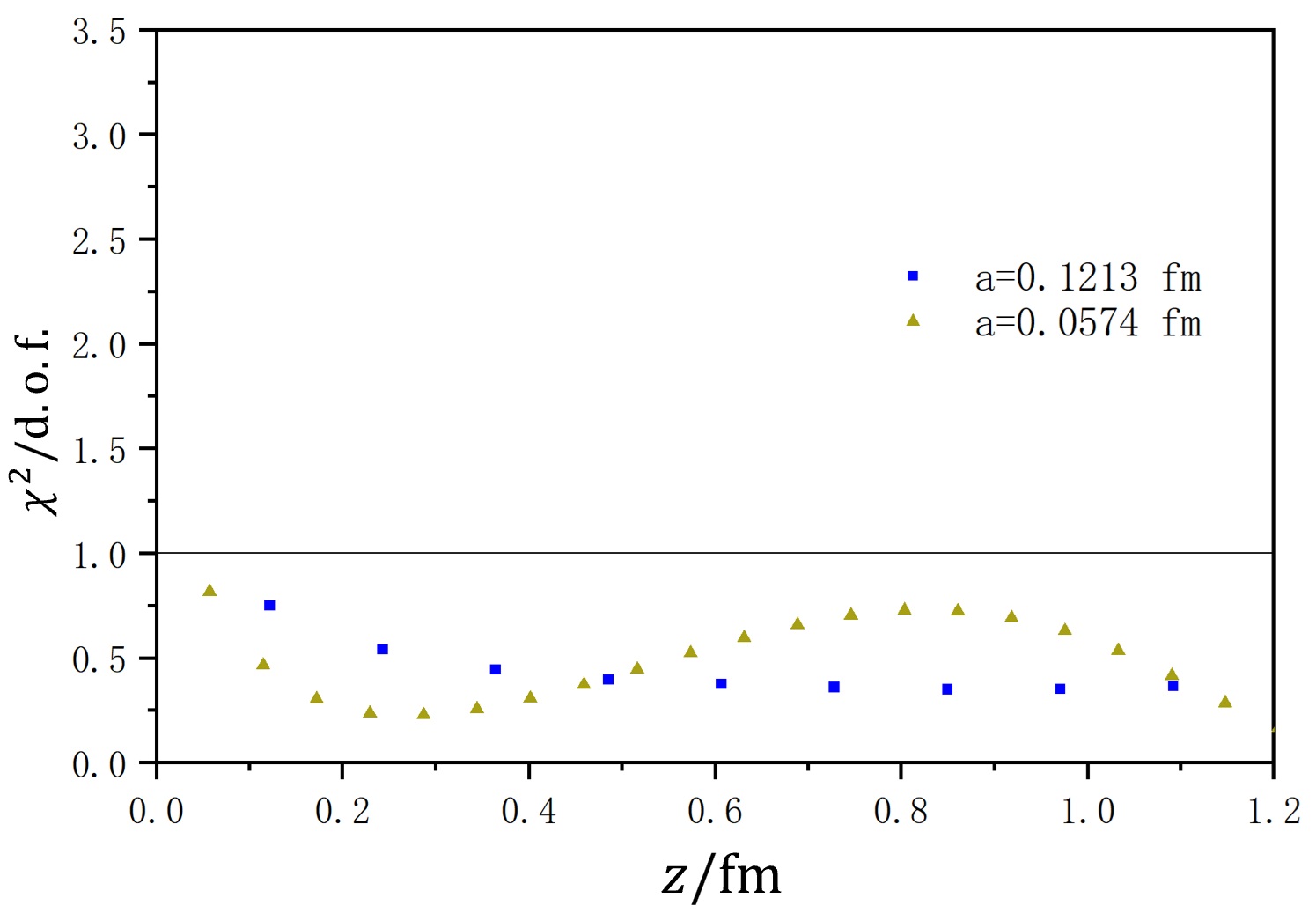}
\caption{
The parameters $n(W(z))$ (top panel) and $c(W(z))$ (middle panel) determined from the fitting ansatz Eq.~(\ref{eq:empirical}) for the trace of Wilson lines defined in Eq.~(\ref{eq:TrLine}) at different $z$, and also the corresponding $\chi^2/$d.o.f. (bottom panel) for a12m310 and a06m310 ensembles.
We use data from gauge fixings with $\delta^F = 10^{-4},10^{-5},10^{-6},10^{-7},10^{-8}$.
One-step HYP smearing has been applied after gauge fixing. The uncertainty is got from bootstrap resampling. The two curves in the middle panel are the fits of $c(W(z))$ { using a polynomial form}.}
\label{fig:empirical}
\end{figure}

As shown in the upper and middle panels of Fig.~\ref{fig:empirical}, $n(W(z))$ is approximately 0.5 and does not show a very strong dependence on $z$ and lattice spacing. Conversely, $c(W(z))$ is much more sensitive to $z$ and lattice spacing. The lower panel indicates that the $\chi^2/$d.o.f. is sufficiently small. Thus, this empirical formula adequately captures the impact of imprecise gauge fixing in the Wilson line case.

The impact of imprecise gauge fixing increases as the Wilson line becomes longer, and it appears that this is primarily due to $c(W(z))$. Therefore, we fit $c(W(z))$ and attempt to model it using a quadratic function $c(W(z)) = c_2z^2+c_1z+c_0$ as shown in the middle panel in Fig.~\ref{fig:empirical}. { For} a12m310, we obtain $c_2=42(3)$, $c_1=7.2(1.3)$, $c_0=-0.38(0.12)$, and $\chi^2/$d.o.f.=0.06. { For} a06m310, the values are $c_2=166(8)$, $c_1=27.6(1.9)$, $c_0=-0.97(0.09)$, and $\chi^2/$d.o.f.=0.51. This result suggests that $c(W(z))$ will increase quadratically as the Wilson line lengthens, allowing us to use a quadratic function to forecast $c(W(z))$ and assess the effect of imprecise gauge fixing. Interestingly, it appears that the coefficient $c_2$ is proportional to $1/a^2$, and this term will become dominant when the Wilson line is sufficiently long.

The Wilson lines with gauge rotations $G_1(x)$ and $G_2(x)$ can be related by the following expression,
\begin{align}\label{eq:RotConver}
&U^{G_1}(x,x+z) \nonumber\\
&= G_1(x)G_2^{-1}(x)U^{G_2}(x,x+z)G_2(x+z)G_1^{-1}(x+z) \nonumber\\
&= [G_1(x)G^{-1}_2(x)]U^{G_2}(x,x+z)[G_1(x+z)G^{-1}_2(x+z)]^{-1},
\end{align}
where $U^G(x,x+z)\equiv G(x)U(x,x+z)G^{\dagger}(x+z)$ and $U^{G_1}$ and $U^{G_2}$ just differ by the relative gauge rotations at both ends of the Wilson line. We can then use the correlation between the relative gauge rotation $\tilde{G}\equiv GG_0^{-1}$ between the gauge fixing with different precision,
\begin{align}\label{eq:RotCorr}
C^G(z) = \frac{1}{3V}\sum_x\text{Tr}\left[[G(x+z)G_0^{-1}(x+z)]^{\dagger}[G(x)G_0^{-1}(x)]\right],
\end{align}
to gain insights into why $W(z)$ depends on the gauge fixing precision. $G$ and $G_0$ are the gauge rotations with certain gauge fixing precision $\delta^F$ and $\delta_0^F=10^{-12}\ll \delta^F$.

\begin{figure}[tbph]
\centering
\includegraphics[width=8cm]{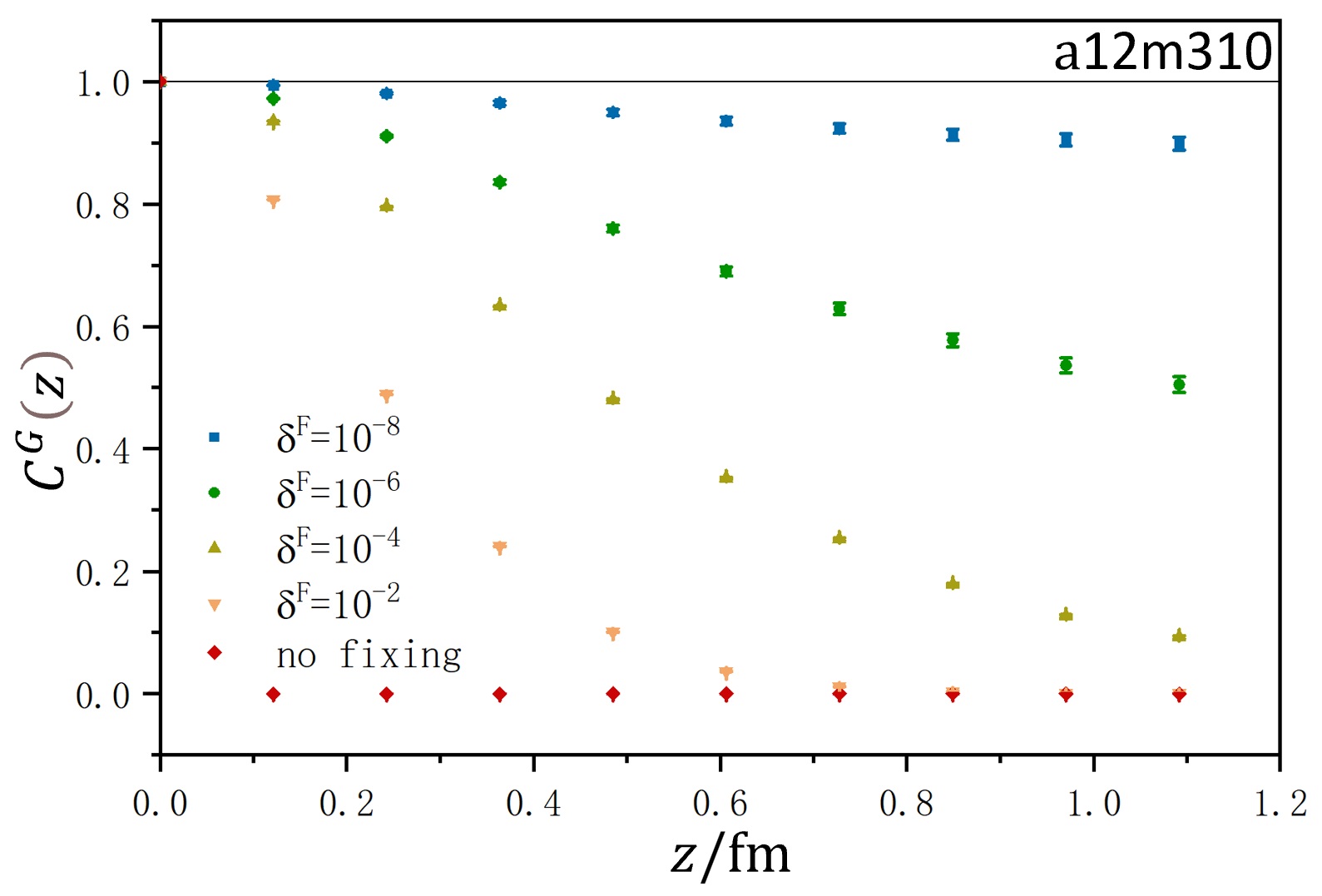}
\includegraphics[width=8cm]{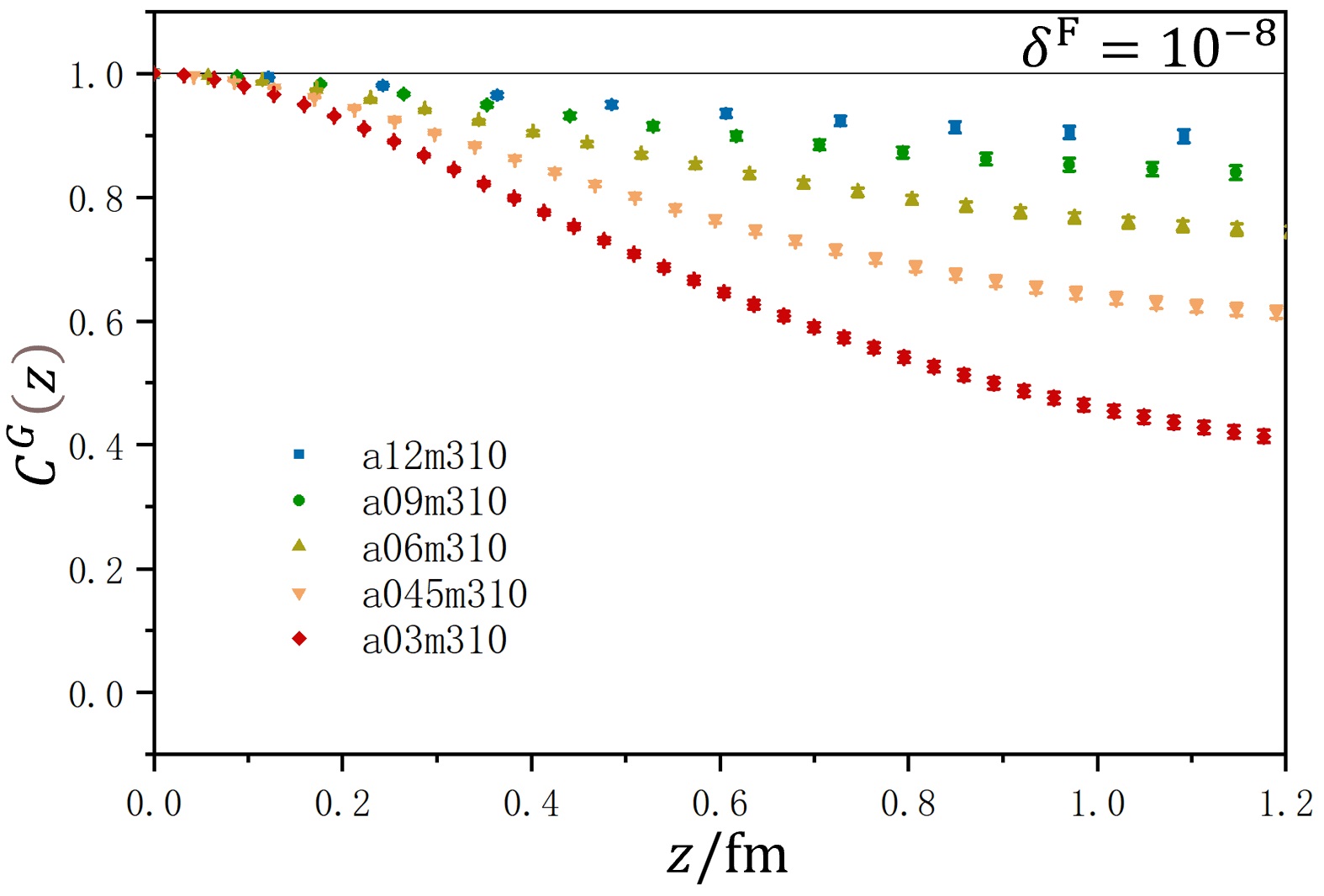}
\caption{The correlation of the relative gauge rotation $\tilde{G}\equiv GG^{-1}_0$ defined in Eq.~(\ref{eq:RotCorr}) as function of the spatial distance $z$ for different gauge fixing precision $\delta^F$ and different ensembles. The gauge rotations $G$ and $G_0$ correspond to gauge fixing with precision $\delta^F$ and $\delta^F_0=10^{-12}\ll \delta^F$.}
\label{fig:RotCorr}
\end{figure}

As shown in the upper panel of Fig.~\ref{fig:RotCorr}, the correlation $C^G(z)$ approaches 1 when the gauge fixing is precise enough, with $\delta^F=\delta^F_0$. On the other hand, $C^G(z)$ deviates from 1 for larger $\delta^F$ and approaches 0 in the limit of no gauge fixing. Such a deviation becomes more significant at large z, while suffering from loop-around effects when z is approximately L/2. These observations are consistent with the idea that gauge fixing creates a correlation between different spatial positions, which will be stronger with more precise gauge fixing.

At the same time, $C^G(z)$ with a given $\delta^F$ but different lattice spacing $a$ also increases with $1/a$, as shown in the lower panel of Fig.~\ref{fig:RotCorr}. This is also understandable since the correlation would depend on the distance in lattice units instead of physical units.

\subsection{Straight Wilson line under $\xi$ gauge}

We also extend our study of Wilson lines from Landau gauge with $\xi=0$ to the general covariant $\xi$ gauge. The gauge condition  of $\xi$ gauge is defined as ~\cite{Giusti:1999im,Zwanziger:1985vi,Giusti:1996kf},
\begin{align}\label{eq:xiGC}
\Delta^G(x) \equiv \sum^4_{\mu}(A^G_{\mu}(x+\frac{a}{2}\hat{\mu})-A^G_{\mu}(x-\frac{a}{2}\hat{\mu})) = \Lambda(x),
\end{align}
where random matrix $\Lambda(x)$ on each lattice site $x$ of each configuration belonging to the $\mathrm{SU}(3)_c$ group are sampled according to the weight $e^{-\frac{1}{2\xi}\text{Tr}\Lambda^2(x)}$.
Note that the Landau gauge is recovered when $\Lambda(x)=0$ at every $x$. { The precision of gauge fixing for} the $\xi$ gauge is defined as,
\begin{align}\label{eq:xiPrec}
\theta^G_{\xi} \equiv \frac{1}{3V}\sum_x\text{Tr}[(\Delta^G(x)-\Lambda(x))^{\dagger}(\Delta^G(x)-\Lambda(x))].
\end{align}
We use this { expression} in $\xi$ gauge { instead of $\delta^F$} in this case.

\begin{figure}[tbph]
\centering
\includegraphics[width=7cm]{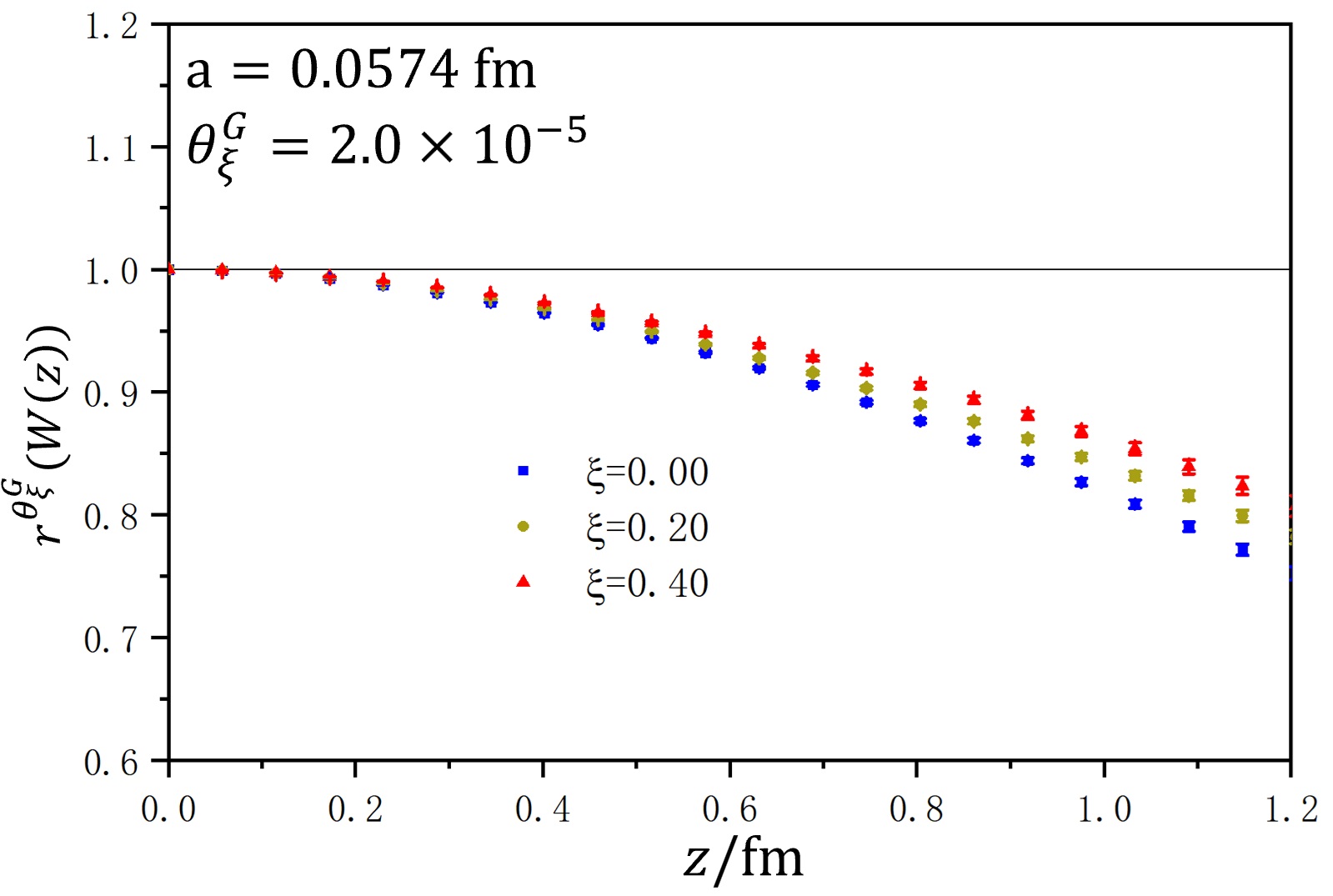}
\includegraphics[width=7cm]{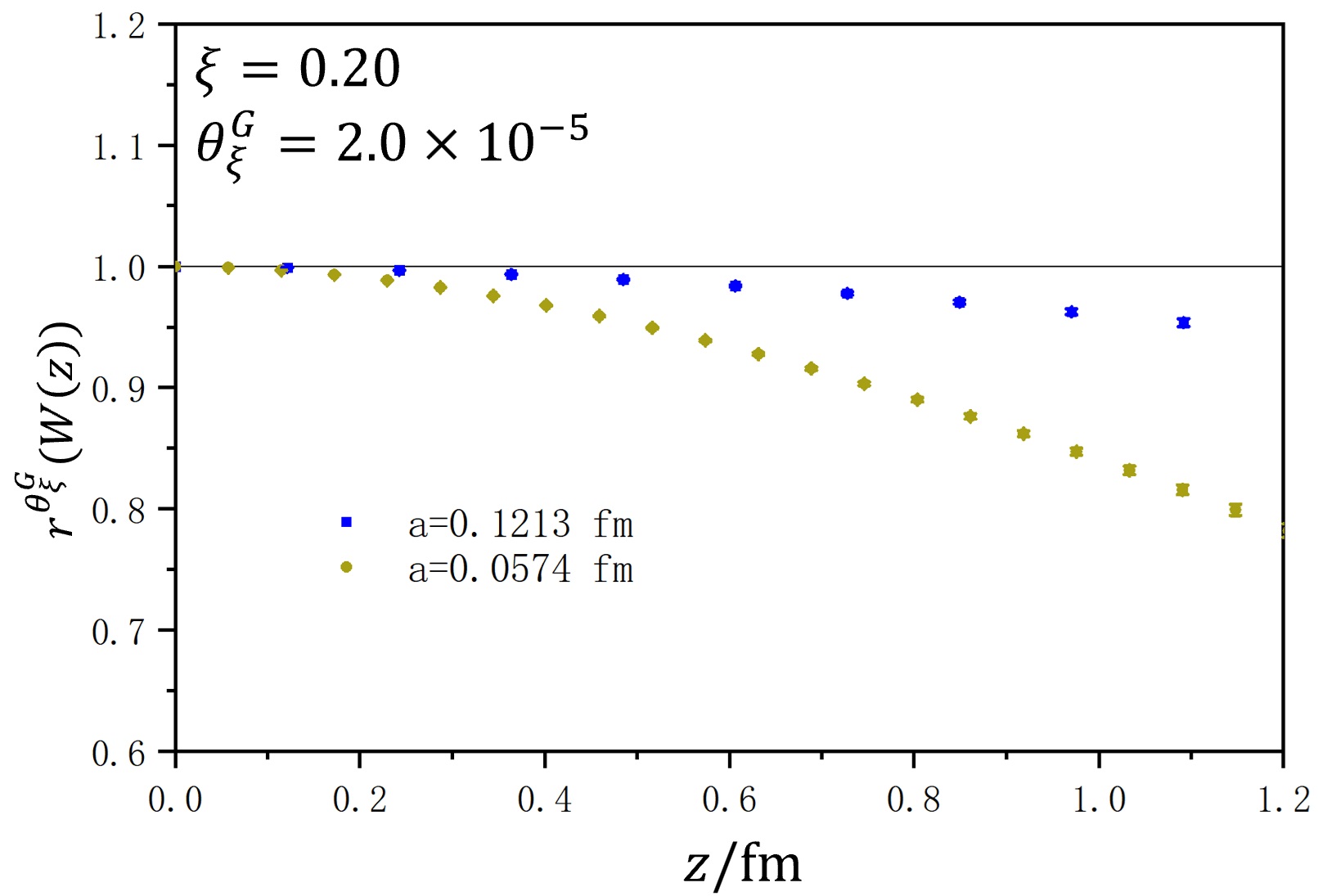}
\includegraphics[width=7cm]{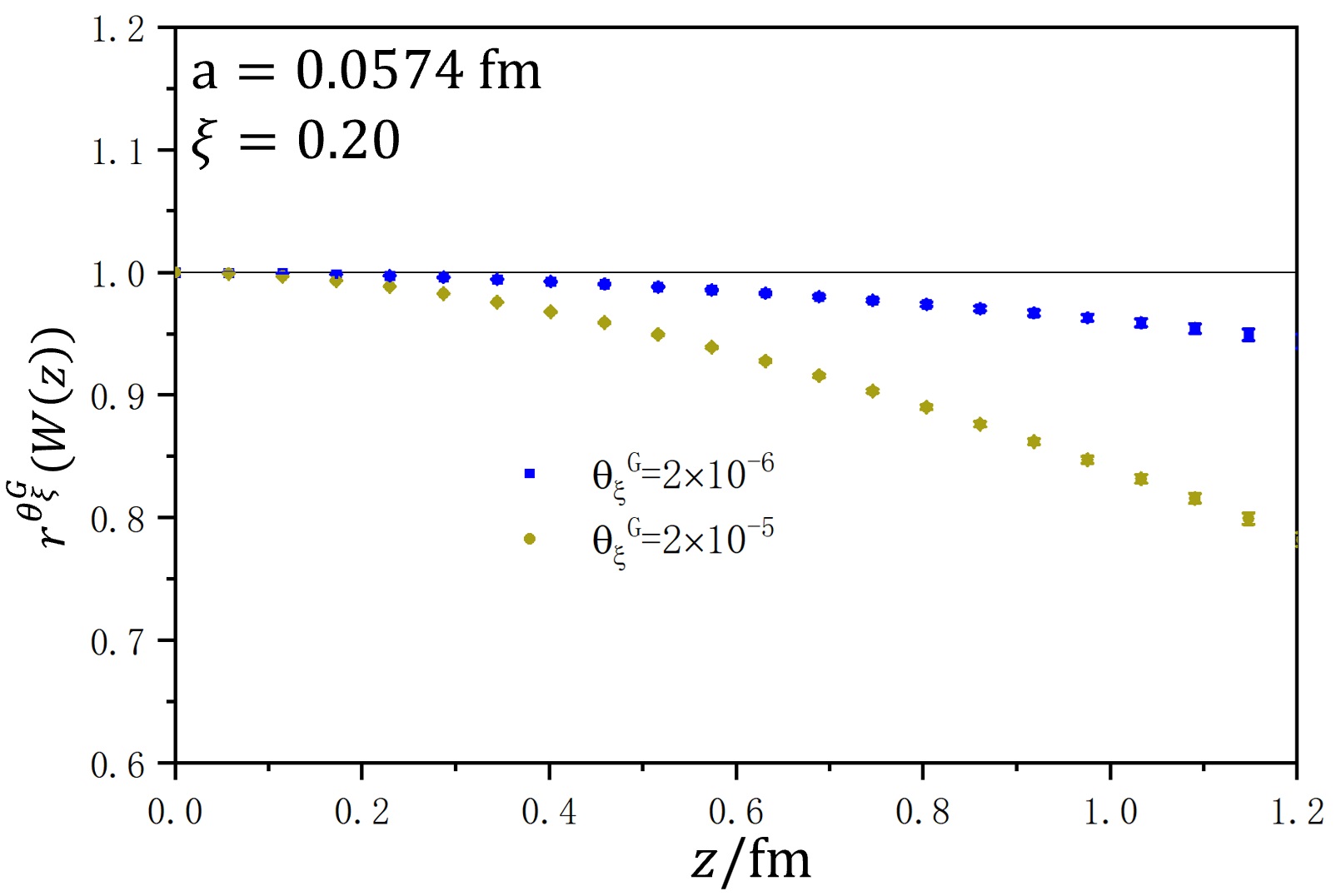}
\caption{The ratio $r^{\theta^{G}_{\xi}}(W(z))$ of the trace of Wilson lines defined in Eq.~(\ref{eq:TrLine}) for different precision with one-step HYP smearing. The result with the precision $\theta_{\xi,0}^{G} = 2\times10^{-7}$ is used as the denominator in the ratio. The impact of imprecise gauge fixing is similar to that in Landau gauge.}
\label{fig:xi}
\end{figure}

As we did for Landau gauge, we plot the ratio 

\begin{align}\label{eq:rGz}
r^{\theta^G_{\xi}}(W) \equiv W^{\theta^G_{\xi}}/W^{\theta^{G}_{\xi,0}}
\end{align}
of the traces of Wilson lines defined in Eq.~(\ref{eq:TrLine}) for different precision of gauge fixing, with $\theta^{G}_{\xi,0}= 2\times10^{-7}$ since it is very hard to reach higher precision here. 
The numerical gauge fixing requires more iteration steps to converge when $\xi$ becomes larger. Additionally, some configurations can not converge within 100,000 iterations if the required precision is too high, particularly for large $\xi$. 

As shown in the upper panel of Fig.~\ref{fig:xi}, it seems that the impact of imprecise gauge fixing becomes slightly smaller as $\xi$ increases. The middle and lower panels show that the effects of lower required precision are larger on finer lattices, at longer distances and for lower average precision. This observation is consistent with our findings in Landau gauge.

\subsection{Staple shaped Wilson line under Landau gauge}

Staple shaped Wilson lines ${\cal W}(b,z,L)$ are also widely used gauge dependent measurement. For example, {a quasi TMD-PDF operator~\cite{Ji:2019sxk} contains a staple shaped Wilson line}. Staple shaped Wilson lines are defined as,

\begin{align}\label{eq:StapleLine}
{\cal W}(b,z,L)&\equiv {\cal P}{\rm exp} \left[ ig_0\int_{-L}^{z} \textrm{d}s\ \hat{n}_z\cdot A^m(b\hat{n}_{\perp}+s\hat{n}_z)T^m\right] \nonumber\\
& \times {\cal P}{\rm exp} \left[ ig_0\int_{0}^{b} \textrm{d}s\ \hat{n}_{\perp}\cdot  A^m(s\hat{n}_{\perp}-L\hat{n}_z)T^m\right] \nonumber\\
& \times {\cal P}{\rm exp} \left[ ig_0\int_{0}^{-L} \textrm{d}s\ \hat{n}_z\cdot  A^m(s\hat{n}_z)T^m\right]
,
\end{align}
where $b=|\vec{b}_{\perp}|$ and $z$ are the separations between the two endpoints of the staple shaped Wilson { line} along the transverse direction $\hat{n}_{\perp}$ and longitudinal direction $\hat{n}_z$, respectively. { Straight Wilson lines are recovered when $z=0, L=0$.} The trace is denoted as $W_{\rm st}(b,z,L)$ as in the Wilson line case.

\begin{figure}[tbph]
\centering
\includegraphics[width=7cm]{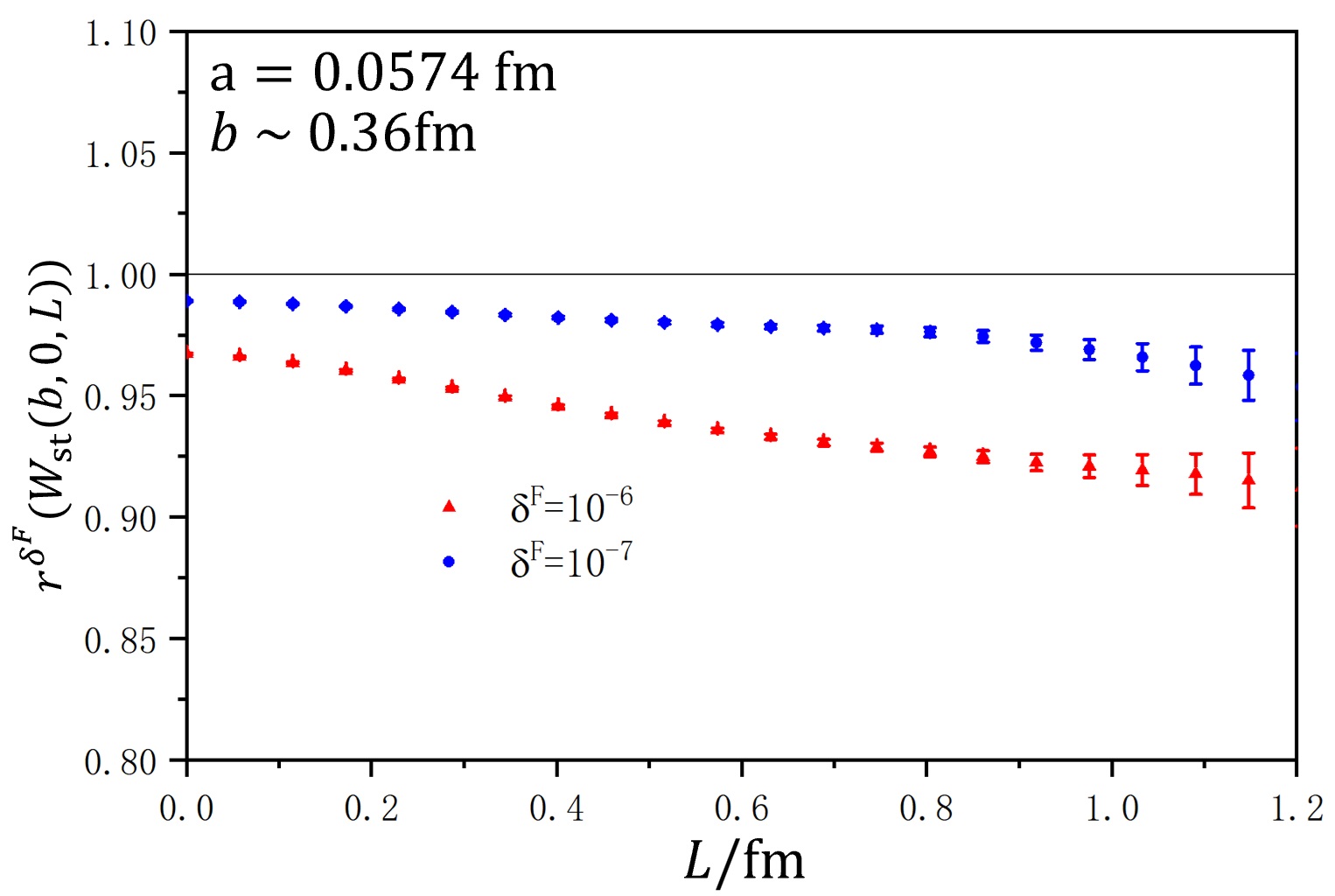}
\includegraphics[width=7cm]{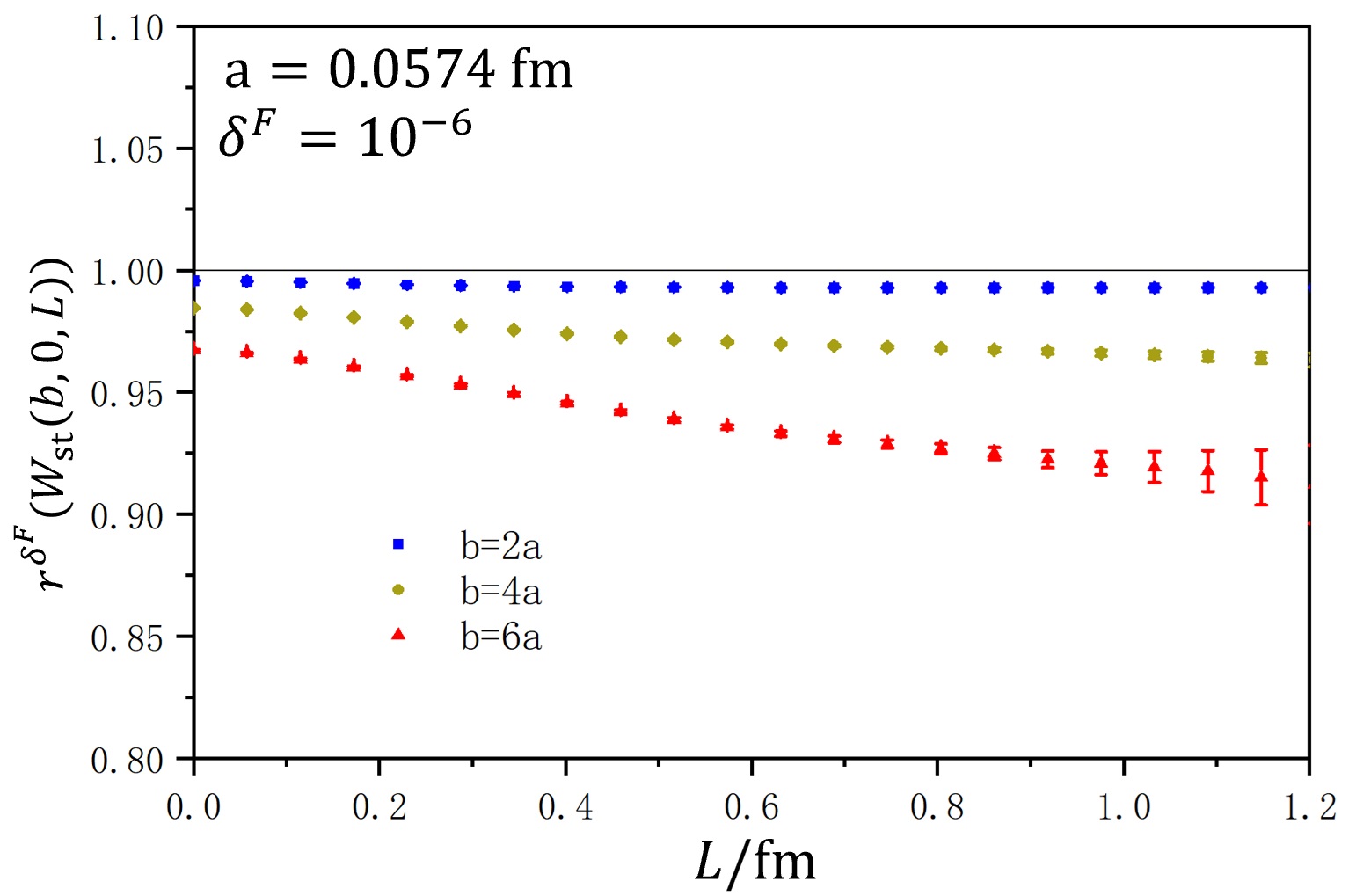}
\includegraphics[width=7cm]{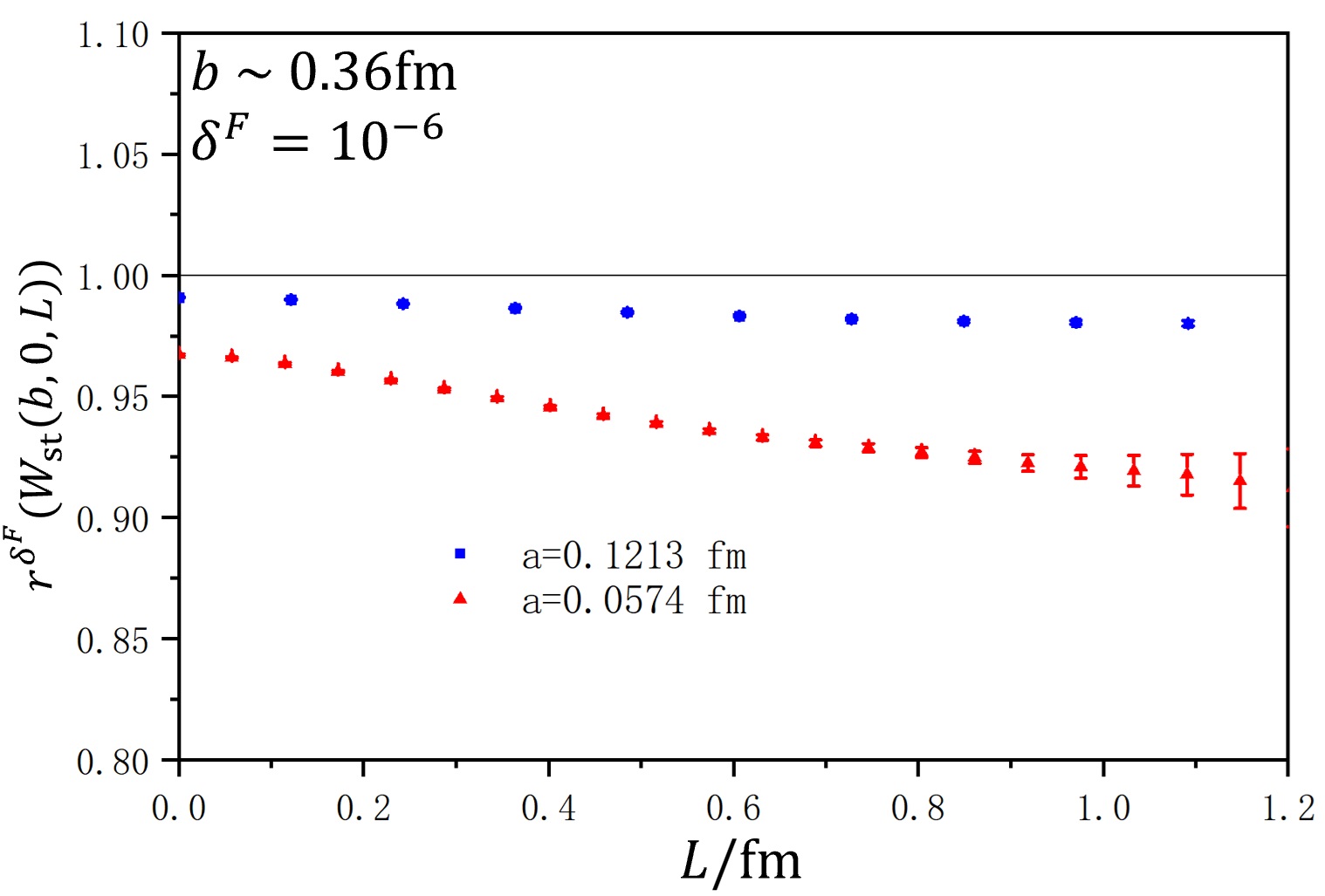}
\caption{The ratio  $r^{\delta^F}(W_{\rm st}(b,z,L))$ 
Eq.(\ref{eq:rGx}) of the trace of staple shaped Wilson lines defined in Eq.~(\ref{eq:StapleLine}) { for} different { precision} with one-step HYP smearing after gauge fixing. We use the $z=0$ case as an { illustration.} The result { for} $\delta^F_0 = 10^{-12}$ is chosen as the denominator. It is observed that the impact of imprecise gauge fixing depends on $L$, but this dependence is much weaker than that on $b$.}
\label{fig:StapleWilsonLine}
\end{figure}

When considering the staple shaped Wilson lines ${\cal W}(b,z,L)$ and its trace $W_{\rm st}(b,z,L)$ for different gauge fixing, the only difference lies in the gauge transformation $G(x)$ on the two endpoints. When $b$ and $z$ remain unchanged, the endpoints are fixed, but $L$ still has an influence on the impact of imprecise gauge fixing, as depicted in Fig.~\ref{fig:StapleWilsonLine}. When $L$ becomes longer, the impact of imprecise gauge fixing is enhanced as shown in the upper panel of Fig.~\ref{fig:StapleWilsonLine}. Furthermore, longer $b$ and smaller lattice spacing make this phenomena more pronounced, as shown in the middle and lower panels of Fig.~\ref{fig:StapleWilsonLine}. As the gauge fixing precision dependence is not very sensitive on $\xi$, we will not delve into the discussion of the staple-shaped Wilson line in the $\xi$ gauge.

\section{Quasi PDF}\label{sec:pdf}

{ A quasi-PDF operator is a quark bilinear operator with a} Wilson line,
\begin{align}\label{eq:QuasiPDF}
O^{\text{GI}}_{\Gamma}(z)=\bar{\psi}(0) \Gamma U(0,z) \psi (z)
.
\end{align}
Its renormalized nucleon matrix element is connected to PDFs through factorization theorems~\cite{Ji:2013dva,Ji:2014gla,Ma:2014jla,Ma:2017pxb,Izubuchi:2018srq}. Its hadron matrix element is
\begin{align}\label{eq:bare}
\tilde{h}^{\text{GI}}_{\chi,\Gamma}(z,P_z;1/a) \equiv\langle \chi(P_z)|O^{\text{GI}}_{\Gamma}(z)|\chi(P_z)\rangle
.
\end{align}
The UV divergence is independent of the hadron state, as shown in Refs.~\cite{Zhang:2020rsx,LatticePartonCollaborationLPC:2021xdx}. { In} the following we will concentrate on the pion matrix element $\tilde{h}^{\text{GI}}_{\pi}(z,P_z;1/a)\equiv\langle \pi(P_z)|O^{\text{GI}}_{\gamma_t}(z)|\pi(P_z)\rangle$ of the quasi PDF operator $O^{\text{GI}}_{\gamma_t}(z)$ {at different lattice spacing $a$} as an illustrative example. The pion matrix element can be extracted from the following ratio
\begin{align}\label{eq:ratio}
&R_{\pi}(t_1,z;a,t_2)\nonumber\\
&\equiv \frac{\langle O_\pi(t_2)\sum_{\vec{x}}O^{\text{GI}}_{\Gamma}(z;(\vec{x},t_1))O_{\pi}^{\dagger}(0)\rangle}{\langle O_\pi(t_2)O_{\pi}^{\dagger}(0)\rangle}
\nonumber\\
&=\langle \pi|O^{\text{GI}}_{\Gamma}(z)|\pi\rangle+{\cal O}(e^{-\Delta m t_1},e^{-\Delta m (t_2-t_1)},e^{-\Delta m t_2})
,
\end{align}
where $\Delta m$ is the mass gap between the pion and its first excited state which is aournd 1 GeV. One-step HYP smearing is applied { to} the Wilson lines to enhance the signal to noise ratio (SNR). The source/sink setup and also the ground state matrix element extraction are similar to that in Refs.~\cite{Zhang:2020rsx,LatticePartonCollaborationLPC:2021xdx,Zhang:2022xuw}.

\subsection{Quasi PDF renormalized using the Wilson line}

In lattice regularization, { one encounters a} linear divergence { of the} quasi PDF operator $O^{\text{GI}}_{\gamma_t}(z)$ defined in Eq.~(\ref{eq:QuasiPDF}). To ensure the existence of a finite continuum limit, it is necessary to remove this linear divergence. Studies in the continuum~\cite{Ji:2015jwa, Ji:2017oey,Ishikawa:2017faj,Green:2017xeu} suggest that the quasi-PDF operator is multiplicatively renormalizable, and the linear divergence just comes from the Wilson line. We plot the ratio of { the pion} matrix element $\tilde{h}^{\text{GI}}_{\pi,\gamma_t}(z,0;1/a)$ { and the trace of the Wilson line $W(z)$  to see whether there is a residual linear divergence on the lattice.}

The central value of the pion matrix element with Coulomb wall source shows {negligible} dependence on the precision of Coulomb gauge { fixing,} as demonstrated in the appendix. So we will concentrate on the precision of Landau gauge fixing used { for the} Wilson line rather than Coulomb gauge fixing used { for the} pion matrix element with wall source.

\begin{figure}[tbph]
\centering
\includegraphics[width=7cm]{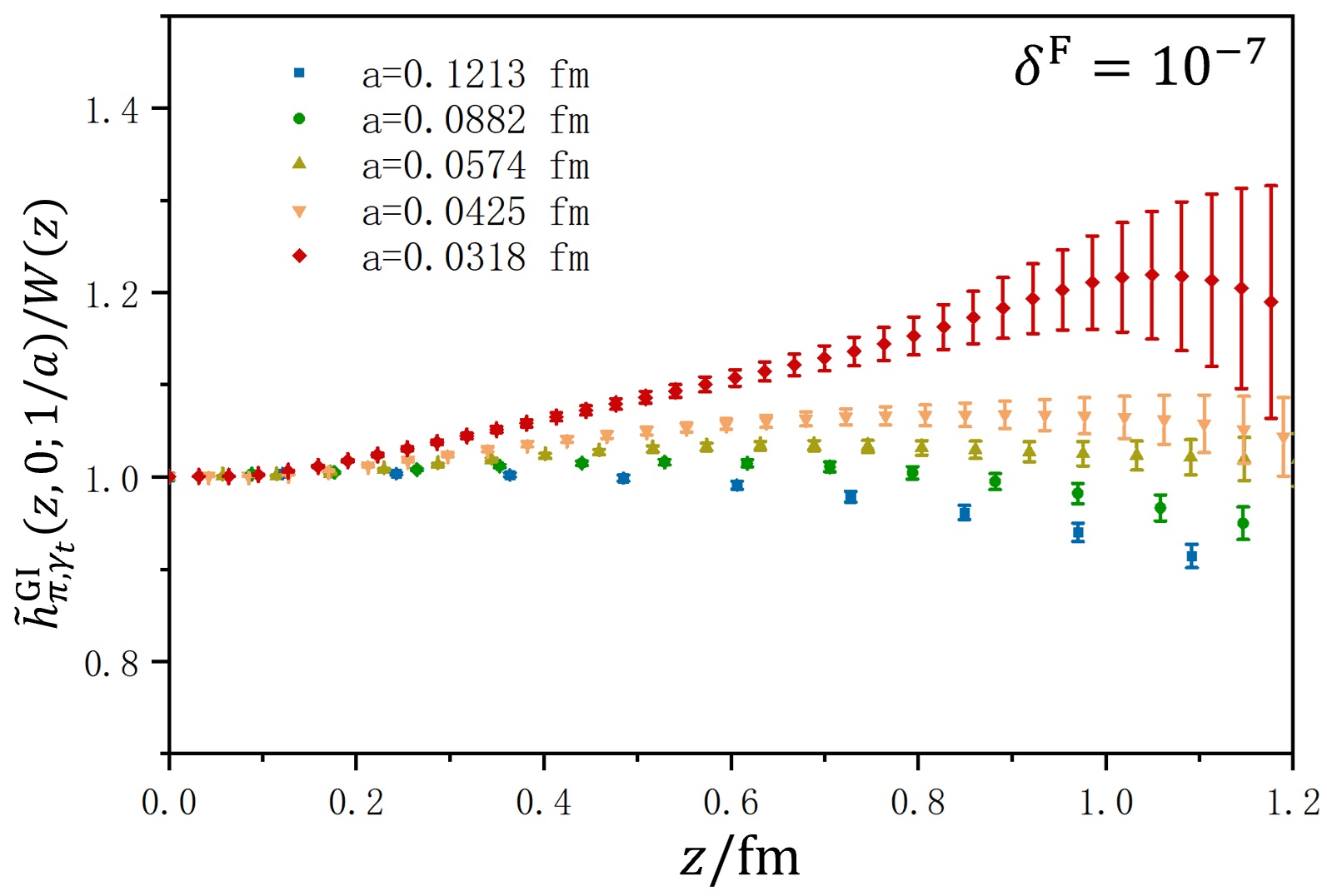}
\includegraphics[width=7cm]{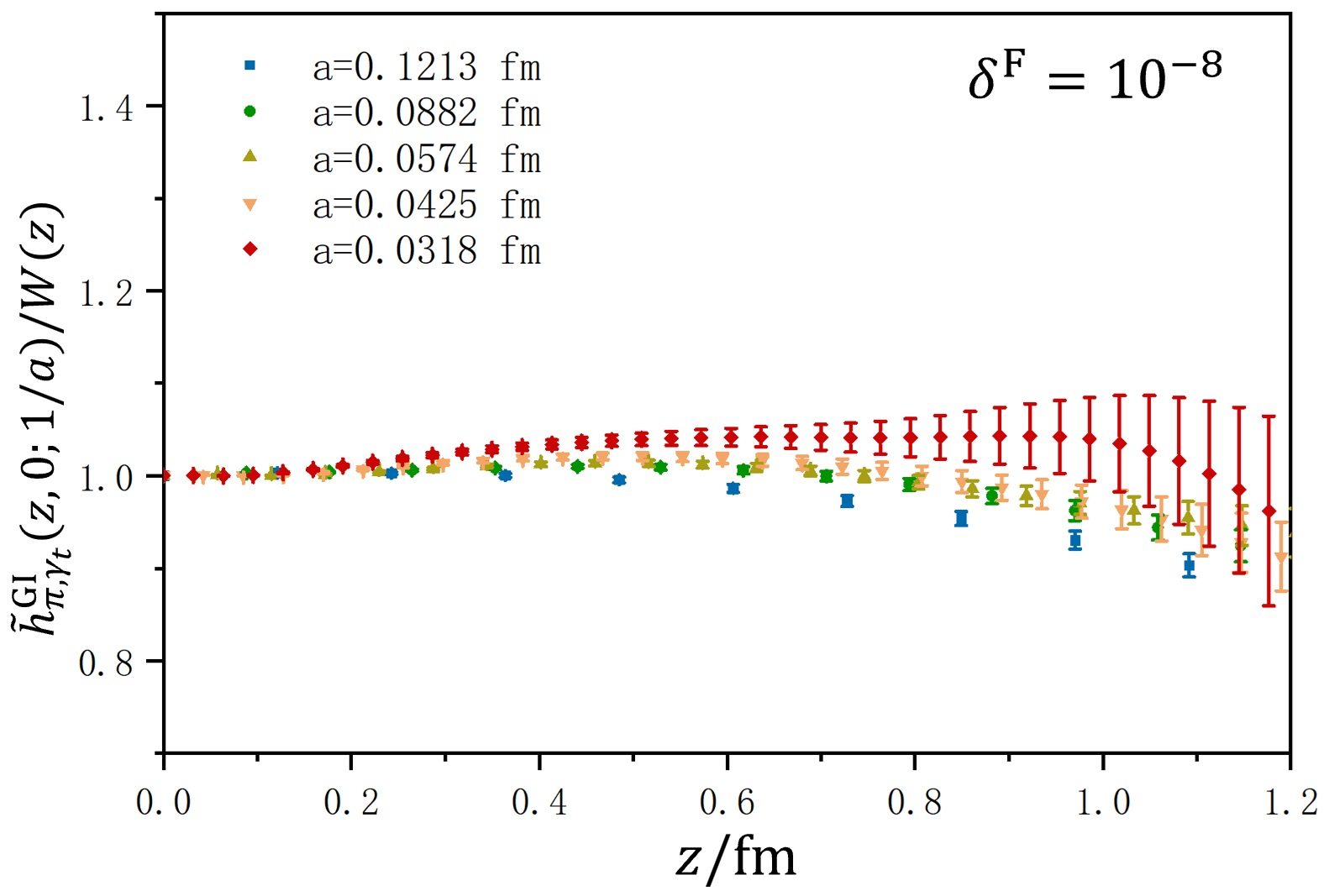}
\includegraphics[width=7cm]{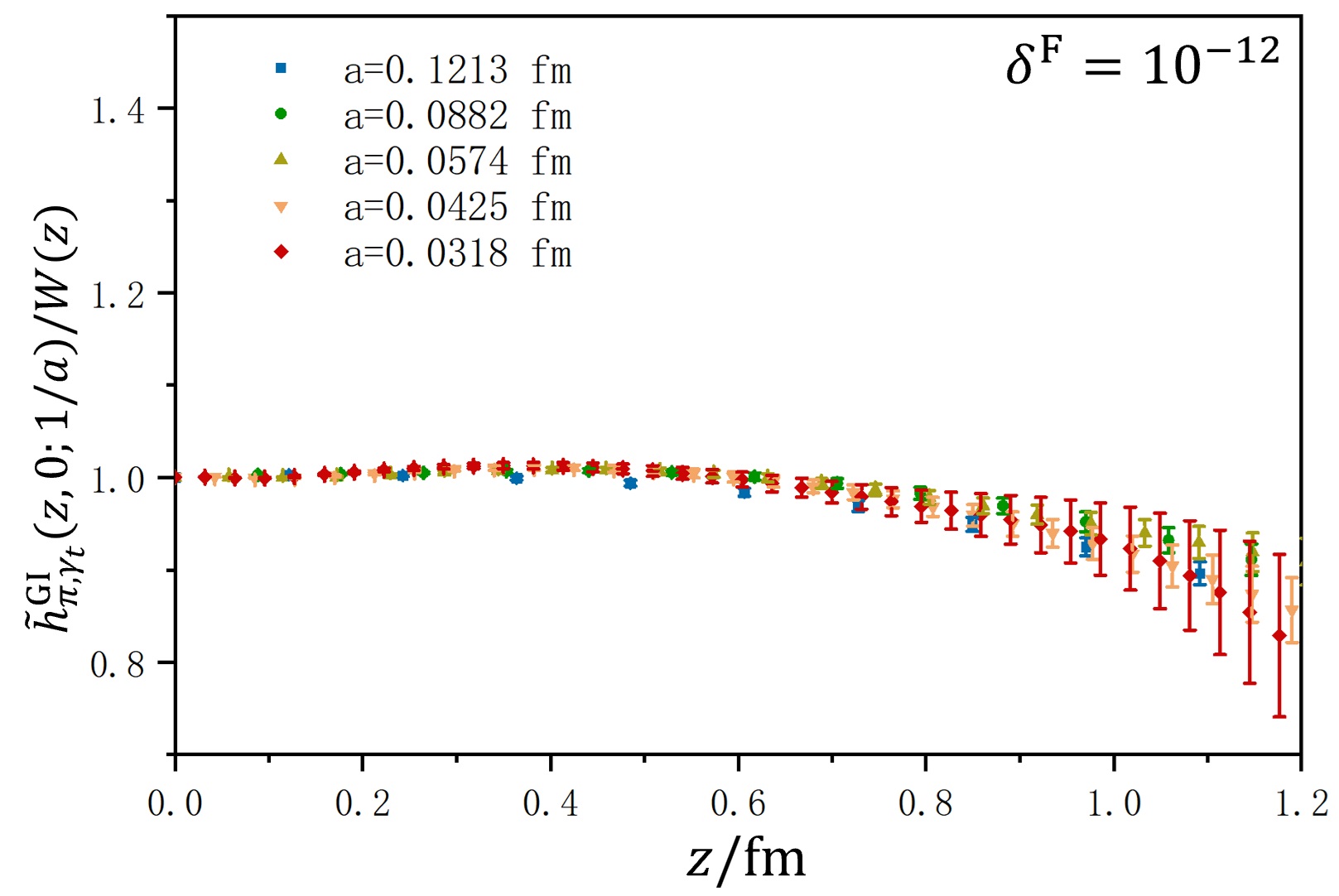}
\caption{Pion matrix element { in lattice regularization} $\tilde{h}^{\text{GI}}_{\pi,\gamma_t}(z,0;1/a)$ defined in Eq.~(\ref{eq:bare}) { divided by} $W(z)$ defined in Eq.~(\ref{eq:TrLine}) { for} different lattice spacings and { different precision} of gauge fixing. One-step HYP smearing has been applied after gauge fixing.}
\label{fig:LineRenorm}
\end{figure}

As shown in Fig.~\ref{fig:LineRenorm}, the results at $\delta^F = 10^{-7}$ { differ} for different lattice spacings. The discrepancy { increases} for finer { lattices and  
longer Wilson lines.} When the precision of gauge fixing is increased, such as { to} $\delta^F = 10^{-8}$, the results show convergence. Moreover, for $\delta^F = 10^{-12}$, the results from different lattice spacings are consistent within uncertainty even { for} the finest lattice and at long distance, indicating the { complete} elimination of the linear divergence. Although at short distance, the results { for} $\delta^F = 10^{-7}$, $\delta^F = 10^{-8}$ and $\delta^F = 10^{-12}$ are very close to each other, the impact of imprecise gauge fixing becomes significant at long distance, especially for smaller lattice spacing. This is due to the unphysical contribution { caused by} imprecise gauge fixing { of} Wilson lines, as mentioned previously.

Since the ratio has removed the linear divergence properly, one can do the perturbative calculation of $W(z)$ to match the hadron matrix element $\tilde{h}^{\text{GI}}_{\pi,\gamma_t}$ to the $\overline{\textrm{MS}}$ scheme.  First define a  "subtracted" quasi PDF renormalized in the short-distance ratio (SDR) scheme which divide the quasi PDF by the Wilson line and normalized with a pion matrix element in the rest frame at a short distance $z_0$,
\begin{align}\label{eq:quasiPDFSDR}
h^{{\rm GI,SDR}}_{\chi,\gamma_t}(z,P_z;1/z_0)=\frac{\tilde{h}^{\text{GI}}_{\chi,\gamma_t}(z,P_z;1/a)/W(z)}{\tilde{h}^{\text{GI}}_{\pi,\gamma_t}(z_0,0;1/a)/W(z_0)}.
\end{align}
Then $h^{{\rm GI,SDR}}_{\chi,\gamma_t}$ can be matched to the $\overline{\textrm{MS}}$ scheme using the perturbative calculation of $W(z)$, $W(z_0)$ and the on-shell $\tilde{h}^{\text{GI}}_{\chi,\gamma_t}$ which is independent of $\chi$,
\begin{align}\label{eq:quasiPDFSDR}
h^{{\rm GI,\overline{\textrm{MS}}}}_{\chi,\gamma_t}(z,P_z;\mu)=&C(z,z_0,\mu)h^{{\rm GI,SDR}}_{\chi,\gamma_t}(z,P_z;z_0),
\end{align}
where 
\begin{align}
&C(z,z_0,\mu)=\frac{\tilde{h}^{\text{GI},\overline{\textrm{MS}}}_{{\rm pert},\gamma_t}(z_0,0;\mu)}{W^{\overline{\textrm{MS}}}_{\rm pert}(z_0,\mu)}W^{\overline{\textrm{MS}}}_{\rm pert}(z,\mu)\nonumber\\
&= \Big\{1+\frac{\alpha_sC_F}{4\pi}\Big[\xi\text{ln}z_0^2\mu^2+5+\xi(2\gamma_E-\text{ln}4)\Big]\Big\}\nonumber\\
&\quad \times\Big\{1+\frac{\alpha_sC_F}{4\pi}(3-\xi)\Big[\text{ln}z^2\mu^2+2\gamma_E-\text{ln}4\Big]\Big\}+{\cal O}(\alpha_s^2)\nonumber\\
&\ _{\overrightarrow{\xi\rightarrow 0}}\ \tilde{h}^{\text{GI},\overline{\textrm{MS}}}_{{\rm pert},\gamma_t}(z,0;\mu)+{\cal O}(\alpha_s^2).
\end{align}
Note that since the long Wilson line does contain non-perturbative physics at large-$z$, the above approach will introduce some unknown non-perturbative effects. The comparison with other renormalization schemes can be found in the following subsection.

\subsection{Quasi PDF renormalized in RI/MOM scheme and self renormalization}

In RI/MOM renormalization, we need to calculate gauge dependent 
quark matrix elements to renormalize hadron matrix elements to get results free of any residual divergences. The definition of the renormalization constant is (e.g.,~\cite{Liu:2018uuj}),
\begin{align}\label{eq:Z_ri}
&Z^{\text{GI}}_{\Gamma}(z,P_z;\mu,1/a)\nonumber\\
&=\frac{Z_q(P_z;\mu,1/a)}{\textrm{Tr}[\Gamma\langle q(p)|O^{\text{GI}}_{\Gamma}(z)|q(p)\rangle]_{p^2=-\mu^2, p_z=p_t=0}},
\end{align}
where $|q(p)\rangle$ is the off-shell quark state with external momentum $p$, and we have assumed a lattice regularization and included { the} $1/a$ dependence in the matrix elements. The setup in the calculation of quark matrix { elements} is the same as that in Ref.~\cite{Zhang:2020rsx}, except for the precision of gauge fixing. $Z_q$ is defined from the hadron $\chi$ matrix element of a local vector current,
\begin{align}\label{eq:Z_q}
&Z_q(P_z;\mu,1/a)\nonumber\\
&=\frac{\textrm{Tr}[\Gamma\langle q(p)|O^{\text{GI}}_{\Gamma}(0)|q(p)\rangle]_{p^2=-\mu^2, p_z=p_t=0}}{\tilde{h}^{\text{GI}}_{\chi,\Gamma}(0,P_z;1/a)}.
\end{align}

When the Landau gauge fixed volume source~\cite{Chen:2017mzz} is used in the calculation, the statistical uncertainty is suppressed by a factor $1/\sqrt{V}$ compared to the point source case. Note that the $Z_q$ definition used here is exactly the same as $\tilde{Z}_q\equiv \textrm{Tr}[p\!\!\!/S^{-1}]/p^2$ from the quark propagator in dimensional regularization~\cite{Gracey:2003yr}, but the discretization effect is much smaller than that for $\tilde{Z}_q$ in lattice regularization as shown in Ref.~\cite{Chang:2021vvx}.

We multiply the pion matrix element $\tilde{h}^{\text{GI}}_{\pi,\gamma_t}(z,Pz;1/a)$ with $Z^{\rm GI}_{\gamma_t}$ to obtain the non-perturbatively renormalized and normalized matrix element at a given RI/MOM scale $\mu$,
\begin{align}\label{eq:hRIMOM}
h^{{\text{GI}},\text{RI/MOM}}_{\pi,\gamma_t}(z,P_z;\mu)=Z^{\text{GI}}_{\gamma_t}(z,P_z;\mu,1/a)\tilde{h}^{\text{GI}}_{\pi,\gamma_t}(z,P_z;1/a).
\end{align}

Since the quasi-PDF operator is multiplicatively renormalizable and the linear divergence just comes from the Wilson line and is independent of the external quark or hadron state, RI/MOM renormalization~\cite{Martinelli:1994ty} is believed to be a good candidate to remove the linear divergence~\cite{Constantinou:2017sej,Green:2017xeu,Chen:2017mzz,Stewart:2017tvs,Alexandrou:2017huk,Alexandrou:2020qtt,Lin:2020fsj}.

\begin{figure}[tbph]
\centering
\includegraphics[width=7cm]{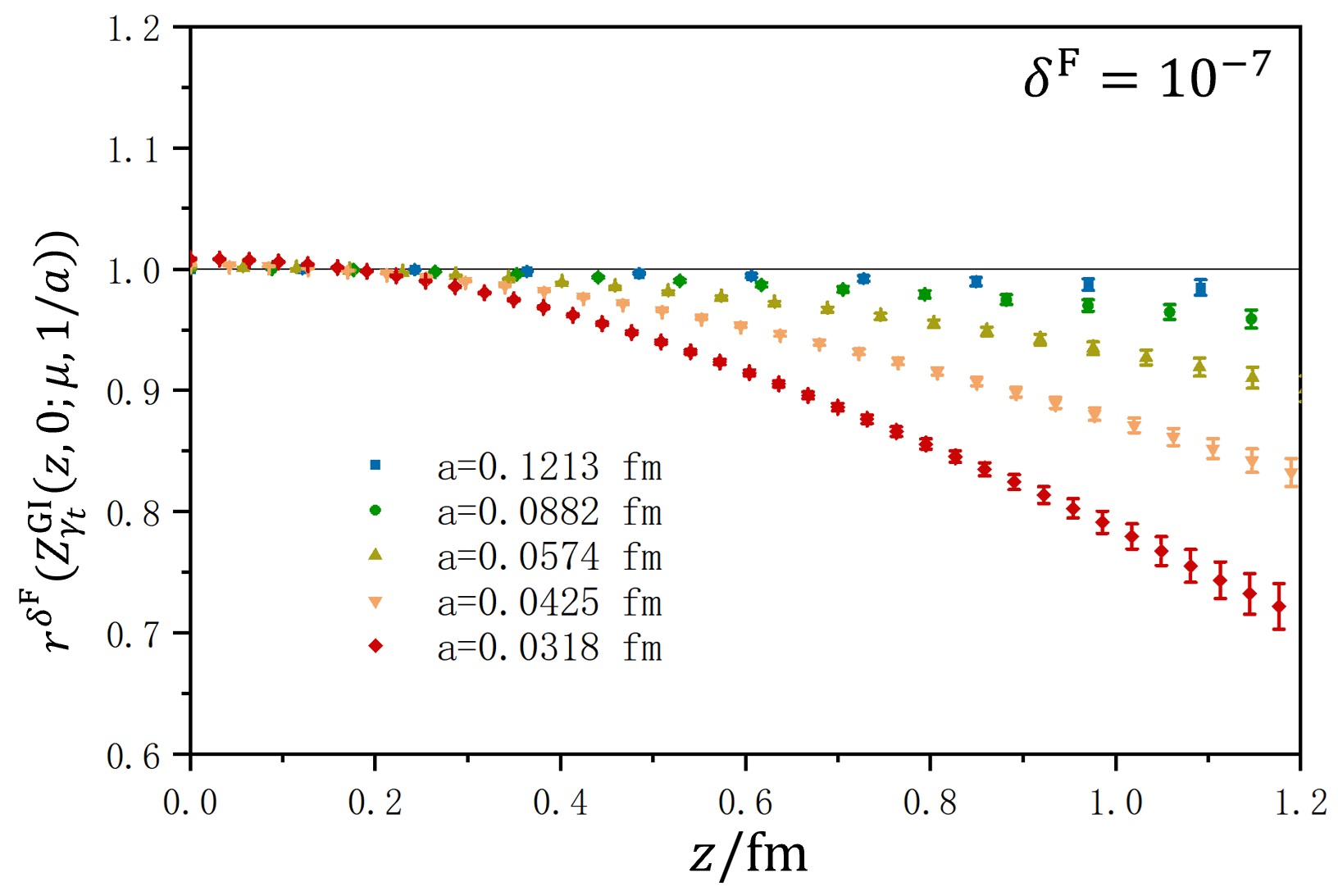}
\includegraphics[width=7cm]{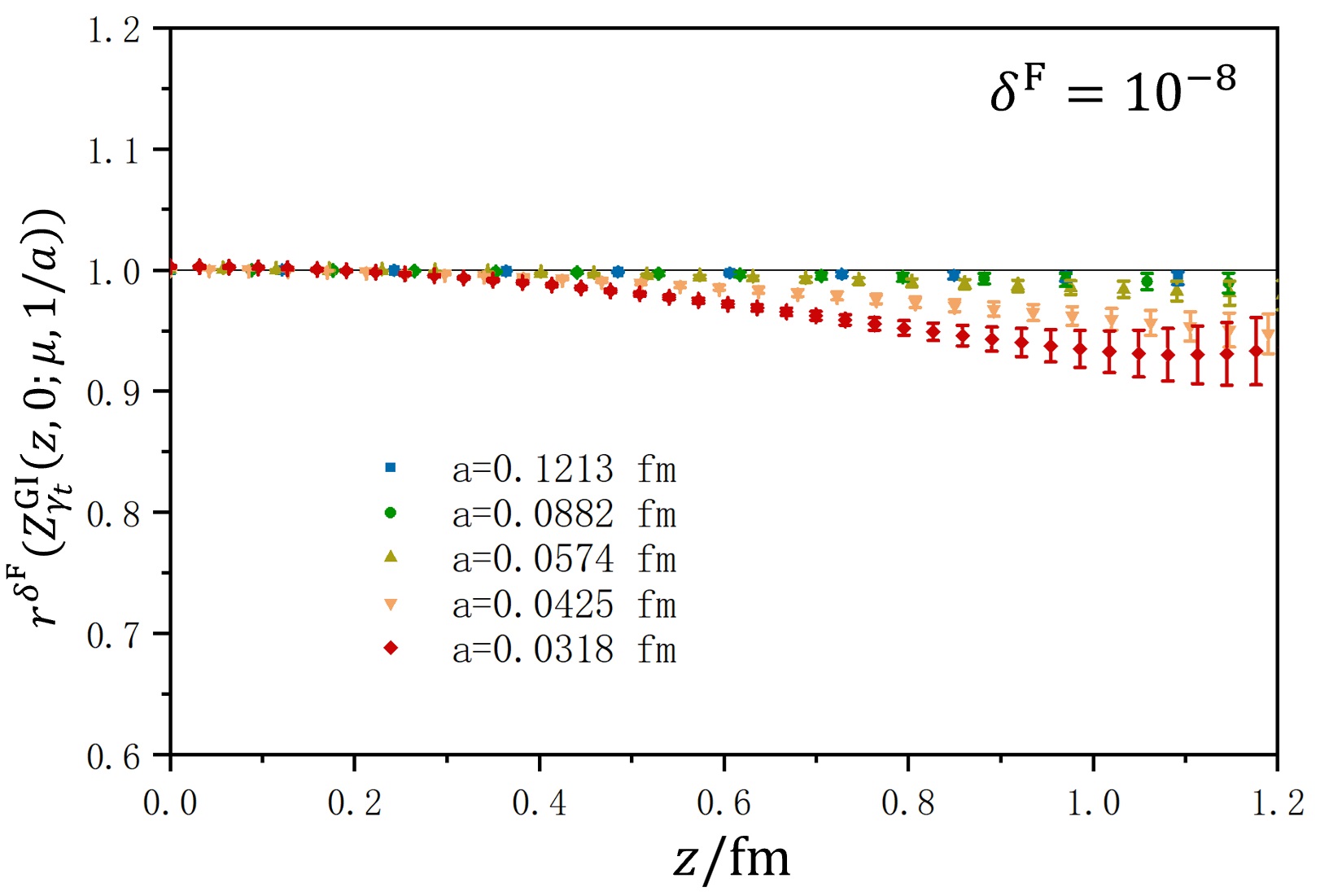}
\caption{The ratio $r^{\delta^F}(Z^{\text{GI}}_{\gamma_t}(z,P_z;\mu,1/a))$ of the renormalization factor defined in Eq.~(\ref{eq:Z_ri}) in RI/MOM scheme { for two values of the required precision, with one-step HYP smearing. The case $\delta^F_0 = 10^{-12}$ is used as denominator in the ratios}.}
\label{fig:RIRatio}
\end{figure}

If the linear divergence just comes from the Wilson line and is independent of the external quark or hadron state, then RI/MOM renormalization~\cite{Martinelli:1994ty} can be a good candidate to remove the linear divergence~\cite{Constantinou:2017sej,Green:2017xeu,Chen:2017mzz,Stewart:2017tvs,Alexandrou:2017huk,Alexandrou:2020qtt,Lin:2020fsj}. But similar to Wilson line, the quasi PDF quark matrix element in RI/MOM scheme is a non-local gauge dependent quantity, and thus it is also highly sensitive to the precision of Landau gauge fixing at long distance on fine lattices as shown in Fig.~\ref{fig:RIRatio}. 

Compared to Fig.~\ref{fig:WilsonLine}, we observe that the impact of imprecise gauge fixing on the quark matrix element is weaker than that on { the} Wilson line. This suggests that the quark external state mitigates the impact of imprecise gauge fixing. For example, the difference of the quark matrix elements at $\delta^F = 10^{-8}$ and $\delta^F = 10^{-12}$ on a06m310 at 1.2 fm is 0.021(0.012) whereas this difference is 0.032(0.003) for the Wilson line. Similarly, on a03m310, the difference is 0.06(0.03) and 0.116(0.09) for the quark matrix element and the Wilson line, respectively. 

\begin{figure}[tbph]
\centering
\includegraphics[width=7cm]{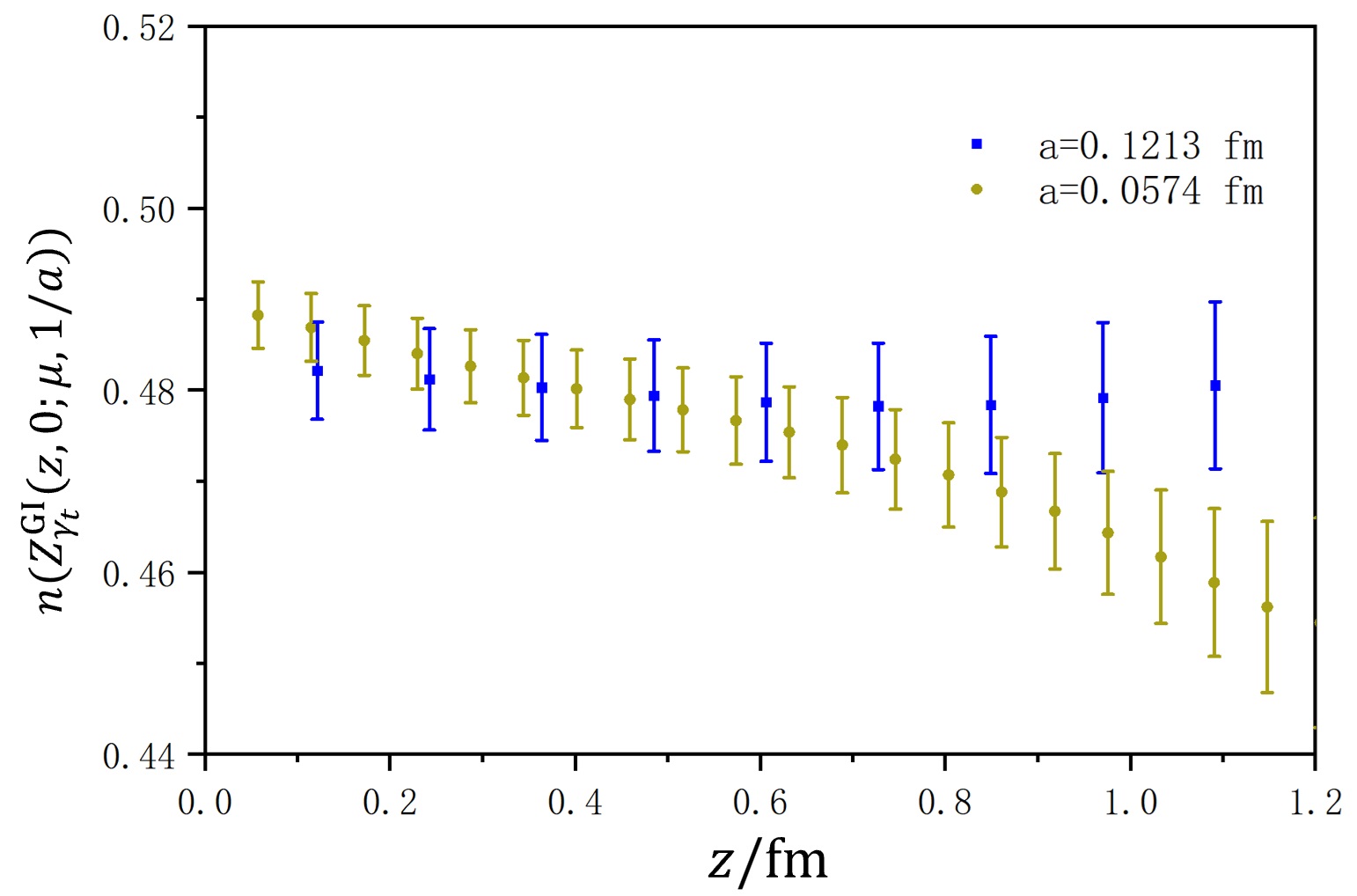}
\includegraphics[width=7cm]{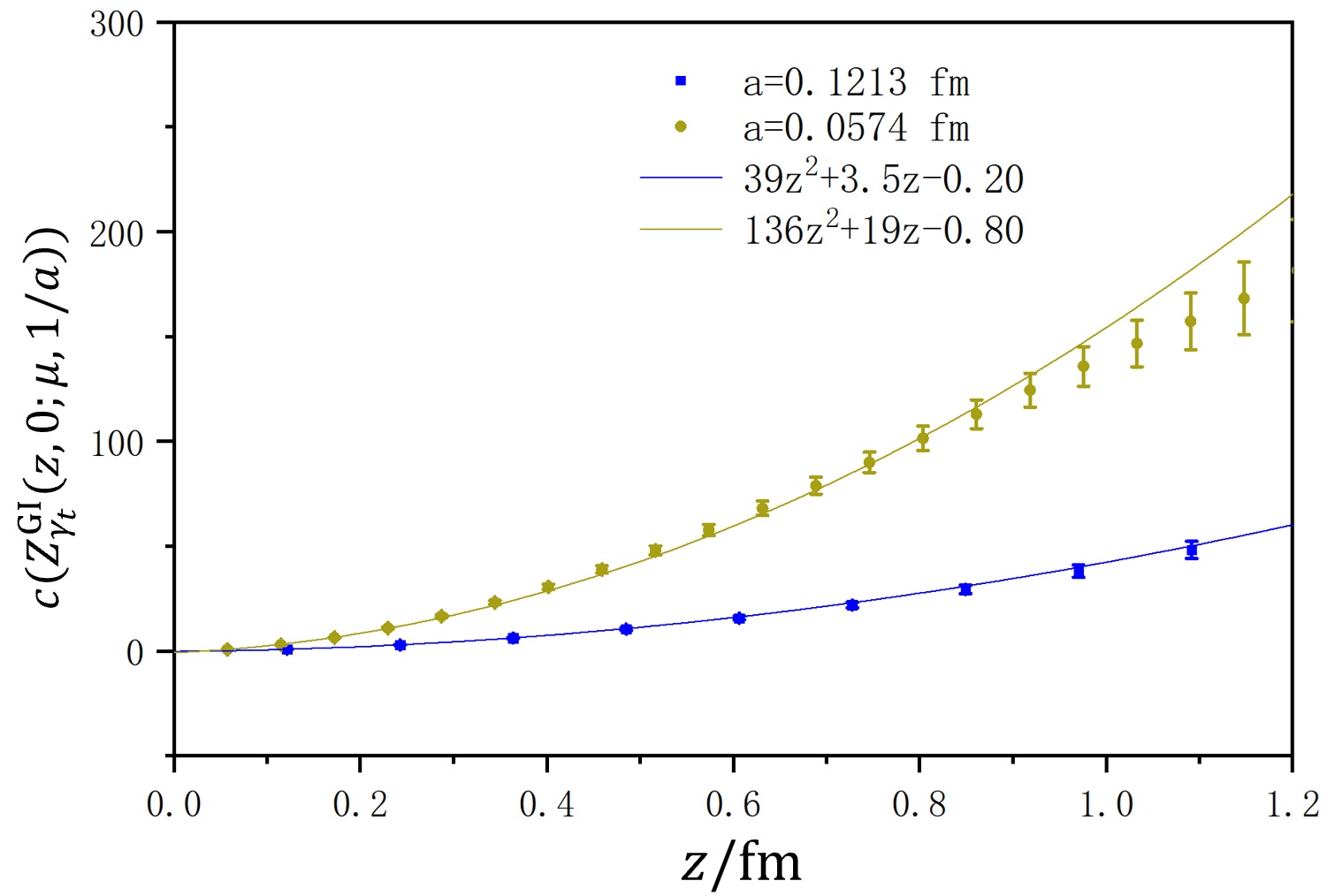}
\includegraphics[width=7cm]{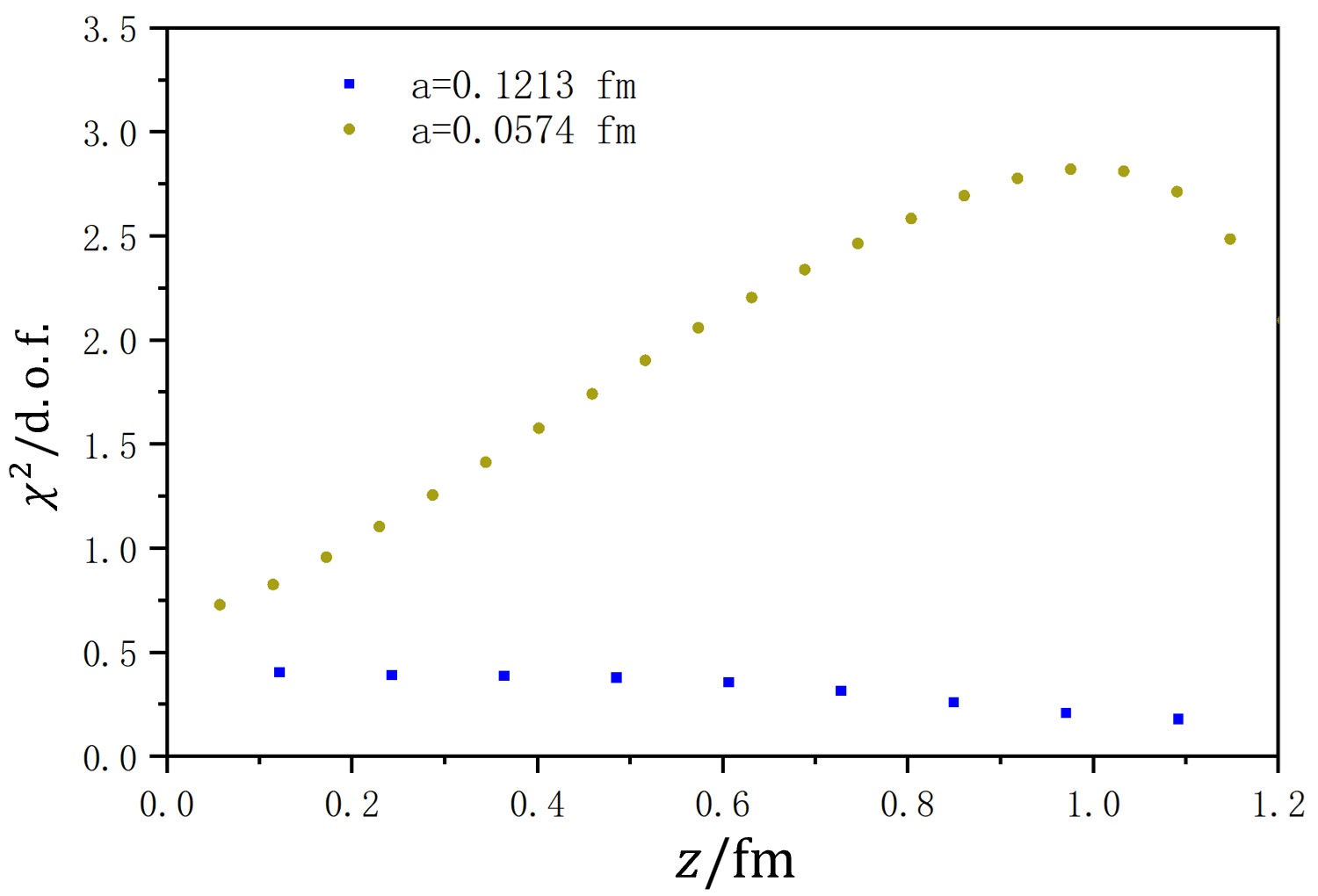}
\caption{The { fitted results for $n(Z^{\text{GI}}_{\gamma_t}(z,0;\mu,1/a))$
and $c(Z^{\text{GI}}_{\gamma_t}(z,0;\mu,1/a))$} defined in Eq.~(\ref{eq:Z_ri}) and Eq.~(\ref{eq:empirical}), along with the corresponding $\chi^2/$d.o.f. on different ensembles. We use data { for gauge fixing with $\delta^F = 10^{-4},10^{-5},10^{-6},10^{-7},10^{-8}$.} one-step HYP smearing has been applied after gauge fixing. The uncertainty is got by bootstrap resampling. The two curves in the middle panel are the { fitted results for} $c(Z^{\text{GI}}_{\gamma_t}(z,0;\mu,1/a))$.}
\label{fig:RIfit}
\end{figure}

It is important to check whether the empirical formula defined in Eq.~(\ref{eq:empirical}) can also describe the impact of imprecise gauge fixing on quark matrix { elements. The fitted} results are presented in Fig.~\ref{fig:RIfit}. Here, $n(Z^{\text{GI}}_{\gamma_t}(z,0;\mu,1/a))$ is about 0.5 and $c(Z^{\text{GI}}_{\gamma_t}(z,0;\mu,1/a))$ increases for longer $z$ and smaller lattice spacing. The lower panel indicates { that} $\chi^2/$d.o.f. is not consistently small and can reach around 3 on a06m310. Nevertheless, we maintain that this empirical formula performs reasonably well. Ultimately, the precision of gauge fixing is unphysical, and there is some flexibility in its definition. Since $\delta^F$ is simply the trace of the Wilson line, it is more suitable for the Wilson line case than for the quark matrix element case when investigating the impact of imprecise gauge fixing. We believe that this empirical formula is sufficient to help us estimate the impact of imprecise gauge fixing for the quark matrix element. Additionally, we find that $c(Z_q)$ is negative on both a12m310 and a06m310, as $n(Z_q(P_z;\mu,1/a))=0.48(0.02)$, $c(Z_q(P_z;\mu,1/a))=-1.1(0.2)$ and $\chi^2/$d.o.f.=0.30 on a12m310, $n(Z_q(P_z;\mu,1/a))=0.507(0.09)$, $c(Z_q(P_z;\mu,1/a))=-6.0(0.5)$ and $\chi^2/$d.o.f.=0.36 on a06m310. The negative value of $c(Z_q)\propto 1/a^2$ implies that $Z_q$ defined in Eq.~(\ref{eq:Z_q}) will { increase} as the precision of Landau gauge fixing decreases{, and { that its} impact can increase significantly at small lattice spacing.}

We can perform a similar fit for $c(Z^{\text{GI}}_{\gamma_t}(z,0;\mu,1/a))$ using a quadratic function $c(Z^{\text{GI}}_{\gamma_t}(z,0;\mu,1/a)) = c_2z^2+c_1z+c_0$ as shown in the middle panel in Fig.~\ref{fig:RIfit}. On a12m310, we obtain $c_2=39(4)$, $c_1=3.5(1.4)$, $c_0=-0.20(0.12)$, with $\chi^2/$d.o.f.=0.15. For a06m310, the values are $c_2=136(10)$, $c_1=19(2)$, $c_0=-0.80(0.10)$, and $\chi^2/$d.o.f.=1.40. This result indicates that $c(Z^{\text{GI}}_{\gamma_t}(z,0;\mu,1/a))$ will grow quadratically as the Wilson line becomes longer. Consequently, we can employ a quadratic function to predict $c(Z^{\text{GI}}_{\gamma_t}(z,0;\mu,1/a))$ and estimate the impact of gauge fixing.

\begin{figure}[tbph]
\centering
\includegraphics[width=7cm]{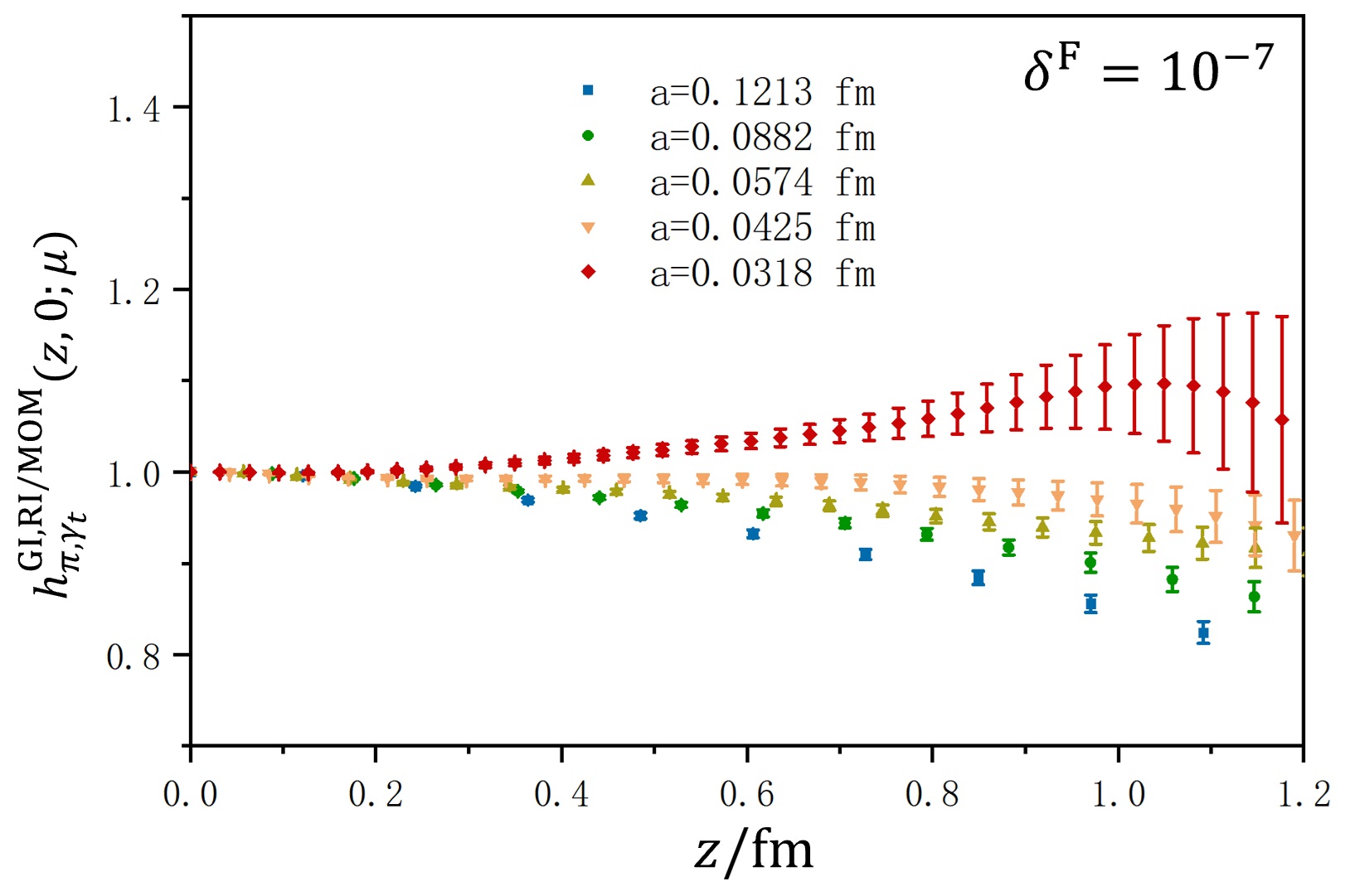}
\includegraphics[width=7cm]{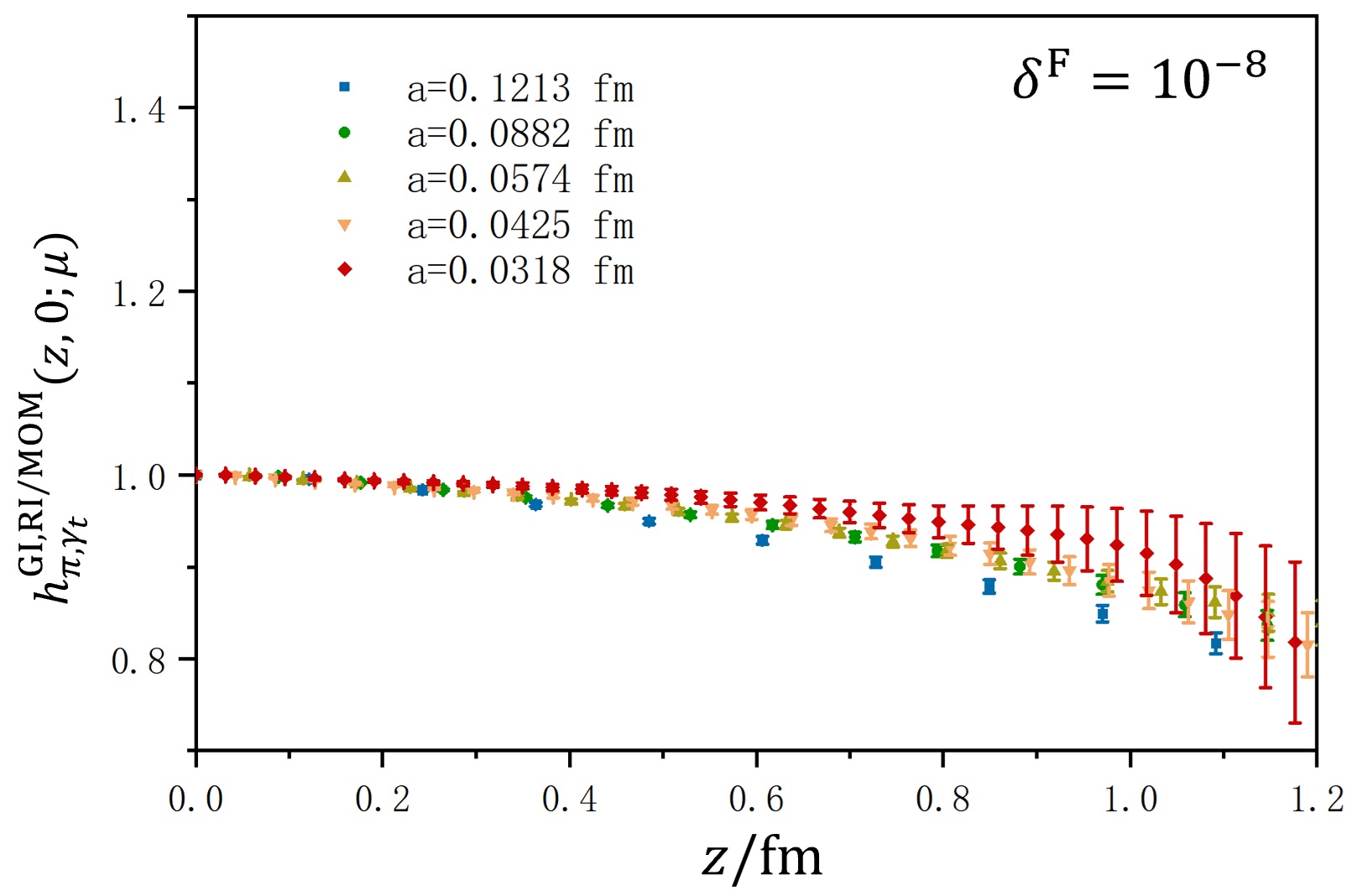}
\includegraphics[width=7cm]{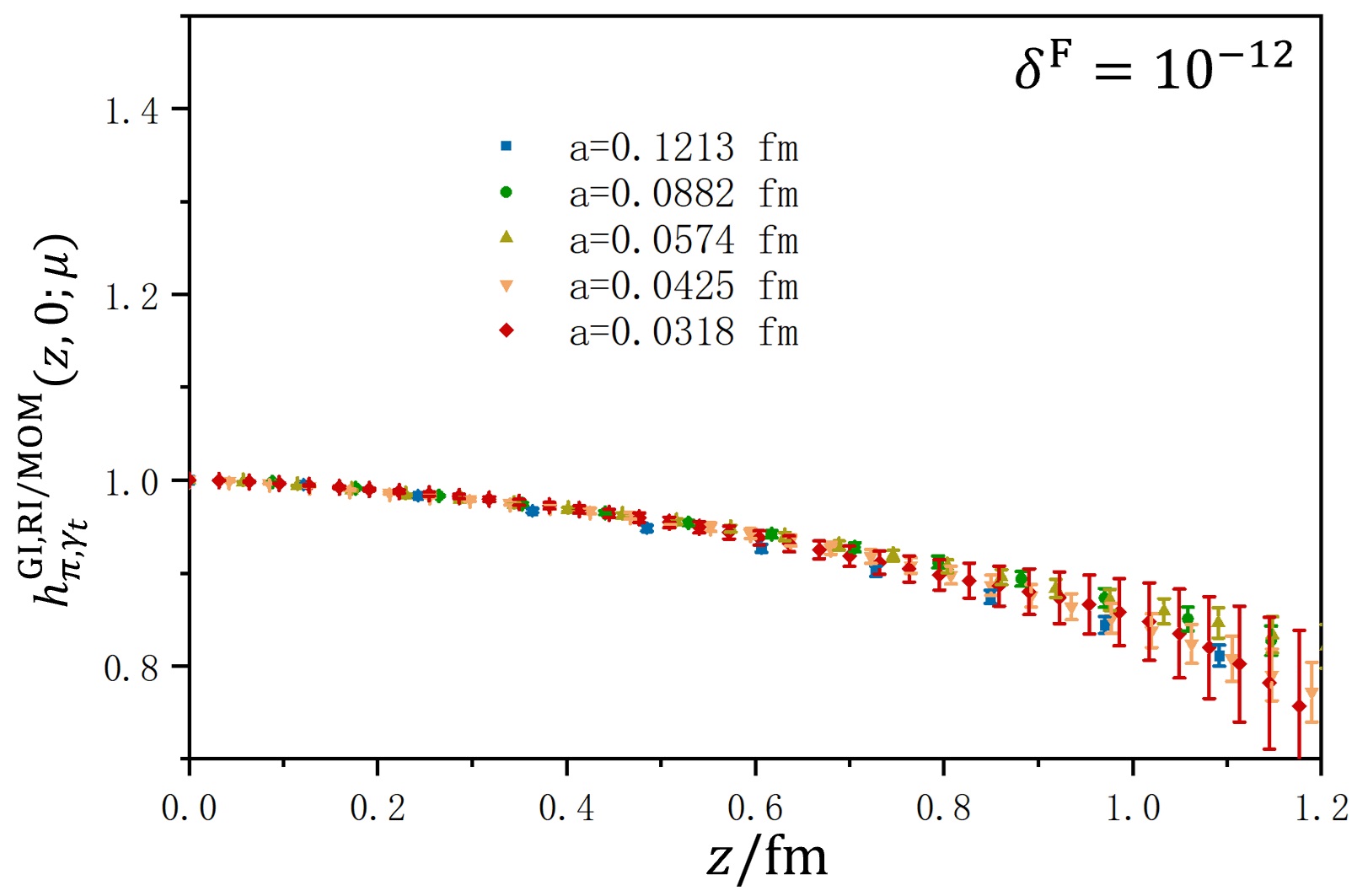}
\caption{Quasi PDF matrix element in RI/MOM scheme defined in Eq.~(\ref{eq:TrLine}) at different lattice spacings and { for different precision} of Landau gauge fixing.}
\label{fig:RIRenorm}
\end{figure}

Thus the imprecise gauge fixing has an significant impact on quasi PDF in RI/MOM scheme. As shown in Fig.~\ref{fig:RIRenorm}, the { outcome} at $\delta^F = 10^{-7}$ { exhibits} inconsistencies across various lattice spacings, with a more pronounced discrepancy observed for finer lattices. Furthermore, the discrepancy amplifies as the line lengthens. However, with an increased precision of gauge fixing, such as $\delta^F = 10^{-8}$, the results { behave better}. Additionally, for $\delta^F = 10^{-12}$, the results { for} different lattice spacings { agree within the uncertainties}, even on the finest lattice and at long distances, indicating the { complete} elimination of the linear divergence. If the precision is not high enough, the divergent behavior may be mistaken for residual linear divergence, as what we did in Ref.~\cite{Zhang:2020rsx}.

Even though the RI/MOM renormalization with precise enough gauge fixing can eliminate the linear divergence, the non-perturbative effect at large $z$ remains. Thus one can also extract divergent renormalization factors of quasi-PDF by parameterizing the results in the rest frame on the lattice, which is called self renormalization~\cite{LatticePartonCollaborationLPC:2021xdx}.

The fit strategy used in Ref.~\cite{LatticePartonCollaborationLPC:2021xdx} can be improved into a joint fit of the results at different lattice spacing $a$ and Wilson line length $z$:
\begin{align}
&\text{ln}\tilde{h}^{\text{GI}}_{\pi,\gamma_t}(z,0;1/a)=\frac{kz}{a\text{ln}[a\Lambda_{\text{QCD}}]}+f(z)a+m_0z\nonumber\\
&\quad +\frac{3C_f}{b_0}\text{ln}\bigg[\frac{\text{ln}(1/a\Lambda_{\text{QCD}})}{\text{ln}(\mu/\Lambda_{\text{QCD}})}\bigg]+\frac{1}{2}\text{ln}\bigg[1+\frac{d}{\text{ln}[a\Lambda_{\text{QCD}}]}\bigg]^2\nonumber \\
&\quad +\begin{cases}
\text{ln}\big[Z_{\overline{{\text{MS}}}}(z,\mu,\Lambda_{\overline{{\text{MS}}}})\big] & \text{if } z_0 \le z \le z_1\\
g(z) & \text{if } z_1<z
\end{cases},\label{eq:self1}
\end{align}
where the linear divergence factor $k$, the effective QCD scale $\Lambda_{\text{QCD}}$, the discretization effect coefficient $f(z)$, the renormalon factor $m_0$, the overall $\alpha_s$ correction coefficient $d$, and the renormalized matrix element $g(z)$ are the parameters to be fitted, and $Z_{\overline{{\text{MS}}}}(z,\mu,\Lambda_{\overline{\text{MS}}})$ represents the quasi-PDF in the $\overline{\text{MS}}$ scheme at 1-loop level. 

It is important for $z_0$ and $z_1$ to be sufficiently large to mitigate discretization effects, yet small enough to ensure the validity of perturbative theory. Additionally, $z_1-z_0$ should not be excessively small. In this work, we choose $z_0=0.06~\text{fm}$ and $z_1=0.15~\text{fm}$, and interpolate the data at different lattice spacing with intervals of 0.03 fm since both $f(z)$ and $g(z)$ depend on $z$.

Based on this strategy, the renormalized quasi-PDF in the rest frame in the $\overline{\text{MS}}$ scheme will be
\begin{align}\label{eq:selfRes}
\tilde{h}^{\text{GI}}_{\pi,\gamma_t}(z,0;1/a)=\begin{cases}
Z_{\overline{{\text{MS}}}}(z,\mu,\Lambda_{\overline{{\text{MS}}}}) & \text{if } z_0 \le z \le z_1\\
\mathrm{Exp}[g(z)] & \text{if } z_1<z
\end{cases}.
\end{align}

\begin{figure}[tbph]
\centering
\includegraphics[width=7cm]{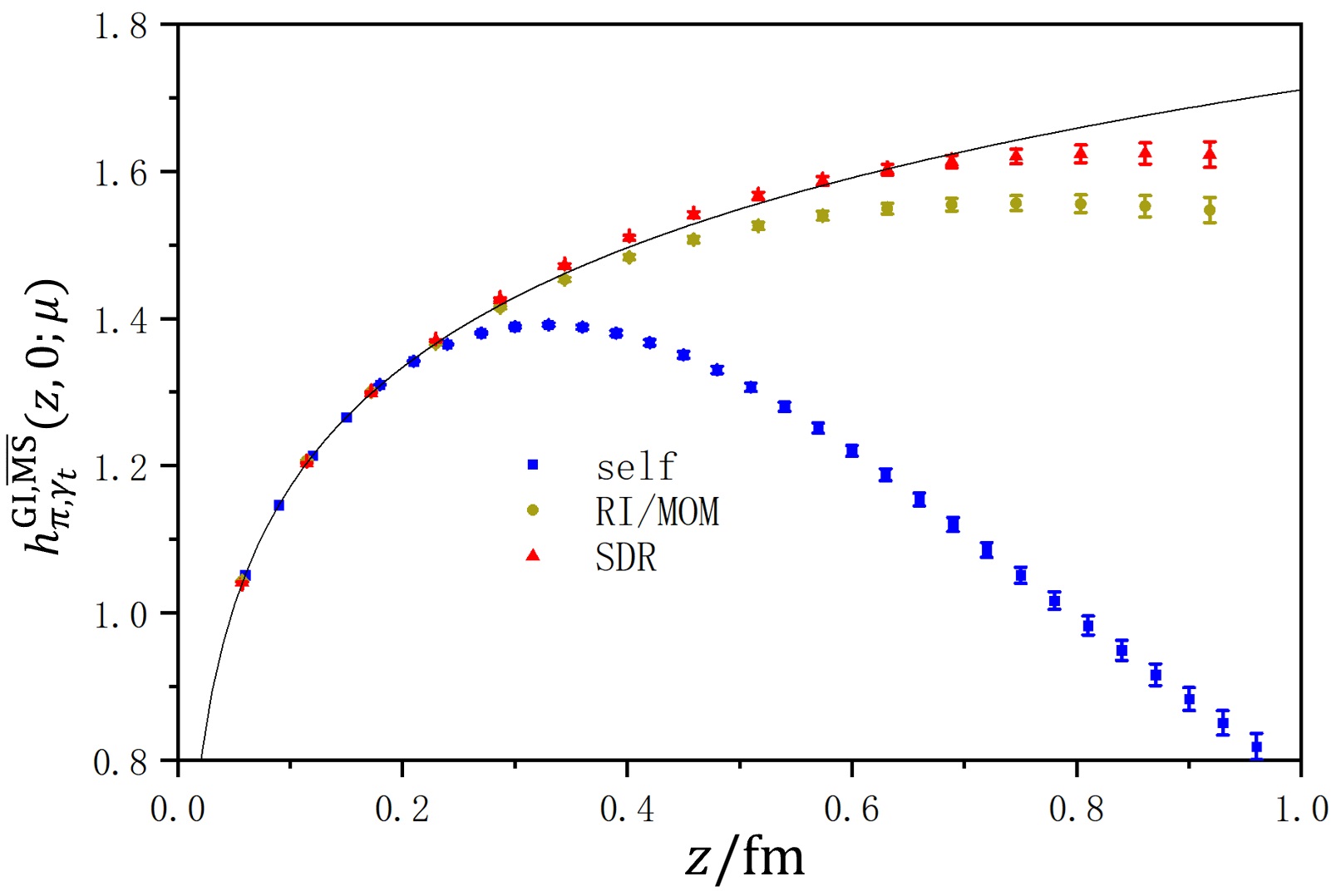}
\caption{Quasi PDF matrix element in $\overline{\text{MS}}$ scheme with scale $\mu = 2\text{GeV}$. The results through the intermediate RI/MOM and SDR schemes are calculated on a06m310 and converted to the $\overline{\text{MS}}$ scheme by perturbative matching at 1-loop level. The result from self renormalization is extracted from joint fits with 5 lattice spacings, and $z$ is interpolated to multiples of 0.03 fm. The curve is the perturbative result at 1-loop level in the $\overline{\text{MS}}$ scheme.}
\label{fig:selfRenorm}
\end{figure}

The conversion factor between RI/MOM and $\overline{\text{MS}}$ { at} 1-loop level can be found in Ref.~\cite{Constantinou:2017sej}. As shown in Fig.~\ref{fig:selfRenorm}, the $\overline{\text{MS}}$ results using the self renormalization, the intermediate RI/MOM  or SDR schemes are consistent with each other and close to { the} perturbative result at short distance, but they deviate from each other at long distance {due to the non-perturbative effect from the intrinsic scale $1/z$} introduced in RI/MOM and SDR schemes. 

\subsection{Coulomb gauge quasi PDF}

The Coulomb gauge quasi PDF~\cite{Gao:2023lny} (CG qPDF) operator is
\begin{align}\label{eq:CGqPDF}
O^{\text{CG}}_{\Gamma}(z)=\bar{\psi}(0) \Gamma \psi (z)
.
\end{align}
At next-to leading order (NLO), its LaMET matching to { PDFs} has been demonstrated in Ref.~\cite{Gao:2023lny}. This operator is { non-local} and its hadron matrix element $\tilde{h}^{\text{CG}}_{\chi,\Gamma}(z,P_z;1/a)$ is gauge dependent because the quark { fields} in the operator are non-local and not linked by Wilson lines. If it is the free of { the} linear divergence, the residual log divergence can be removed by { a} short distance matrix element,
\begin{align}\label{eq:renormCG}
h^{\text{CG}}_{\chi,\Gamma}(z,P_z;1/a) \equiv \tilde{h}^{\text{CG}}_{\chi,\Gamma}(z,P_z;1/a) / \tilde{h}^{\text{CG}}_{\chi,\Gamma}(z_0,P_z;1/a)
.
\end{align}
The hadron matrix element is non-local and gauge dependent, similar to the quark matrix { elements of quasi PDFs and quasi TMD-PDFs}. Therefore, it is expected to be sensitive to the precision of gauge fixing.

\begin{figure}[tbph]
\centering
\includegraphics[width=7cm]{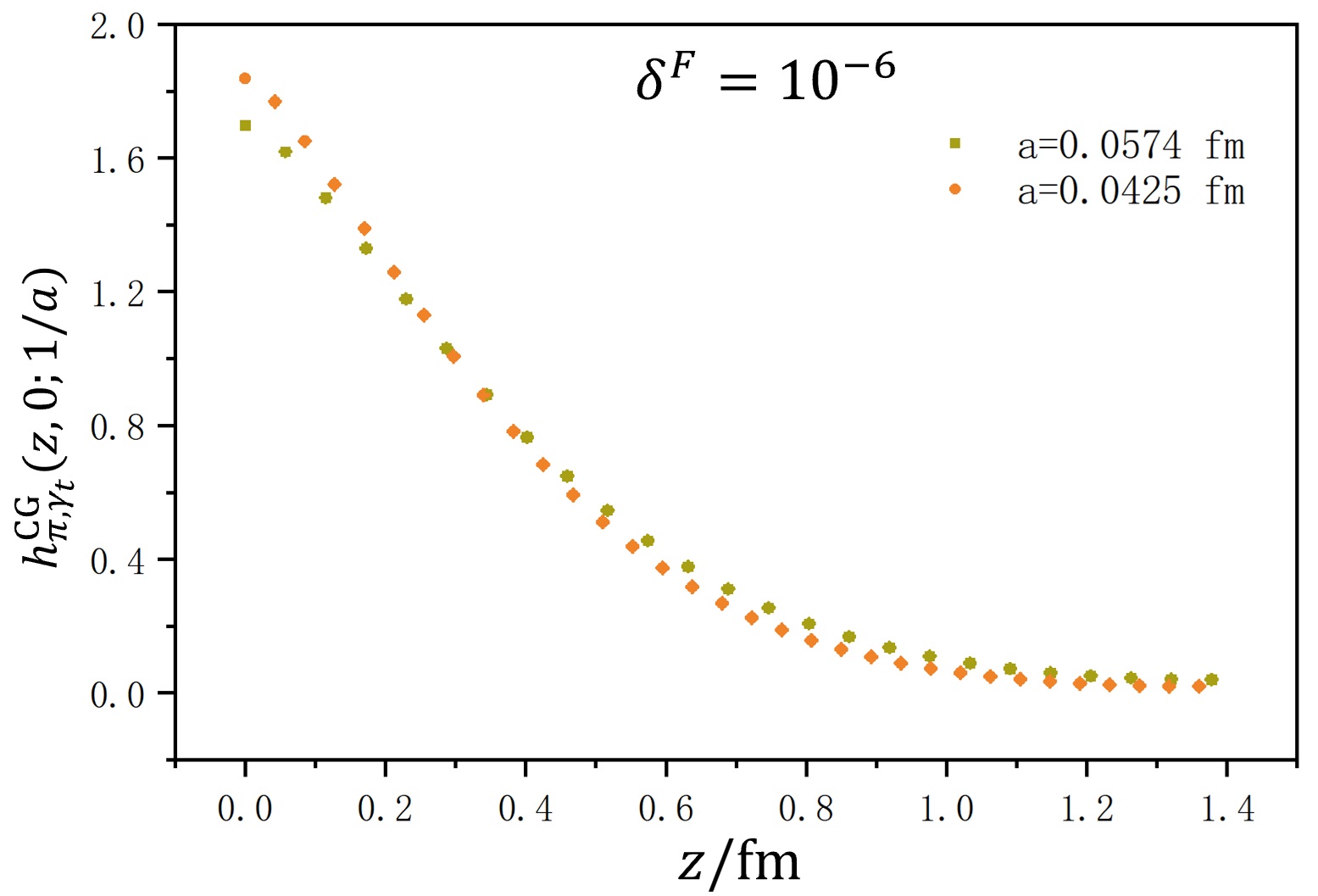}
\includegraphics[width=7cm]{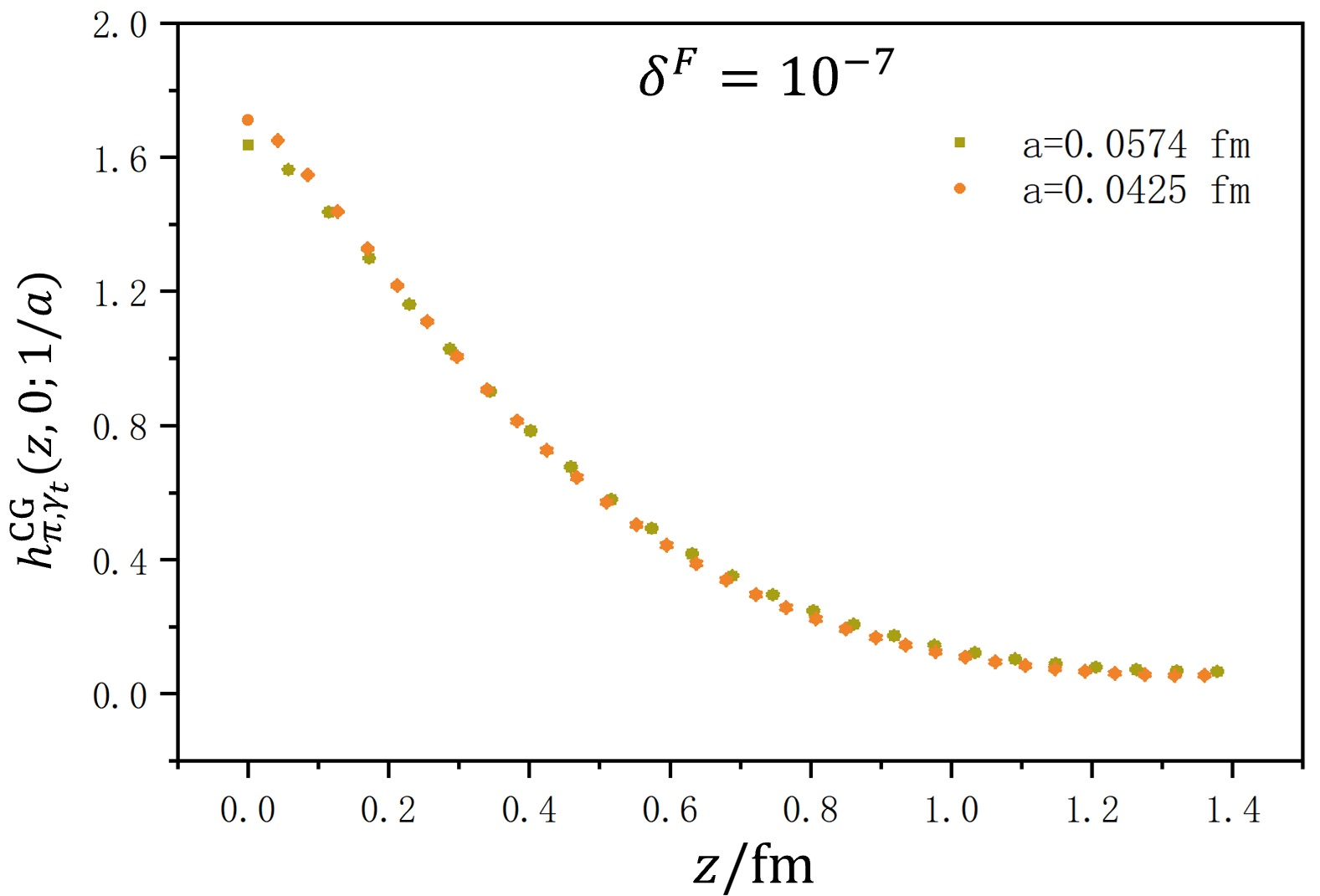}
\caption{Renormalized Coulomb gauge quasi PDF $h^{\text{CG}}_{\pi,\gamma_t}(z,0;1/a)$ { for different required precision} of gauge fixing and on different ensembles. We { studied the unpolarized case ($\Gamma=\gamma_t$) for the pion in the rest frame as an illustration}. $z_0$ is interpolated to 0.3 fm in this work.}
\label{fig:CGQPDF}
\end{figure}

We performed calculations on the MILC ensembles a06m310 and a045m310, with lattice spacings close to 0.06 fm and 0.04 fm, respectively, as used in Ref.~\cite{Gao:2023lny}. As shown in Fig.~\ref{fig:CGQPDF}, the results are not consistent with each other when the precision of Coulomb gauge fixing is set to $\delta^F = 10^{-6}$. If we improve the precision to $\delta^F = 10^{-7}$, the results are more consistent. This suggests  absence of linear divergences or residual logarithm divergences in normalized CG qPDFs.

\section{Quasi TMD-PDF}\label{sec:tmd}

As a natural generalization of the collinear PDFs, the transverse-momentum-dependent (TMD) PDFs 
provide a useful description of the transverse structure of hadrons. In LaMET, the calculation of TMD-PDFs starts from the unsubtracted quasi TMD-PDF~\cite{Ji:2019sxk,Ebert:2019okf,Ji:2020ect}. { The quasi} TMD-PDF operator is a quark bilinear operator with staple-shaped Wilson line,
\begin{align}\label{eq:QuasiTMD-PDF}
O^{\text{TMD}}_{\Gamma}(b,z,L)&\equiv \bar{\psi}(\vec{0}_{\perp},0) \Gamma {\cal W}(b,z,L) \psi (\vec{b}_{\perp},z).
\end{align}
$L$ should be large enough to approximate the infinitely long Wilson line in the continuum. Similar as for the quasi PDF case, we focus on the pion matrix element and the unpolarized case ($\Gamma=\gamma_t$). Details in the RI/MOM scheme can be found in Ref.~\cite{Zhang:2022xuw}. Similar to quasi PDFs, quasi TMD-PDF hadron matrix elements are denoted as $\tilde{h}^{\text{TMD}}_{\chi,\Gamma}(b,z,P_z;1/a)$, and renormalized matrix elements in the RI/MOM scheme are denoted as $h^{{\text{TMD}},\text{RI/MOM}}_{\chi,\Gamma}(b,z,P_z;\mu)$.

\begin{figure}[tbph]
\centering
\includegraphics[width=7cm]{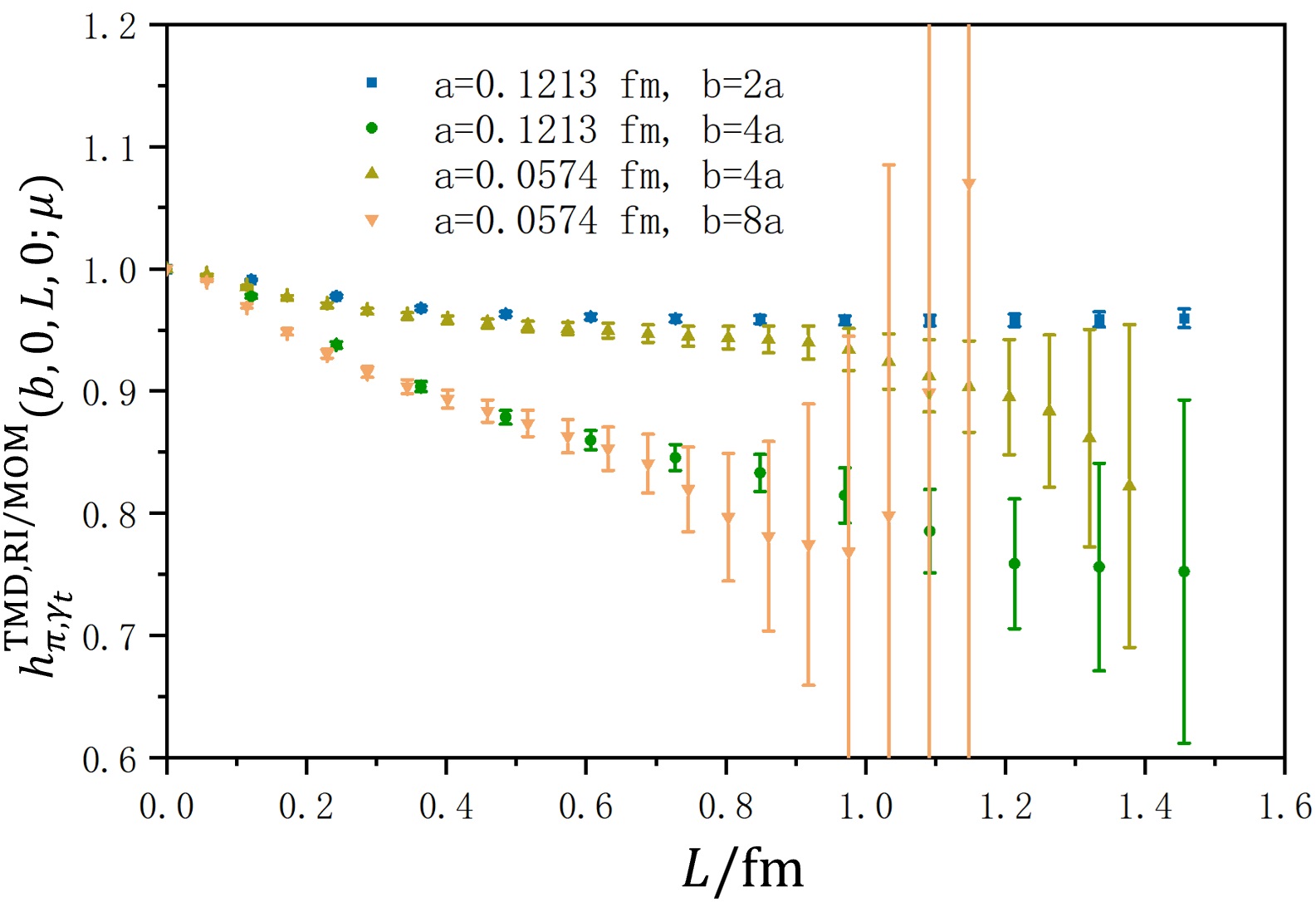}
\includegraphics[width=7cm]{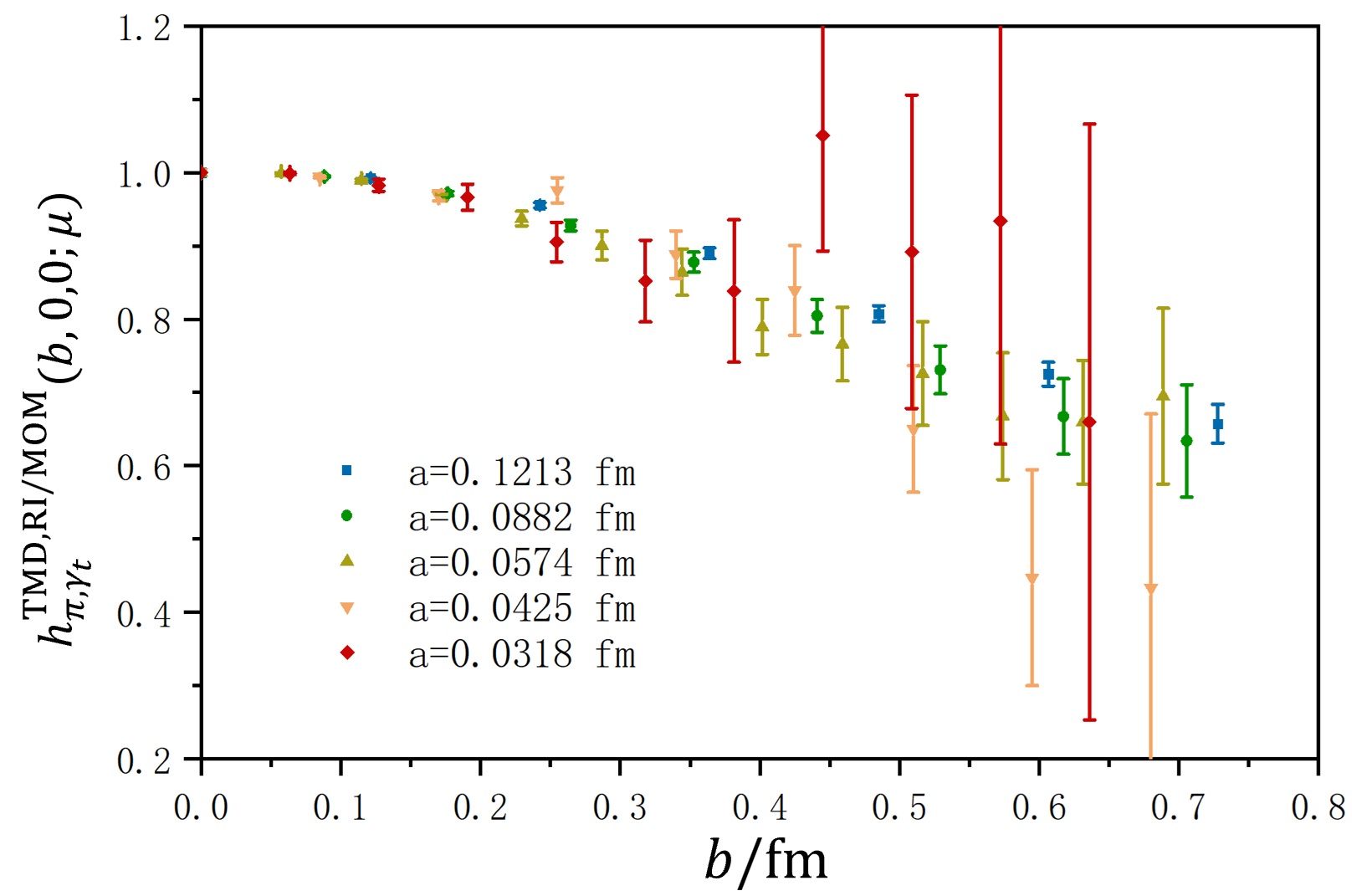}
\caption{Quasi TMD-PDF matrix element in RI/MOM scheme. The operator $O^{\text{TMD}}_{\Gamma}(b,z,L)$ is defined in Eq.~(\ref{eq:QuasiTMD-PDF}). The upper panel shows the $L$ dependence { for} different $b$ on a06m310. The lower panel shows the $b$ dependence on five ensembles. All the quark matrix elements are calculated at $\delta^F = 10^{-8}$.}
\label{fig:TMDRIRenorm}
\end{figure}

As shown in the upper panel of Fig.~\ref{fig:TMDRIRenorm}, the results from different $b$ all saturate to a plateau. This indicates that the pinch pole singularity has been eliminated by the quark matrix element on { the} lattice. For a well-defined quasi-TMD operator, $L$ should be { larger than the spatial parton distribution and independent of $L$ after proper subtraction. However,} the statistical uncertainty will be larger at longer $L$. In practice, we can choose a proper $L$ to balance the statistical uncertainty and systematic uncertainty, as we did in Ref.~\cite{Zhang:2022xuw}.

As shown  in the lower panel of Fig.~\ref{fig:TMDRIRenorm}, the results at $\delta^F = 10^{-8}$ from different lattice spacings exhibit consistency, similar to the behavior observed in { the} quasi PDF case with $\delta^F = 10^{-12}$. This suggests that the elimination of linear divergence, logarithmic divergence and cusp divergence have been achieved, and the precision $\delta^F=10^{-8}$ is adequate in this case. If the precision of Landau gauge fixing is $\delta^F = 10^{-6}$ as { as in Ref.~\cite{Zhang:2022xuw},  convergent} behavior cannot be observed in the RI/MOM scheme.

The conversion factor between RI/MOM and $\overline{\text{MS}}$ can be found in Ref.~\cite{Constantinou:2019vyb,Ebert:2019tvc}. The reason why their perturbative results do not converge is that there is { a pinch} pole singularity in quasi TMD-PDFs. As $L$ increases, the conversion factor does not converge, resulting in a huge value. One method that can be used to remove the pinch pole singularity~\cite{Ji:2020ect} is dividing quasi TMD-PDF matrix { elements} by the square root of a { rectangular} Wilson loop. The calculation on { a} lattice will not be affected because the contributions of { a Wilson loop in the quark matrix element and the hadron matrix element} cancel each other out. This procedure has also been mentioned in other work, such as Ref.~\cite{Alexandrou:2023ucc,Spanoudes:2024kpb}.{ What we} really need to do is adding a Wilson loop term to the conversion factor.

\begin{figure}[tbph]
\centering
\includegraphics[width=7cm]{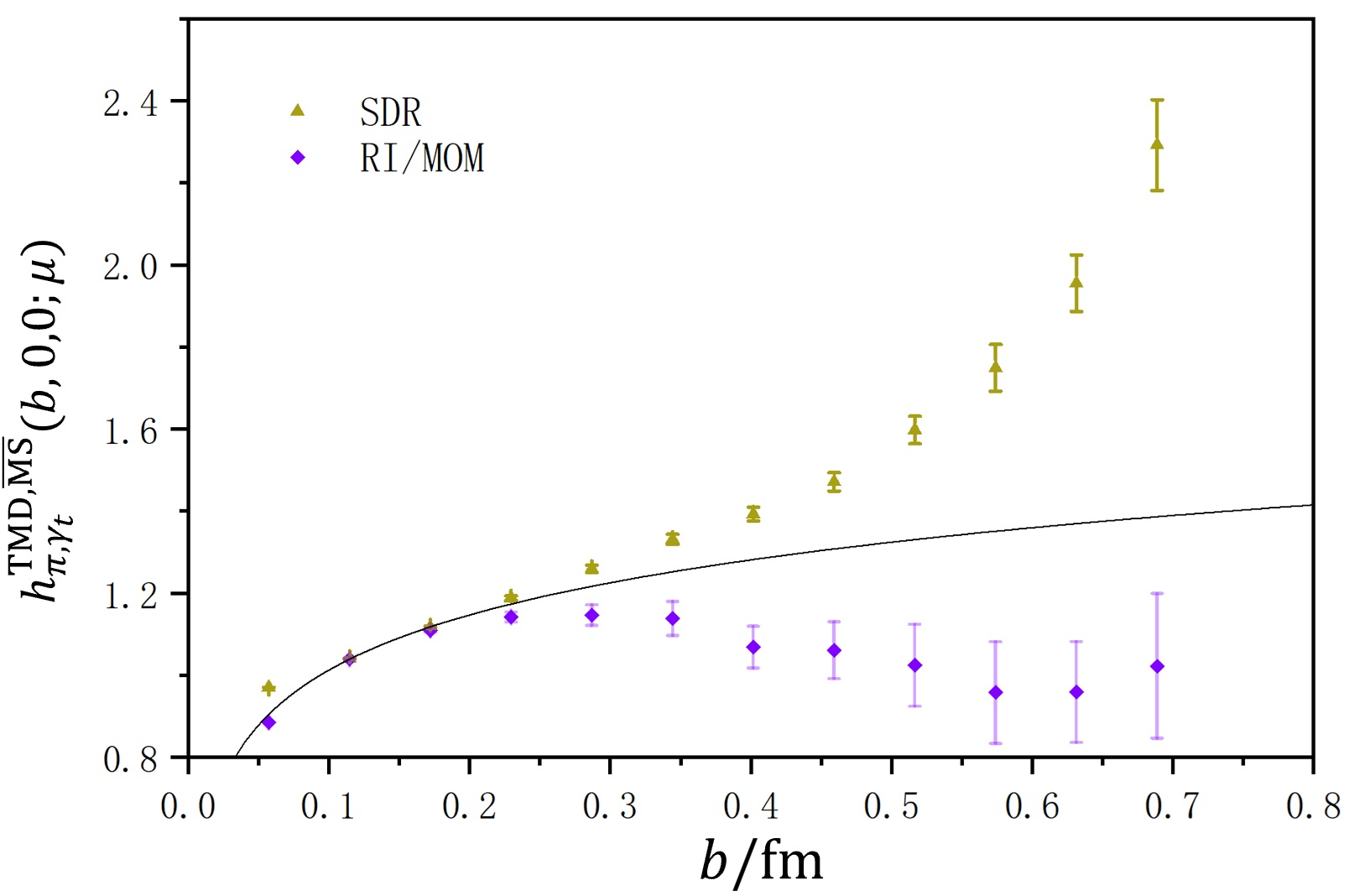}
\caption{Quasi TMD-PDF matrix element on a06m310 in $\overline{\text{MS}}$ scheme, $\mu = \text{2GeV}$. The result from { the} RI/MOM scheme is converted to { the} $\overline{\text{MS}}$ scheme by perturbative matching { at} 1-loop level. The curve is the perturbative result { at} 1-loop level in { the} $\overline{\text{MS}}$ scheme.}
\label{fig:TMDMS}
\end{figure}

On the other hand, Ref.~\cite{Zhang:2022xuw} proposed the short distance ratio scheme which is gauge invariant and can also be converted to { the} $\overline{\text{MS}}$ scheme through perturbative matching. To get rid of the pinch pole singularity and the linear divergence from the Wilson link self-energy, a more convenient ``subtracted" quasi-TMD-PDF is then formed~\cite{Ji:2019sxk},
\begin{align}\label{eq:quasiTMDcorr}
h^{\text{TMD}}_{\chi,\gamma_t}(b,z,P_z;1/a)&=\lim_{L\to\infty}\frac{\tilde{h}^{\text{TMD}}_{\chi,\gamma_t}(b,z,L,P_z;1/a)}{\sqrt{Z_E(b,2L+z;1/a)}},
\end{align}
We write the fully renormalized quasi-TMD-PDF as
\begin{align}\label{eq:quasiTMDcorrSDR}
h^{\text{TMD},{\rm SDR}}_{\chi,\gamma_t}(b,z,P_z;\frac{1}{b_0})=\frac{h^{\text{TMD}}_{\chi,\gamma_t}(b,z,P_z;1/a)}{h^{\text{TMD}}_{\pi,\gamma_t}(b_0,z_0=0,0,1/a)},
\end{align}
where we also chosen the pion matrix element in the denominator (denoted by the subscript $\pi$) likes the quasi-PDF case. For simplicity, we have also chosen $z_0=0$. The singular dependence on $a$ on the r.h.s. of Eq. (\ref{eq:quasiTMDcorrSDR}) has been cancelled, leaving a dependence on the perturbative short scale $b_0$. 
To perform the renormalization, the pion matrix element $h_{\pi,\gamma_t}(b_0,0,0, 1/a)$ needs to be calculated non-perturbatively on the lattice. 
In order to match the renormalized quasi-TMD-PDF to the standard TMD-PDF, we also need the perturbative results of the above matrix elements, for which we can choose on-shell quark external states. The numerator 
of the r.h.s. of Eq.~(\ref{eq:quasiTMDcorrSDR}) has been calculated previously in dimensional regularization (DR) and $\overline{\rm MS}$ scheme in Ref.~\cite{Ebert:2019okf}, whereas the denominator is independent of $\chi$ and reads
\begin{align}\label{eq:hMSbar} 
&h^{\text{TMD},\overline{\rm MS}}_{{\rm pert},\gamma_t}(b_0,z_0,0;\mu)=1+\frac{\alpha_s C_F}{2\pi}\bigg\{\frac{1}{2}+3\gamma_E-3\text{ln}2\nonumber\\
&\quad \quad +\frac{3}{2}\text{ln}[\mu^2(b_0^2+z_0^2)]-2\frac{z_0}{b_0}\text{arctan}\frac{z_0}{b_0}\bigg\}+{\cal O}(\alpha_s^2).
\end{align}
With this, we can also convert the SDR result to the $\overline{\rm MS}$ scheme via
\begin{align}\label{eq:quasiTMDcorr1}
&h^{\text{TMD},\overline{\rm MS}}_{\chi,\gamma_t}(b,z,P_z;\mu)= \nonumber\\
&h^{\text{TMD},\overline{\rm MS}}_{{\rm pert},\gamma_t}(b_0,0,0;\mu)h^{\text{TMD},{\rm SDR}}_{\chi,\gamma_t}(b,z,P_z;\frac{1}{b_0}),
\end{align}
where the $b_0$ dependence cancels.

As shown in Fig.~\ref{fig:TMDMS}, both results are consistent with each other and are close to  the perturbative result at short distance, and they deviate from each other at long distance due to the non-perturbative effects from the intrinsic scales $1/z$ and/or $1/b$ introduced in RI/MOM scheme.

\section{Summary}\label{sec:summary}

We have investigated the gauge fixing precision dependence for several measurements, using 5 lattice spacings, in Landau gauge, Coulomb gauge and $\xi$ gauge. Our results indicate that the linear divergences in quasi PDF operators and quasi TMD-PDFs all comes from Wilson lines, and are independent of the external state {when the gauge fixing is precise enough}. There is also no linear divergence in Coulomb gauge quasi PDFs renormalized by short distance matrix elements. 

Thus, we verified that self-renormalization, which fits the hadron matrix elements at different lattice spacings, RI/MOM renormalization, which uses the quark matrix element and precise enough Landau gauge fixing, and the short-distance renormalization, which divides by the corresponding Wilson line or square root of the Wilson loop, can fully eliminate the UV divergence of the quasi-PDF and quasi-TMD matrix elements, and make the renormalized ones finite in the continuum limit. Obviously the combinations of RI/MOM and SDR schemes (e.g. the short distance RI/MOM scheme~\cite{Alexandrou:2023ucc}) should also have a finite continuum limit. 

At the same time, even though all the approaches are consistent with the perturbative results at short distances up to the discretization error, they can be significantly different at long distances due to introduction of spurious infrared physics in RI/MOM and similar schemes. Therefore, the proper removal of the non-perturbative effects, such as self-renormalization, should be crucial to obtain reliable matrix elements at long distances.

When the gauge fixing precision is restricted, the dependence of the results on the gauge fixing residual $\delta^F$ can be roughly approximated by $e^{-c\sqrt{\delta^F}}$, where $c$ grows quadratically with the Wilson line length $z/a$ for both Wilson lines and quark matrix elements. Since the accurate gauge fixing with $\delta^F$ can be extremely costly on large lattices, this empirical formula can aid in estimating the systematic uncertainty stemming from imprecise gauge fixing. For example, if we requires the systematic uncertainty from the imprecise gauge fixing to be 1\% at $a=0.05$ fm and $z\sim 1$ fm $=20a$, then the $\delta^F$ should be around $(10^{-2}/(z/a)^2)^2\sim 6\times 10^{-10}$.

\section*{Acknowledgement}
We thank the MILC collaborations for providing us their gauge configurations, and Xiang Gao, Xiangyu Jiang, Peter Petreczky, 
and Yong Zhao for useful information and discussions. The calculations were performed using the Chroma software suite~\cite{Edwards:2004sx} with QUDA~\cite{Clark:2009wm,Babich:2011np,Clark:2016rdz} through HIP programming model~\cite{Bi:2020wpt}. The numerical calculation were carried out on the ORISE Supercomputer, and HPC Cluster of ITP-CAS. This work is supported in part by NSFC grants No. 12293060, 12293062, 12293065 and 12047503, the science and education integration young faculty project of University of Chinese Academy of Sciences, the Strategic Priority Research Program of Chinese Academy of Sciences, Grant No.\ XDB34030303 and YSBR-101, and also a NSFC-DFG joint grant under Grant No.\ 12061131006 and SCHA 458/22.

\bibliography{ref}

\begin{thebibliography}{46}%
\makeatletter
\providecommand \@ifxundefined [1]{%
 \@ifx{#1\undefined}
}%
\providecommand \@ifnum [1]{%
 \ifnum #1\expandafter \@firstoftwo
 \else \expandafter \@secondoftwo
 \fi
}%
\providecommand \@ifx [1]{%
 \ifx #1\expandafter \@firstoftwo
 \else \expandafter \@secondoftwo
 \fi
}%
\providecommand \natexlab [1]{#1}%
\providecommand \enquote  [1]{``#1''}%
\providecommand \bibnamefont  [1]{#1}%
\providecommand \bibfnamefont [1]{#1}%
\providecommand \citenamefont [1]{#1}%
\providecommand \href@noop [0]{\@secondoftwo}%
\providecommand \href [0]{\begingroup \@sanitize@url \@href}%
\providecommand \@href[1]{\@@startlink{#1}\@@href}%
\providecommand \@@href[1]{\endgroup#1\@@endlink}%
\providecommand \@sanitize@url [0]{\catcode `\\12\catcode `\$12\catcode
  `\&12\catcode `\#12\catcode `\^12\catcode `\_12\catcode `\%12\relax}%
\providecommand \@@startlink[1]{}%
\providecommand \@@endlink[0]{}%
\providecommand \url  [0]{\begingroup\@sanitize@url \@url }%
\providecommand \@url [1]{\endgroup\@href {#1}{\urlprefix }}%
\providecommand \urlprefix  [0]{URL }%
\providecommand \Eprint [0]{\href }%
\providecommand \doibase [0]{http://dx.doi.org/}%
\providecommand \selectlanguage [0]{\@gobble}%
\providecommand \bibinfo  [0]{\@secondoftwo}%
\providecommand \bibfield  [0]{\@secondoftwo}%
\providecommand \translation [1]{[#1]}%
\providecommand \BibitemOpen [0]{}%
\providecommand \bibitemStop [0]{}%
\providecommand \bibitemNoStop [0]{.\EOS\space}%
\providecommand \EOS [0]{\spacefactor3000\relax}%
\providecommand \BibitemShut  [1]{\csname bibitem#1\endcsname}%
\let\auto@bib@innerbib\@empty
\bibitem [{\citenamefont {Martinelli}\ \emph {et~al.}(1995)\citenamefont
  {Martinelli}, \citenamefont {Pittori}, \citenamefont {Sachrajda},
  \citenamefont {Testa},\ and\ \citenamefont {Vladikas}}]{Martinelli:1994ty}%
  \BibitemOpen
  \bibfield  {author} {\bibinfo {author} {\bibfnamefont {G.}~\bibnamefont
  {Martinelli}}, \bibinfo {author} {\bibfnamefont {C.}~\bibnamefont {Pittori}},
  \bibinfo {author} {\bibfnamefont {C.~T.}\ \bibnamefont {Sachrajda}}, \bibinfo
  {author} {\bibfnamefont {M.}~\bibnamefont {Testa}}, \ and\ \bibinfo {author}
  {\bibfnamefont {A.}~\bibnamefont {Vladikas}},\ }\href {\doibase
  10.1016/0550-3213(95)00126-D} {\bibfield  {journal} {\bibinfo  {journal}
  {Nucl. Phys.}\ }\textbf {\bibinfo {volume} {B445}},\ \bibinfo {pages} {81}
  (\bibinfo {year} {1995})},\ \Eprint {http://arxiv.org/abs/hep-lat/9411010}
  {arXiv:hep-lat/9411010 [hep-lat]} \BibitemShut {NoStop}%
\bibitem [{\citenamefont {Mandula}(1999)}]{Mandula:1999nj}%
  \BibitemOpen
  \bibfield  {author} {\bibinfo {author} {\bibfnamefont {J.~E.}\ \bibnamefont
  {Mandula}},\ }\href {\doibase 10.1016/S0370-1573(99)00027-7} {\bibfield
  {journal} {\bibinfo  {journal} {Phys. Rept.}\ }\textbf {\bibinfo {volume}
  {315}},\ \bibinfo {pages} {273} (\bibinfo {year} {1999})},\ \Eprint
  {http://arxiv.org/abs/hep-lat/9907020} {arXiv:hep-lat/9907020} \BibitemShut
  {NoStop}%
\bibitem [{\citenamefont {Wilson}(1974)}]{Wilson:1974sk}%
  \BibitemOpen
  \bibfield  {author} {\bibinfo {author} {\bibfnamefont {K.~G.}\ \bibnamefont
  {Wilson}},\ }\href {\doibase 10.1103/PhysRevD.10.2445} {\bibfield  {journal}
  {\bibinfo  {journal} {Phys. Rev. D}\ }\textbf {\bibinfo {volume} {10}},\
  \bibinfo {pages} {2445} (\bibinfo {year} {1974})}\BibitemShut {NoStop}%
\bibitem [{\citenamefont {Mandula}\ and\ \citenamefont
  {Ogilvie}(1987)}]{Mandula:1987rh}%
  \BibitemOpen
  \bibfield  {author} {\bibinfo {author} {\bibfnamefont {J.~E.}\ \bibnamefont
  {Mandula}}\ and\ \bibinfo {author} {\bibfnamefont {M.}~\bibnamefont
  {Ogilvie}},\ }\href {\doibase 10.1016/0370-2693(87)91541-3} {\bibfield
  {journal} {\bibinfo  {journal} {Phys. Lett. B}\ }\textbf {\bibinfo {volume}
  {185}},\ \bibinfo {pages} {127} (\bibinfo {year} {1987})}\BibitemShut
  {NoStop}%
\bibitem [{\citenamefont {Katz}\ \emph {et~al.}(1988)\citenamefont {Katz},
  \citenamefont {Batrouni}, \citenamefont {Davies}, \citenamefont {Kronfeld},
  \citenamefont {Lepage}, \citenamefont {Rossi}, \citenamefont {Svetitsky},\
  and\ \citenamefont {Wilson}}]{Katz:1987ti}%
  \BibitemOpen
  \bibfield  {author} {\bibinfo {author} {\bibfnamefont {G.}~\bibnamefont
  {Katz}}, \bibinfo {author} {\bibfnamefont {G.}~\bibnamefont {Batrouni}},
  \bibinfo {author} {\bibfnamefont {C.}~\bibnamefont {Davies}}, \bibinfo
  {author} {\bibfnamefont {A.~S.}\ \bibnamefont {Kronfeld}}, \bibinfo {author}
  {\bibfnamefont {P.}~\bibnamefont {Lepage}}, \bibinfo {author} {\bibfnamefont
  {P.}~\bibnamefont {Rossi}}, \bibinfo {author} {\bibfnamefont
  {B.}~\bibnamefont {Svetitsky}}, \ and\ \bibinfo {author} {\bibfnamefont
  {K.}~\bibnamefont {Wilson}},\ }\href {\doibase 10.1103/PhysRevD.37.1589}
  {\bibfield  {journal} {\bibinfo  {journal} {Phys. Rev. D}\ }\textbf {\bibinfo
  {volume} {37}},\ \bibinfo {pages} {1589} (\bibinfo {year}
  {1988})}\BibitemShut {NoStop}%
\bibitem [{\citenamefont {Giusti}\ \emph
  {et~al.}(2001{\natexlab{a}})\citenamefont {Giusti}, \citenamefont {Paciello},
  \citenamefont {Parrinello}, \citenamefont {Petrarca},\ and\ \citenamefont
  {Taglienti}}]{Giusti:2001xf}%
  \BibitemOpen
  \bibfield  {author} {\bibinfo {author} {\bibfnamefont {L.}~\bibnamefont
  {Giusti}}, \bibinfo {author} {\bibfnamefont {M.~L.}\ \bibnamefont
  {Paciello}}, \bibinfo {author} {\bibfnamefont {C.}~\bibnamefont
  {Parrinello}}, \bibinfo {author} {\bibfnamefont {S.}~\bibnamefont
  {Petrarca}}, \ and\ \bibinfo {author} {\bibfnamefont {B.}~\bibnamefont
  {Taglienti}},\ }\href {\doibase 10.1142/S0217751X01004281} {\bibfield
  {journal} {\bibinfo  {journal} {Int. J. Mod. Phys. A}\ }\textbf {\bibinfo
  {volume} {16}},\ \bibinfo {pages} {3487} (\bibinfo {year}
  {2001}{\natexlab{a}})},\ \Eprint {http://arxiv.org/abs/hep-lat/0104012}
  {arXiv:hep-lat/0104012} \BibitemShut {NoStop}%
\bibitem [{\citenamefont {Lytle}\ \emph {et~al.}(2018)\citenamefont {Lytle},
  \citenamefont {Davies}, \citenamefont {Hatton}, \citenamefont {Lepage},\ and\
  \citenamefont {Sturm}}]{Lytle:2018evc}%
  \BibitemOpen
  \bibfield  {author} {\bibinfo {author} {\bibfnamefont {A.~T.}\ \bibnamefont
  {Lytle}}, \bibinfo {author} {\bibfnamefont {C.~T.~H.}\ \bibnamefont
  {Davies}}, \bibinfo {author} {\bibfnamefont {D.}~\bibnamefont {Hatton}},
  \bibinfo {author} {\bibfnamefont {G.~P.}\ \bibnamefont {Lepage}}, \ and\
  \bibinfo {author} {\bibfnamefont {C.}~\bibnamefont {Sturm}} (\bibinfo
  {collaboration} {HPQCD}),\ }\href {\doibase 10.1103/PhysRevD.98.014513}
  {\bibfield  {journal} {\bibinfo  {journal} {Phys. Rev. D}\ }\textbf {\bibinfo
  {volume} {98}},\ \bibinfo {pages} {014513} (\bibinfo {year} {2018})},\
  \Eprint {http://arxiv.org/abs/1805.06225} {arXiv:1805.06225 [hep-lat]}
  \BibitemShut {NoStop}%
\bibitem [{\citenamefont {Ji}(2013)}]{Ji:2013dva}%
  \BibitemOpen
  \bibfield  {author} {\bibinfo {author} {\bibfnamefont {X.}~\bibnamefont
  {Ji}},\ }\href {\doibase 10.1103/PhysRevLett.110.262002} {\bibfield
  {journal} {\bibinfo  {journal} {Phys. Rev. Lett.}\ }\textbf {\bibinfo
  {volume} {110}},\ \bibinfo {pages} {262002} (\bibinfo {year} {2013})},\
  \Eprint {http://arxiv.org/abs/1305.1539} {arXiv:1305.1539 [hep-ph]}
  \BibitemShut {NoStop}%
\bibitem [{\citenamefont {Ji}\ \emph {et~al.}(2021)\citenamefont {Ji},
  \citenamefont {Liu}, \citenamefont {Liu}, \citenamefont {Zhang},\ and\
  \citenamefont {Zhao}}]{Ji:2020ect}%
  \BibitemOpen
  \bibfield  {author} {\bibinfo {author} {\bibfnamefont {X.}~\bibnamefont
  {Ji}}, \bibinfo {author} {\bibfnamefont {Y.-S.}\ \bibnamefont {Liu}},
  \bibinfo {author} {\bibfnamefont {Y.}~\bibnamefont {Liu}}, \bibinfo {author}
  {\bibfnamefont {J.-H.}\ \bibnamefont {Zhang}}, \ and\ \bibinfo {author}
  {\bibfnamefont {Y.}~\bibnamefont {Zhao}},\ }\href {\doibase
  10.1103/RevModPhys.93.035005} {\bibfield  {journal} {\bibinfo  {journal}
  {Rev. Mod. Phys.}\ }\textbf {\bibinfo {volume} {93}},\ \bibinfo {pages}
  {035005} (\bibinfo {year} {2021})},\ \Eprint
  {http://arxiv.org/abs/2004.03543} {arXiv:2004.03543 [hep-ph]} \BibitemShut
  {NoStop}%
\bibitem [{\citenamefont {Gao}\ \emph {et~al.}(2023)\citenamefont {Gao},
  \citenamefont {Liu},\ and\ \citenamefont {Zhao}}]{Gao:2023lny}%
  \BibitemOpen
  \bibfield  {author} {\bibinfo {author} {\bibfnamefont {X.}~\bibnamefont
  {Gao}}, \bibinfo {author} {\bibfnamefont {W.-Y.}\ \bibnamefont {Liu}}, \ and\
  \bibinfo {author} {\bibfnamefont {Y.}~\bibnamefont {Zhao}},\ }\href@noop {}
  {\  (\bibinfo {year} {2023})},\ \Eprint {http://arxiv.org/abs/2306.14960}
  {arXiv:2306.14960 [hep-ph]} \BibitemShut {NoStop}%
\bibitem [{\citenamefont {Hart}\ \emph {et~al.}(2009)\citenamefont {Hart},
  \citenamefont {von Hippel},\ and\ \citenamefont {Horgan}}]{Hart:2008sq}%
  \BibitemOpen
  \bibfield  {author} {\bibinfo {author} {\bibfnamefont {A.}~\bibnamefont
  {Hart}}, \bibinfo {author} {\bibfnamefont {G.~M.}\ \bibnamefont {von
  Hippel}}, \ and\ \bibinfo {author} {\bibfnamefont {R.~R.}\ \bibnamefont
  {Horgan}} (\bibinfo {collaboration} {HPQCD}),\ }\href {\doibase
  10.1103/PhysRevD.79.074008} {\bibfield  {journal} {\bibinfo  {journal} {Phys.
  Rev. D}\ }\textbf {\bibinfo {volume} {79}},\ \bibinfo {pages} {074008}
  (\bibinfo {year} {2009})},\ \Eprint {http://arxiv.org/abs/0812.0503}
  {arXiv:0812.0503 [hep-lat]} \BibitemShut {NoStop}%
\bibitem [{\citenamefont {Bazavov}\ \emph {et~al.}(2013)\citenamefont {Bazavov}
  \emph {et~al.}}]{Bazavov:2012xda}%
  \BibitemOpen
  \bibfield  {author} {\bibinfo {author} {\bibfnamefont {A.}~\bibnamefont
  {Bazavov}} \emph {et~al.} (\bibinfo {collaboration} {MILC}),\ }\href
  {\doibase 10.1103/PhysRevD.87.054505} {\bibfield  {journal} {\bibinfo
  {journal} {Phys. Rev.}\ }\textbf {\bibinfo {volume} {D87}},\ \bibinfo {pages}
  {054505} (\bibinfo {year} {2013})},\ \Eprint {http://arxiv.org/abs/1212.4768}
  {arXiv:1212.4768 [hep-lat]} \BibitemShut {NoStop}%
\bibitem [{\citenamefont {Edwards}\ and\ \citenamefont
  {Joo}(2005)}]{Edwards:2004sx}%
  \BibitemOpen
  \bibfield  {author} {\bibinfo {author} {\bibfnamefont {R.~G.}\ \bibnamefont
  {Edwards}}\ and\ \bibinfo {author} {\bibfnamefont {B.}~\bibnamefont {Joo}}
  (\bibinfo {collaboration} {SciDAC, LHPC, UKQCD}),\ }\bibfield  {booktitle}
  {\emph {\bibinfo {booktitle} {{Lattice field theory. Proceedings, 22nd
  International Symposium, Lattice 2004, Batavia, USA, June 21-26, 2004}}},\
  }\href {\doibase 10.1016/j.nuclphysbps.2004.11.254} {\bibfield  {journal}
  {\bibinfo  {journal} {Nucl. Phys. Proc. Suppl.}\ }\textbf {\bibinfo {volume}
  {140}},\ \bibinfo {pages} {832} (\bibinfo {year} {2005})},\ \bibinfo {note}
  {[,832(2004)]},\ \Eprint {http://arxiv.org/abs/hep-lat/0409003}
  {arXiv:hep-lat/0409003 [hep-lat]} \BibitemShut {NoStop}%
\bibitem [{\citenamefont {Giusti}\ \emph
  {et~al.}(2001{\natexlab{b}})\citenamefont {Giusti}, \citenamefont {Paciello},
  \citenamefont {Petrarca},\ and\ \citenamefont {Taglienti}}]{Giusti:1999im}%
  \BibitemOpen
  \bibfield  {author} {\bibinfo {author} {\bibfnamefont {L.}~\bibnamefont
  {Giusti}}, \bibinfo {author} {\bibfnamefont {M.~L.}\ \bibnamefont
  {Paciello}}, \bibinfo {author} {\bibfnamefont {S.}~\bibnamefont {Petrarca}},
  \ and\ \bibinfo {author} {\bibfnamefont {B.}~\bibnamefont {Taglienti}},\
  }\href {\doibase 10.1103/PhysRevD.63.014501} {\bibfield  {journal} {\bibinfo
  {journal} {Phys. Rev. D}\ }\textbf {\bibinfo {volume} {63}},\ \bibinfo
  {pages} {014501} (\bibinfo {year} {2001}{\natexlab{b}})},\ \Eprint
  {http://arxiv.org/abs/hep-lat/9911038} {arXiv:hep-lat/9911038} \BibitemShut
  {NoStop}%
\bibitem [{\citenamefont {Zwanziger}(1986)}]{Zwanziger:1985vi}%
  \BibitemOpen
  \bibfield  {author} {\bibinfo {author} {\bibfnamefont {D.}~\bibnamefont
  {Zwanziger}},\ }\href@noop {} {\bibfield  {journal} {\bibinfo  {journal}
  {NATO Sci. Ser. B}\ }\textbf {\bibinfo {volume} {141}},\ \bibinfo {pages}
  {345} (\bibinfo {year} {1986})}\BibitemShut {NoStop}%
\bibitem [{\citenamefont {Giusti}(1997)}]{Giusti:1996kf}%
  \BibitemOpen
  \bibfield  {author} {\bibinfo {author} {\bibfnamefont {L.}~\bibnamefont
  {Giusti}},\ }\href {\doibase 10.1016/S0550-3213(97)00273-3} {\bibfield
  {journal} {\bibinfo  {journal} {Nucl. Phys. B}\ }\textbf {\bibinfo {volume}
  {498}},\ \bibinfo {pages} {331} (\bibinfo {year} {1997})},\ \Eprint
  {http://arxiv.org/abs/hep-lat/9605032} {arXiv:hep-lat/9605032} \BibitemShut
  {NoStop}%
\bibitem [{\citenamefont {Ji}\ \emph {et~al.}(2020)\citenamefont {Ji},
  \citenamefont {Liu},\ and\ \citenamefont {Liu}}]{Ji:2019sxk}%
  \BibitemOpen
  \bibfield  {author} {\bibinfo {author} {\bibfnamefont {X.}~\bibnamefont
  {Ji}}, \bibinfo {author} {\bibfnamefont {Y.}~\bibnamefont {Liu}}, \ and\
  \bibinfo {author} {\bibfnamefont {Y.-S.}\ \bibnamefont {Liu}},\ }\href
  {\doibase 10.1016/j.nuclphysb.2020.115054} {\bibfield  {journal} {\bibinfo
  {journal} {Nucl. Phys. B}\ }\textbf {\bibinfo {volume} {955}},\ \bibinfo
  {pages} {115054} (\bibinfo {year} {2020})},\ \Eprint
  {http://arxiv.org/abs/1910.11415} {arXiv:1910.11415 [hep-ph]} \BibitemShut
  {NoStop}%
\bibitem [{\citenamefont {Ji}(2014)}]{Ji:2014gla}%
  \BibitemOpen
  \bibfield  {author} {\bibinfo {author} {\bibfnamefont {X.}~\bibnamefont
  {Ji}},\ }\href {\doibase 10.1007/s11433-014-5492-3} {\bibfield  {journal}
  {\bibinfo  {journal} {Sci. China Phys. Mech. Astron.}\ }\textbf {\bibinfo
  {volume} {57}},\ \bibinfo {pages} {1407} (\bibinfo {year} {2014})},\ \Eprint
  {http://arxiv.org/abs/1404.6680} {arXiv:1404.6680 [hep-ph]} \BibitemShut
  {NoStop}%
\bibitem [{\citenamefont {Ma}\ and\ \citenamefont
  {Qiu}(2018{\natexlab{a}})}]{Ma:2014jla}%
  \BibitemOpen
  \bibfield  {author} {\bibinfo {author} {\bibfnamefont {Y.-Q.}\ \bibnamefont
  {Ma}}\ and\ \bibinfo {author} {\bibfnamefont {J.-W.}\ \bibnamefont {Qiu}},\
  }\href {\doibase 10.1103/PhysRevD.98.074021} {\bibfield  {journal} {\bibinfo
  {journal} {Phys. Rev.}\ }\textbf {\bibinfo {volume} {D98}},\ \bibinfo {pages}
  {074021} (\bibinfo {year} {2018}{\natexlab{a}})},\ \Eprint
  {http://arxiv.org/abs/1404.6860} {arXiv:1404.6860 [hep-ph]} \BibitemShut
  {NoStop}%
\bibitem [{\citenamefont {Ma}\ and\ \citenamefont
  {Qiu}(2018{\natexlab{b}})}]{Ma:2017pxb}%
  \BibitemOpen
  \bibfield  {author} {\bibinfo {author} {\bibfnamefont {Y.-Q.}\ \bibnamefont
  {Ma}}\ and\ \bibinfo {author} {\bibfnamefont {J.-W.}\ \bibnamefont {Qiu}},\
  }\href {\doibase 10.1103/PhysRevLett.120.022003} {\bibfield  {journal}
  {\bibinfo  {journal} {Phys. Rev. Lett.}\ }\textbf {\bibinfo {volume} {120}},\
  \bibinfo {pages} {022003} (\bibinfo {year} {2018}{\natexlab{b}})},\ \Eprint
  {http://arxiv.org/abs/1709.03018} {arXiv:1709.03018 [hep-ph]} \BibitemShut
  {NoStop}%
\bibitem [{\citenamefont {Izubuchi}\ \emph {et~al.}(2018)\citenamefont
  {Izubuchi}, \citenamefont {Ji}, \citenamefont {Jin}, \citenamefont
  {Stewart},\ and\ \citenamefont {Zhao}}]{Izubuchi:2018srq}%
  \BibitemOpen
  \bibfield  {author} {\bibinfo {author} {\bibfnamefont {T.}~\bibnamefont
  {Izubuchi}}, \bibinfo {author} {\bibfnamefont {X.}~\bibnamefont {Ji}},
  \bibinfo {author} {\bibfnamefont {L.}~\bibnamefont {Jin}}, \bibinfo {author}
  {\bibfnamefont {I.~W.}\ \bibnamefont {Stewart}}, \ and\ \bibinfo {author}
  {\bibfnamefont {Y.}~\bibnamefont {Zhao}},\ }\href {\doibase
  10.1103/PhysRevD.98.056004} {\bibfield  {journal} {\bibinfo  {journal} {Phys.
  Rev.}\ }\textbf {\bibinfo {volume} {D98}},\ \bibinfo {pages} {056004}
  (\bibinfo {year} {2018})},\ \Eprint {http://arxiv.org/abs/1801.03917}
  {arXiv:1801.03917 [hep-ph]} \BibitemShut {NoStop}%
\bibitem [{\citenamefont {Zhang}\ \emph {et~al.}(2021)\citenamefont {Zhang},
  \citenamefont {Li}, \citenamefont {Huo}, \citenamefont {Sch\"afer},
  \citenamefont {Sun},\ and\ \citenamefont {Yang}}]{Zhang:2020rsx}%
  \BibitemOpen
  \bibfield  {author} {\bibinfo {author} {\bibfnamefont {K.}~\bibnamefont
  {Zhang}}, \bibinfo {author} {\bibfnamefont {Y.-Y.}\ \bibnamefont {Li}},
  \bibinfo {author} {\bibfnamefont {Y.-K.}\ \bibnamefont {Huo}}, \bibinfo
  {author} {\bibfnamefont {A.}~\bibnamefont {Sch\"afer}}, \bibinfo {author}
  {\bibfnamefont {P.}~\bibnamefont {Sun}}, \ and\ \bibinfo {author}
  {\bibfnamefont {Y.-B.}\ \bibnamefont {Yang}} (\bibinfo {collaboration}
  {\ensuremath{\chi}QCD}),\ }\href {\doibase 10.1103/PhysRevD.104.074501}
  {\bibfield  {journal} {\bibinfo  {journal} {Phys. Rev. D}\ }\textbf {\bibinfo
  {volume} {104}},\ \bibinfo {pages} {074501} (\bibinfo {year} {2021})},\
  \Eprint {http://arxiv.org/abs/2012.05448} {arXiv:2012.05448 [hep-lat]}
  \BibitemShut {NoStop}%
\bibitem [{\citenamefont {Huo}\ \emph {et~al.}(2021)\citenamefont {Huo} \emph
  {et~al.}}]{LatticePartonCollaborationLPC:2021xdx}%
  \BibitemOpen
  \bibfield  {author} {\bibinfo {author} {\bibfnamefont {Y.-K.}\ \bibnamefont
  {Huo}} \emph {et~al.} (\bibinfo {collaboration} {Lattice Parton Collaboration
  (LPC)}),\ }\href {\doibase 10.1016/j.nuclphysb.2021.115443} {\bibfield
  {journal} {\bibinfo  {journal} {Nucl. Phys. B}\ }\textbf {\bibinfo {volume}
  {969}},\ \bibinfo {pages} {115443} (\bibinfo {year} {2021})},\ \Eprint
  {http://arxiv.org/abs/2103.02965} {arXiv:2103.02965 [hep-lat]} \BibitemShut
  {NoStop}%
\bibitem [{\citenamefont {Zhang}\ \emph {et~al.}(2022)\citenamefont {Zhang},
  \citenamefont {Ji}, \citenamefont {Yang}, \citenamefont {Yao},\ and\
  \citenamefont {Zhang}}]{Zhang:2022xuw}%
  \BibitemOpen
  \bibfield  {author} {\bibinfo {author} {\bibfnamefont {K.}~\bibnamefont
  {Zhang}}, \bibinfo {author} {\bibfnamefont {X.}~\bibnamefont {Ji}}, \bibinfo
  {author} {\bibfnamefont {Y.-B.}\ \bibnamefont {Yang}}, \bibinfo {author}
  {\bibfnamefont {F.}~\bibnamefont {Yao}}, \ and\ \bibinfo {author}
  {\bibfnamefont {J.-H.}\ \bibnamefont {Zhang}} (\bibinfo {collaboration}
  {[Lattice Parton Collaboration (LPC)]}),\ }\href {\doibase
  10.1103/PhysRevLett.129.082002} {\bibfield  {journal} {\bibinfo  {journal}
  {Phys. Rev. Lett.}\ }\textbf {\bibinfo {volume} {129}},\ \bibinfo {pages}
  {082002} (\bibinfo {year} {2022})},\ \Eprint
  {http://arxiv.org/abs/2205.13402} {arXiv:2205.13402 [hep-lat]} \BibitemShut
  {NoStop}%
\bibitem [{\citenamefont {Ji}\ and\ \citenamefont {Zhang}(2015)}]{Ji:2015jwa}%
  \BibitemOpen
  \bibfield  {author} {\bibinfo {author} {\bibfnamefont {X.}~\bibnamefont
  {Ji}}\ and\ \bibinfo {author} {\bibfnamefont {J.-H.}\ \bibnamefont {Zhang}},\
  }\href {\doibase 10.1103/PhysRevD.92.034006} {\bibfield  {journal} {\bibinfo
  {journal} {Phys. Rev.}\ }\textbf {\bibinfo {volume} {D92}},\ \bibinfo {pages}
  {034006} (\bibinfo {year} {2015})},\ \Eprint
  {http://arxiv.org/abs/1505.07699} {arXiv:1505.07699 [hep-ph]} \BibitemShut
  {NoStop}%
\bibitem [{\citenamefont {Ji}\ \emph {et~al.}(2018)\citenamefont {Ji},
  \citenamefont {Zhang},\ and\ \citenamefont {Zhao}}]{Ji:2017oey}%
  \BibitemOpen
  \bibfield  {author} {\bibinfo {author} {\bibfnamefont {X.}~\bibnamefont
  {Ji}}, \bibinfo {author} {\bibfnamefont {J.-H.}\ \bibnamefont {Zhang}}, \
  and\ \bibinfo {author} {\bibfnamefont {Y.}~\bibnamefont {Zhao}},\ }\href
  {\doibase 10.1103/PhysRevLett.120.112001} {\bibfield  {journal} {\bibinfo
  {journal} {Phys. Rev. Lett.}\ }\textbf {\bibinfo {volume} {120}},\ \bibinfo
  {pages} {112001} (\bibinfo {year} {2018})},\ \Eprint
  {http://arxiv.org/abs/1706.08962} {arXiv:1706.08962 [hep-ph]} \BibitemShut
  {NoStop}%
\bibitem [{\citenamefont {Ishikawa}\ \emph {et~al.}(2017)\citenamefont
  {Ishikawa}, \citenamefont {Ma}, \citenamefont {Qiu},\ and\ \citenamefont
  {Yoshida}}]{Ishikawa:2017faj}%
  \BibitemOpen
  \bibfield  {author} {\bibinfo {author} {\bibfnamefont {T.}~\bibnamefont
  {Ishikawa}}, \bibinfo {author} {\bibfnamefont {Y.-Q.}\ \bibnamefont {Ma}},
  \bibinfo {author} {\bibfnamefont {J.-W.}\ \bibnamefont {Qiu}}, \ and\
  \bibinfo {author} {\bibfnamefont {S.}~\bibnamefont {Yoshida}},\ }\href
  {\doibase 10.1103/PhysRevD.96.094019} {\bibfield  {journal} {\bibinfo
  {journal} {Phys. Rev.}\ }\textbf {\bibinfo {volume} {D96}},\ \bibinfo {pages}
  {094019} (\bibinfo {year} {2017})},\ \Eprint
  {http://arxiv.org/abs/1707.03107} {arXiv:1707.03107 [hep-ph]} \BibitemShut
  {NoStop}%
\bibitem [{\citenamefont {Green}\ \emph {et~al.}(2018)\citenamefont {Green},
  \citenamefont {Jansen},\ and\ \citenamefont {Steffens}}]{Green:2017xeu}%
  \BibitemOpen
  \bibfield  {author} {\bibinfo {author} {\bibfnamefont {J.}~\bibnamefont
  {Green}}, \bibinfo {author} {\bibfnamefont {K.}~\bibnamefont {Jansen}}, \
  and\ \bibinfo {author} {\bibfnamefont {F.}~\bibnamefont {Steffens}},\ }\href
  {\doibase 10.1103/PhysRevLett.121.022004} {\bibfield  {journal} {\bibinfo
  {journal} {Phys. Rev. Lett.}\ }\textbf {\bibinfo {volume} {121}},\ \bibinfo
  {pages} {022004} (\bibinfo {year} {2018})},\ \Eprint
  {http://arxiv.org/abs/1707.07152} {arXiv:1707.07152 [hep-lat]} \BibitemShut
  {NoStop}%
\bibitem [{\citenamefont {Liu}\ \emph {et~al.}(2020)\citenamefont {Liu} \emph
  {et~al.}}]{Liu:2018uuj}%
  \BibitemOpen
  \bibfield  {author} {\bibinfo {author} {\bibfnamefont {Y.-S.}\ \bibnamefont
  {Liu}} \emph {et~al.} (\bibinfo {collaboration} {Lattice Parton}),\ }\href
  {\doibase 10.1103/PhysRevD.101.034020} {\bibfield  {journal} {\bibinfo
  {journal} {Phys. Rev.}\ }\textbf {\bibinfo {volume} {D101}},\ \bibinfo
  {pages} {034020} (\bibinfo {year} {2020})},\ \Eprint
  {http://arxiv.org/abs/1807.06566} {arXiv:1807.06566 [hep-lat]} \BibitemShut
  {NoStop}%
\bibitem [{\citenamefont {Chen}\ \emph {et~al.}(2018)\citenamefont {Chen},
  \citenamefont {Ishikawa}, \citenamefont {Jin}, \citenamefont {Lin},
  \citenamefont {Yang}, \citenamefont {Zhang},\ and\ \citenamefont
  {Zhao}}]{Chen:2017mzz}%
  \BibitemOpen
  \bibfield  {author} {\bibinfo {author} {\bibfnamefont {J.-W.}\ \bibnamefont
  {Chen}}, \bibinfo {author} {\bibfnamefont {T.}~\bibnamefont {Ishikawa}},
  \bibinfo {author} {\bibfnamefont {L.}~\bibnamefont {Jin}}, \bibinfo {author}
  {\bibfnamefont {H.-W.}\ \bibnamefont {Lin}}, \bibinfo {author} {\bibfnamefont
  {Y.-B.}\ \bibnamefont {Yang}}, \bibinfo {author} {\bibfnamefont {J.-H.}\
  \bibnamefont {Zhang}}, \ and\ \bibinfo {author} {\bibfnamefont
  {Y.}~\bibnamefont {Zhao}},\ }\href {\doibase 10.1103/PhysRevD.97.014505}
  {\bibfield  {journal} {\bibinfo  {journal} {Phys. Rev.}\ }\textbf {\bibinfo
  {volume} {D97}},\ \bibinfo {pages} {014505} (\bibinfo {year} {2018})},\
  \Eprint {http://arxiv.org/abs/1706.01295} {arXiv:1706.01295 [hep-lat]}
  \BibitemShut {NoStop}%
\bibitem [{\citenamefont {Gracey}(2003)}]{Gracey:2003yr}%
  \BibitemOpen
  \bibfield  {author} {\bibinfo {author} {\bibfnamefont {J.~A.}\ \bibnamefont
  {Gracey}},\ }\href {\doibase 10.1016/S0550-3213(03)00335-3} {\bibfield
  {journal} {\bibinfo  {journal} {Nucl. Phys.}\ }\textbf {\bibinfo {volume}
  {B662}},\ \bibinfo {pages} {247} (\bibinfo {year} {2003})},\ \Eprint
  {http://arxiv.org/abs/hep-ph/0304113} {arXiv:hep-ph/0304113 [hep-ph]}
  \BibitemShut {NoStop}%
\bibitem [{\citenamefont {Chang}\ \emph {et~al.}(2021)\citenamefont {Chang},
  \citenamefont {Liu}, \citenamefont {Raya}, \citenamefont
  {Rodríguez-Quintero},\ and\ \citenamefont {Yang}}]{Chang:2021vvx}%
  \BibitemOpen
  \bibfield  {author} {\bibinfo {author} {\bibfnamefont {L.}~\bibnamefont
  {Chang}}, \bibinfo {author} {\bibfnamefont {Y.-B.}\ \bibnamefont {Liu}},
  \bibinfo {author} {\bibfnamefont {K.}~\bibnamefont {Raya}}, \bibinfo {author}
  {\bibfnamefont {J.}~\bibnamefont {Rodríguez-Quintero}}, \ and\ \bibinfo
  {author} {\bibfnamefont {Y.-B.}\ \bibnamefont {Yang}},\ }\href@noop {} {\
  (\bibinfo {year} {2021})},\ \Eprint {http://arxiv.org/abs/2105.06596}
  {arXiv:2105.06596 [hep-lat]} \BibitemShut {NoStop}%
\bibitem [{\citenamefont {Constantinou}\ and\ \citenamefont
  {Panagopoulos}(2017)}]{Constantinou:2017sej}%
  \BibitemOpen
  \bibfield  {author} {\bibinfo {author} {\bibfnamefont {M.}~\bibnamefont
  {Constantinou}}\ and\ \bibinfo {author} {\bibfnamefont {H.}~\bibnamefont
  {Panagopoulos}},\ }\href {\doibase 10.1103/PhysRevD.96.054506} {\bibfield
  {journal} {\bibinfo  {journal} {Phys. Rev. D}\ }\textbf {\bibinfo {volume}
  {96}},\ \bibinfo {pages} {054506} (\bibinfo {year} {2017})},\ \Eprint
  {http://arxiv.org/abs/1705.11193} {arXiv:1705.11193 [hep-lat]} \BibitemShut
  {NoStop}%
\bibitem [{\citenamefont {Stewart}\ and\ \citenamefont
  {Zhao}(2018)}]{Stewart:2017tvs}%
  \BibitemOpen
  \bibfield  {author} {\bibinfo {author} {\bibfnamefont {I.~W.}\ \bibnamefont
  {Stewart}}\ and\ \bibinfo {author} {\bibfnamefont {Y.}~\bibnamefont {Zhao}},\
  }\href {\doibase 10.1103/PhysRevD.97.054512} {\bibfield  {journal} {\bibinfo
  {journal} {Phys. Rev.}\ }\textbf {\bibinfo {volume} {D97}},\ \bibinfo {pages}
  {054512} (\bibinfo {year} {2018})},\ \Eprint
  {http://arxiv.org/abs/1709.04933} {arXiv:1709.04933 [hep-ph]} \BibitemShut
  {NoStop}%
\bibitem [{\citenamefont {Alexandrou}\ \emph {et~al.}(2017)\citenamefont
  {Alexandrou}, \citenamefont {Cichy}, \citenamefont {Constantinou},
  \citenamefont {Hadjiyiannakou}, \citenamefont {Jansen}, \citenamefont
  {Panagopoulos},\ and\ \citenamefont {Steffens}}]{Alexandrou:2017huk}%
  \BibitemOpen
  \bibfield  {author} {\bibinfo {author} {\bibfnamefont {C.}~\bibnamefont
  {Alexandrou}}, \bibinfo {author} {\bibfnamefont {K.}~\bibnamefont {Cichy}},
  \bibinfo {author} {\bibfnamefont {M.}~\bibnamefont {Constantinou}}, \bibinfo
  {author} {\bibfnamefont {K.}~\bibnamefont {Hadjiyiannakou}}, \bibinfo
  {author} {\bibfnamefont {K.}~\bibnamefont {Jansen}}, \bibinfo {author}
  {\bibfnamefont {H.}~\bibnamefont {Panagopoulos}}, \ and\ \bibinfo {author}
  {\bibfnamefont {F.}~\bibnamefont {Steffens}},\ }\href {\doibase
  10.1016/j.nuclphysb.2017.08.012} {\bibfield  {journal} {\bibinfo  {journal}
  {Nucl. Phys.}\ }\textbf {\bibinfo {volume} {B923}},\ \bibinfo {pages} {394}
  (\bibinfo {year} {2017})},\ \Eprint {http://arxiv.org/abs/1706.00265}
  {arXiv:1706.00265 [hep-lat]} \BibitemShut {NoStop}%
\bibitem [{\citenamefont {Alexandrou}\ \emph {et~al.}(2020)\citenamefont
  {Alexandrou}, \citenamefont {Cichy}, \citenamefont {Constantinou},
  \citenamefont {Green}, \citenamefont {Hadjiyiannakou}, \citenamefont
  {Jansen}, \citenamefont {Manigrasso}, \citenamefont {Scapellato},\ and\
  \citenamefont {Steffens}}]{Alexandrou:2020qtt}%
  \BibitemOpen
  \bibfield  {author} {\bibinfo {author} {\bibfnamefont {C.}~\bibnamefont
  {Alexandrou}}, \bibinfo {author} {\bibfnamefont {K.}~\bibnamefont {Cichy}},
  \bibinfo {author} {\bibfnamefont {M.}~\bibnamefont {Constantinou}}, \bibinfo
  {author} {\bibfnamefont {J.~R.}\ \bibnamefont {Green}}, \bibinfo {author}
  {\bibfnamefont {K.}~\bibnamefont {Hadjiyiannakou}}, \bibinfo {author}
  {\bibfnamefont {K.}~\bibnamefont {Jansen}}, \bibinfo {author} {\bibfnamefont
  {F.}~\bibnamefont {Manigrasso}}, \bibinfo {author} {\bibfnamefont
  {A.}~\bibnamefont {Scapellato}}, \ and\ \bibinfo {author} {\bibfnamefont
  {F.}~\bibnamefont {Steffens}},\ }\href@noop {} {\  (\bibinfo {year}
  {2020})},\ \Eprint {http://arxiv.org/abs/2011.00964} {arXiv:2011.00964
  [hep-lat]} \BibitemShut {NoStop}%
\bibitem [{\citenamefont {Lin}\ \emph {et~al.}(2020)\citenamefont {Lin},
  \citenamefont {Chen},\ and\ \citenamefont {Zhang}}]{Lin:2020fsj}%
  \BibitemOpen
  \bibfield  {author} {\bibinfo {author} {\bibfnamefont {H.-W.}\ \bibnamefont
  {Lin}}, \bibinfo {author} {\bibfnamefont {J.-W.}\ \bibnamefont {Chen}}, \
  and\ \bibinfo {author} {\bibfnamefont {R.}~\bibnamefont {Zhang}},\
  }\href@noop {} {\  (\bibinfo {year} {2020})},\ \Eprint
  {http://arxiv.org/abs/2011.14971} {arXiv:2011.14971 [hep-lat]} \BibitemShut
  {NoStop}%
\bibitem [{\citenamefont {Ebert}\ \emph {et~al.}(2019)\citenamefont {Ebert},
  \citenamefont {Stewart},\ and\ \citenamefont {Zhao}}]{Ebert:2019okf}%
  \BibitemOpen
  \bibfield  {author} {\bibinfo {author} {\bibfnamefont {M.~A.}\ \bibnamefont
  {Ebert}}, \bibinfo {author} {\bibfnamefont {I.~W.}\ \bibnamefont {Stewart}},
  \ and\ \bibinfo {author} {\bibfnamefont {Y.}~\bibnamefont {Zhao}},\ }\href
  {\doibase 10.1007/JHEP09(2019)037} {\bibfield  {journal} {\bibinfo  {journal}
  {JHEP}\ }\textbf {\bibinfo {volume} {09}},\ \bibinfo {pages} {037} (\bibinfo
  {year} {2019})},\ \Eprint {http://arxiv.org/abs/1901.03685} {arXiv:1901.03685
  [hep-ph]} \BibitemShut {NoStop}%
\bibitem [{\citenamefont {Constantinou}\ \emph {et~al.}(2019)\citenamefont
  {Constantinou}, \citenamefont {Panagopoulos},\ and\ \citenamefont
  {Spanoudes}}]{Constantinou:2019vyb}%
  \BibitemOpen
  \bibfield  {author} {\bibinfo {author} {\bibfnamefont {M.}~\bibnamefont
  {Constantinou}}, \bibinfo {author} {\bibfnamefont {H.}~\bibnamefont
  {Panagopoulos}}, \ and\ \bibinfo {author} {\bibfnamefont {G.}~\bibnamefont
  {Spanoudes}},\ }\href {\doibase 10.1103/PhysRevD.99.074508} {\bibfield
  {journal} {\bibinfo  {journal} {Phys. Rev. D}\ }\textbf {\bibinfo {volume}
  {99}},\ \bibinfo {pages} {074508} (\bibinfo {year} {2019})},\ \Eprint
  {http://arxiv.org/abs/1901.03862} {arXiv:1901.03862 [hep-lat]} \BibitemShut
  {NoStop}%
\bibitem [{\citenamefont {Ebert}\ \emph {et~al.}(2020)\citenamefont {Ebert},
  \citenamefont {Stewart},\ and\ \citenamefont {Zhao}}]{Ebert:2019tvc}%
  \BibitemOpen
  \bibfield  {author} {\bibinfo {author} {\bibfnamefont {M.~A.}\ \bibnamefont
  {Ebert}}, \bibinfo {author} {\bibfnamefont {I.~W.}\ \bibnamefont {Stewart}},
  \ and\ \bibinfo {author} {\bibfnamefont {Y.}~\bibnamefont {Zhao}},\ }\href
  {\doibase 10.1007/JHEP03(2020)099} {\bibfield  {journal} {\bibinfo  {journal}
  {JHEP}\ }\textbf {\bibinfo {volume} {03}},\ \bibinfo {pages} {099} (\bibinfo
  {year} {2020})},\ \Eprint {http://arxiv.org/abs/1910.08569} {arXiv:1910.08569
  [hep-ph]} \BibitemShut {NoStop}%
\bibitem [{\citenamefont {Alexandrou}\ \emph {et~al.}(2023)\citenamefont
  {Alexandrou} \emph {et~al.}}]{Alexandrou:2023ucc}%
  \BibitemOpen
  \bibfield  {author} {\bibinfo {author} {\bibfnamefont {C.}~\bibnamefont
  {Alexandrou}} \emph {et~al.},\ }\href {\doibase 10.1103/PhysRevD.108.114503}
  {\bibfield  {journal} {\bibinfo  {journal} {Phys. Rev. D}\ }\textbf {\bibinfo
  {volume} {108}},\ \bibinfo {pages} {114503} (\bibinfo {year} {2023})},\
  \Eprint {http://arxiv.org/abs/2305.11824} {arXiv:2305.11824 [hep-lat]}
  \BibitemShut {NoStop}%
\bibitem [{\citenamefont {Spanoudes}\ \emph {et~al.}(2024)\citenamefont
  {Spanoudes}, \citenamefont {Constantinou},\ and\ \citenamefont
  {Panagopoulos}}]{Spanoudes:2024kpb}%
  \BibitemOpen
  \bibfield  {author} {\bibinfo {author} {\bibfnamefont {G.}~\bibnamefont
  {Spanoudes}}, \bibinfo {author} {\bibfnamefont {M.}~\bibnamefont
  {Constantinou}}, \ and\ \bibinfo {author} {\bibfnamefont {H.}~\bibnamefont
  {Panagopoulos}},\ }\href@noop {} {\  (\bibinfo {year} {2024})},\ \Eprint
  {http://arxiv.org/abs/2401.01182} {arXiv:2401.01182 [hep-lat]} \BibitemShut
  {NoStop}%
\bibitem [{\citenamefont {Clark}\ \emph {et~al.}(2010)\citenamefont {Clark},
  \citenamefont {Babich}, \citenamefont {Barros}, \citenamefont {Brower},\ and\
  \citenamefont {Rebbi}}]{Clark:2009wm}%
  \BibitemOpen
  \bibfield  {author} {\bibinfo {author} {\bibfnamefont {M.~A.}\ \bibnamefont
  {Clark}}, \bibinfo {author} {\bibfnamefont {R.}~\bibnamefont {Babich}},
  \bibinfo {author} {\bibfnamefont {K.}~\bibnamefont {Barros}}, \bibinfo
  {author} {\bibfnamefont {R.~C.}\ \bibnamefont {Brower}}, \ and\ \bibinfo
  {author} {\bibfnamefont {C.}~\bibnamefont {Rebbi}},\ }\href {\doibase
  10.1016/j.cpc.2010.05.002} {\bibfield  {journal} {\bibinfo  {journal}
  {Comput. Phys. Commun.}\ }\textbf {\bibinfo {volume} {181}},\ \bibinfo
  {pages} {1517} (\bibinfo {year} {2010})},\ \Eprint
  {http://arxiv.org/abs/0911.3191} {arXiv:0911.3191 [hep-lat]} \BibitemShut
  {NoStop}%
\bibitem [{\citenamefont {Babich}\ \emph {et~al.}(2011)\citenamefont {Babich},
  \citenamefont {Clark}, \citenamefont {Joo}, \citenamefont {Shi},
  \citenamefont {Brower},\ and\ \citenamefont {Gottlieb}}]{Babich:2011np}%
  \BibitemOpen
  \bibfield  {author} {\bibinfo {author} {\bibfnamefont {R.}~\bibnamefont
  {Babich}}, \bibinfo {author} {\bibfnamefont {M.~A.}\ \bibnamefont {Clark}},
  \bibinfo {author} {\bibfnamefont {B.}~\bibnamefont {Joo}}, \bibinfo {author}
  {\bibfnamefont {G.}~\bibnamefont {Shi}}, \bibinfo {author} {\bibfnamefont
  {R.~C.}\ \bibnamefont {Brower}}, \ and\ \bibinfo {author} {\bibfnamefont
  {S.}~\bibnamefont {Gottlieb}},\ }in\ \href {\doibase 10.1145/2063384.2063478}
  {\emph {\bibinfo {booktitle} {{SC11 International Conference for High
  Performance Computing, Networking, Storage and Analysis Seattle, Washington,
  November 12-18, 2011}}}}\ (\bibinfo {year} {2011})\ \Eprint
  {http://arxiv.org/abs/1109.2935} {arXiv:1109.2935 [hep-lat]} \BibitemShut
  {NoStop}%
\bibitem [{\citenamefont {Clark}\ \emph {et~al.}(2016)\citenamefont {Clark},
  \citenamefont {Jo}, \citenamefont {Strelchenko}, \citenamefont {Cheng},
  \citenamefont {Gambhir},\ and\ \citenamefont {Brower}}]{Clark:2016rdz}%
  \BibitemOpen
  \bibfield  {author} {\bibinfo {author} {\bibfnamefont {M.~A.}\ \bibnamefont
  {Clark}}, \bibinfo {author} {\bibfnamefont {B.}~\bibnamefont {Jo}}, \bibinfo
  {author} {\bibfnamefont {A.}~\bibnamefont {Strelchenko}}, \bibinfo {author}
  {\bibfnamefont {M.}~\bibnamefont {Cheng}}, \bibinfo {author} {\bibfnamefont
  {A.}~\bibnamefont {Gambhir}}, \ and\ \bibinfo {author} {\bibfnamefont
  {R.}~\bibnamefont {Brower}},\ }\href@noop {} {\  (\bibinfo {year} {2016})},\
  \Eprint {http://arxiv.org/abs/1612.07873} {arXiv:1612.07873 [hep-lat]}
  \BibitemShut {NoStop}%
\bibitem [{\citenamefont {Bi}\ \emph {et~al.}(2020)\citenamefont {Bi},
  \citenamefont {Xiao}, \citenamefont {Gong}, \citenamefont {Guo},
  \citenamefont {Sun}, \citenamefont {Xu},\ and\ \citenamefont
  {Yang}}]{Bi:2020wpt}%
  \BibitemOpen
  \bibfield  {author} {\bibinfo {author} {\bibfnamefont {Y.-J.}\ \bibnamefont
  {Bi}}, \bibinfo {author} {\bibfnamefont {Y.}~\bibnamefont {Xiao}}, \bibinfo
  {author} {\bibfnamefont {M.}~\bibnamefont {Gong}}, \bibinfo {author}
  {\bibfnamefont {W.-Y.}\ \bibnamefont {Guo}}, \bibinfo {author} {\bibfnamefont
  {P.}~\bibnamefont {Sun}}, \bibinfo {author} {\bibfnamefont {S.}~\bibnamefont
  {Xu}}, \ and\ \bibinfo {author} {\bibfnamefont {Y.-B.}\ \bibnamefont
  {Yang}},\ }\bibfield  {booktitle} {\emph {\bibinfo {booktitle} {{Proceedings,
  37th International Symposium on Lattice Field Theory (Lattice 2019): Wuhan,
  China, June 16-22 2019}}},\ }\href {\doibase 10.22323/1.363.0286} {\bibfield
  {journal} {\bibinfo  {journal} {PoS}\ }\textbf {\bibinfo {volume}
  {LATTICE2019}},\ \bibinfo {pages} {286} (\bibinfo {year} {2020})},\ \Eprint
  {http://arxiv.org/abs/2001.05706} {arXiv:2001.05706 [hep-lat]} \BibitemShut
  {NoStop}%
\end{thebibliography}%

\clearpage

\begin{widetext}

\section*{Appendix}

\subsection{The distribution of $\Delta^G(x)$}

The quantity $\Delta^G(x)$, defined in Eq.~(\ref{eq:delta}) in Landau gauge, does not reach { exactly} zero at every lattice site due to imperfect gauge fixing. As illustrated in Fig.~\ref{fig:DeltaDis}, the distribution is symmetrical and exhibits a long tail. It is important to note that the { scale of the abscissa} is smaller for higher precision. With a precision that is 100 times greater, the tail becomes approximately 10 times shorter. However, the tail does not display a strong dependence on lattice spacing. Additionally, we compute $\theta^G$, as defined in Eq.~(\ref{eq:theta}), on the same configuration { for comparison}. The magnitude of $\theta^G$ is directly proportional to that of $\delta^F$ in the { part} of the ensemble that we are investigating.

\begin{figure}[tbph]
\centering
\includegraphics[width=8cm]{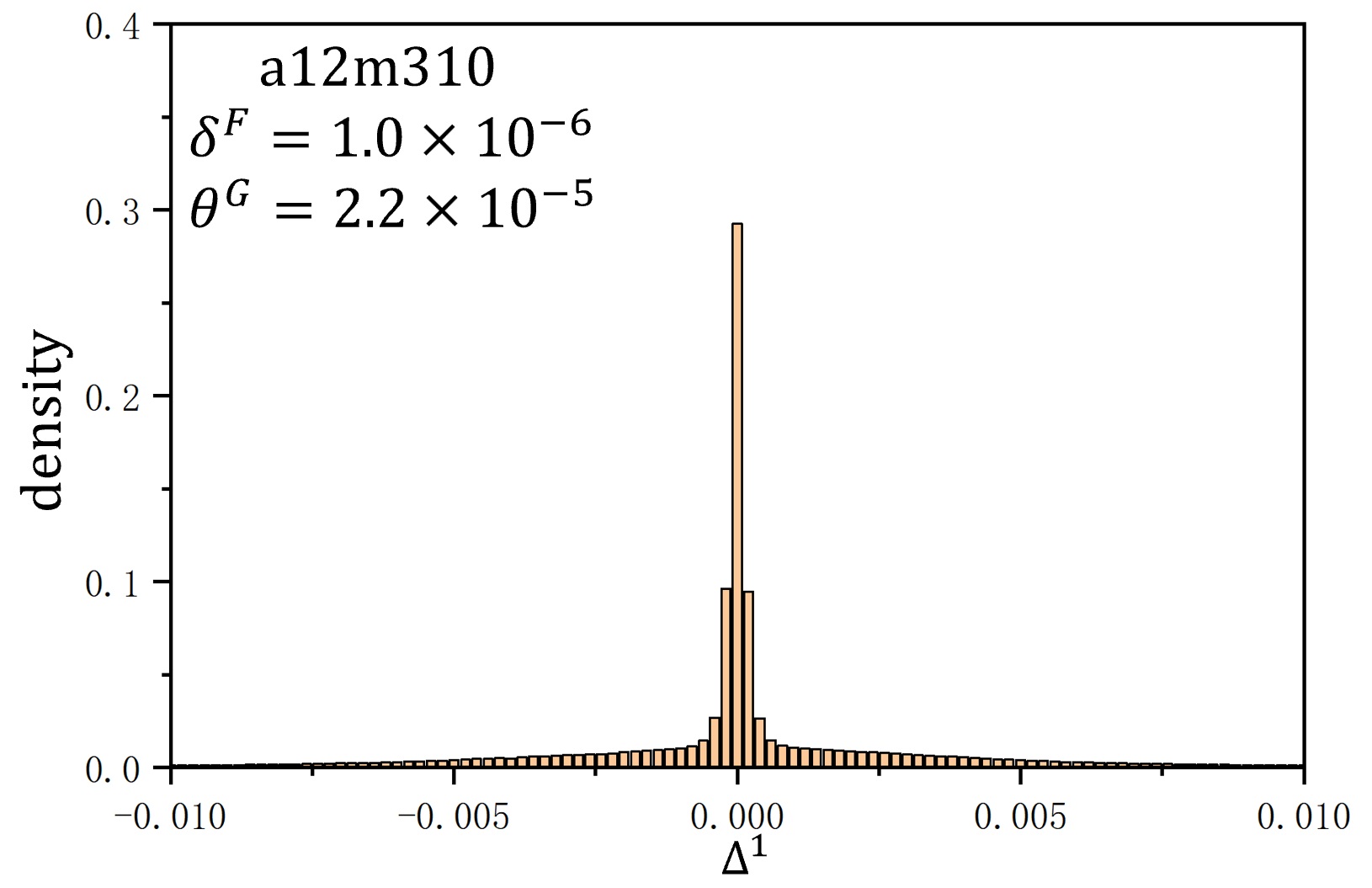}
\includegraphics[width=8cm]{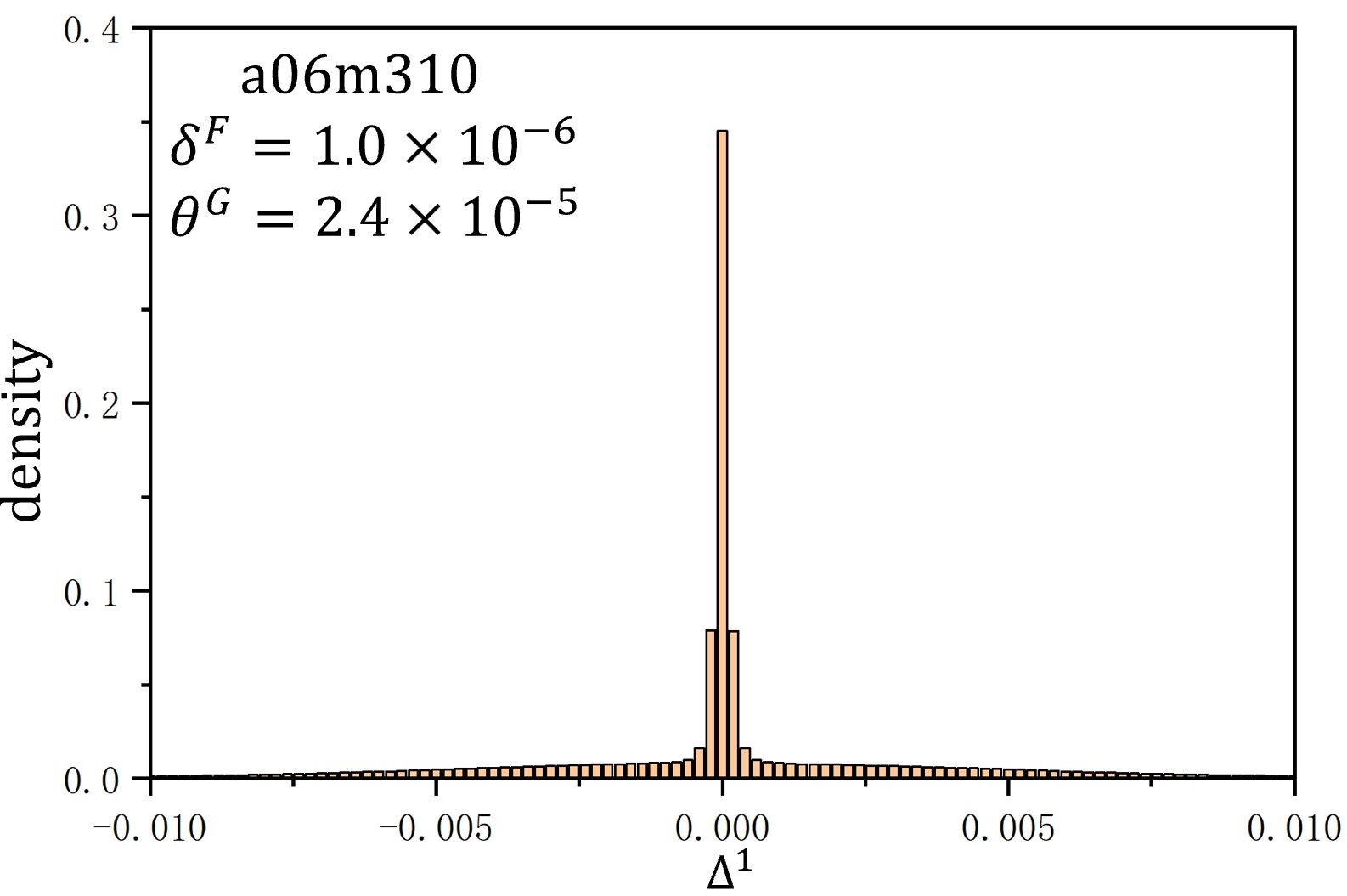}
\includegraphics[width=8cm]{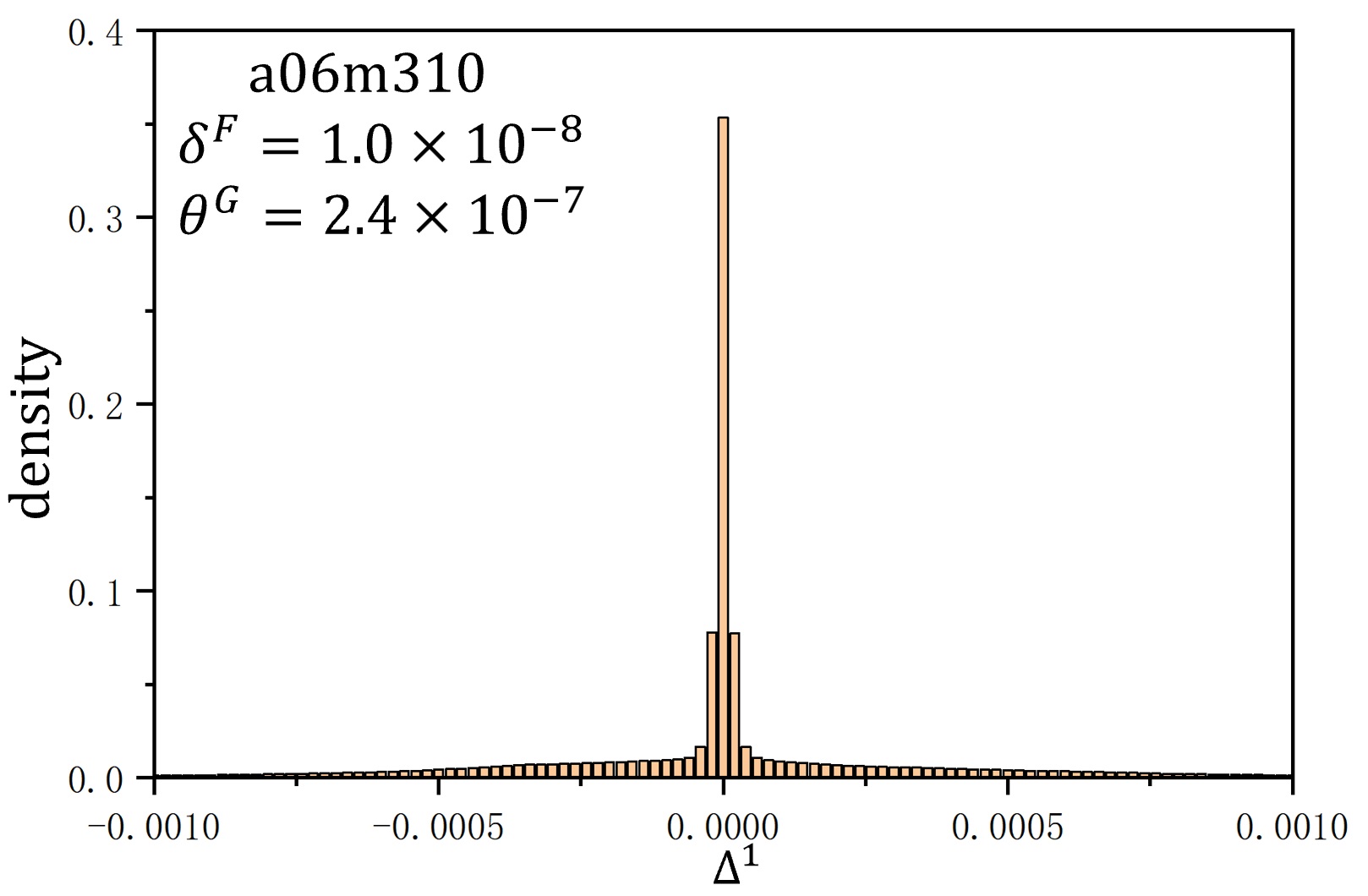}
\includegraphics[width=8cm]{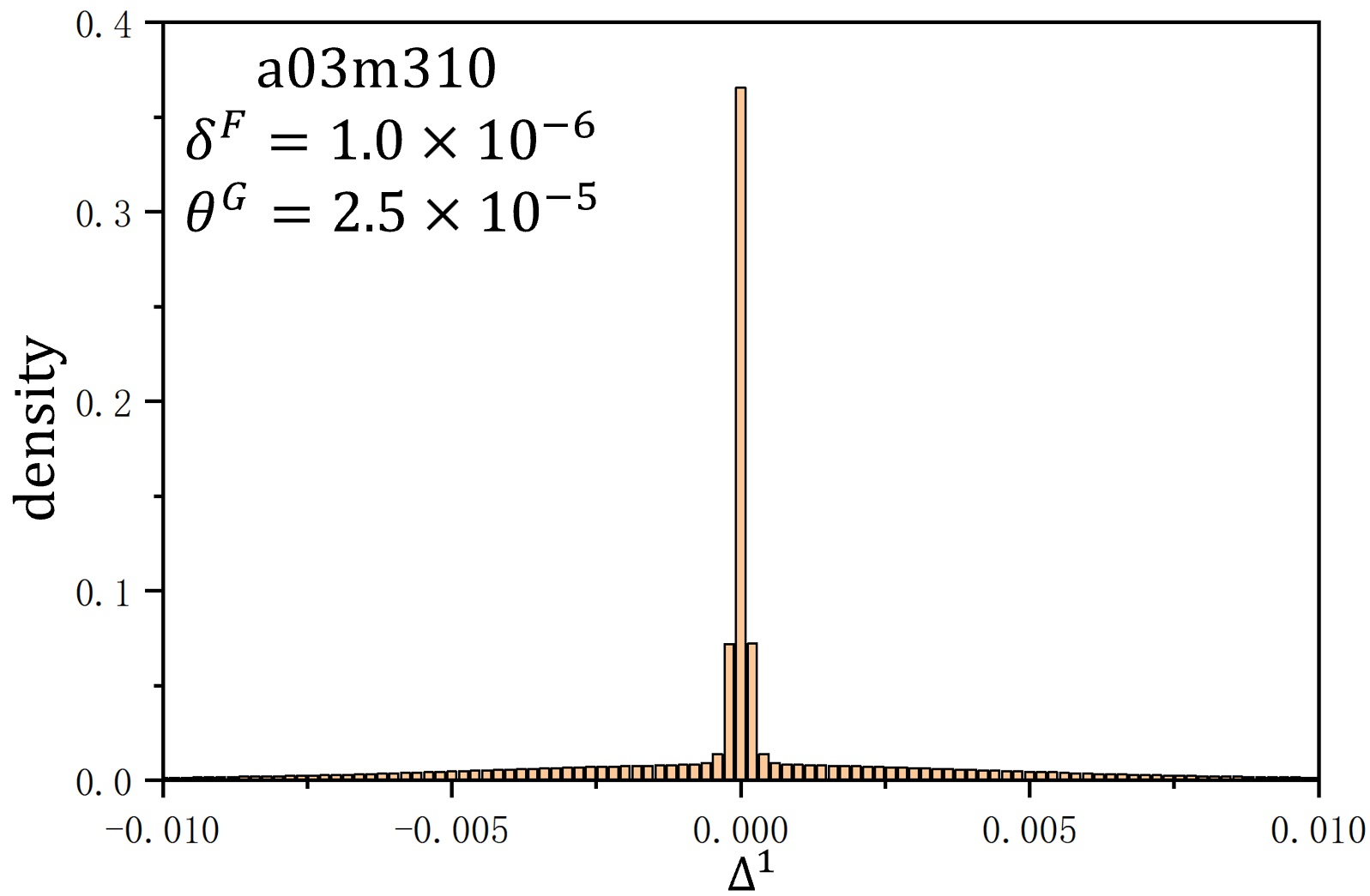}
\caption{$\Delta(x) = \Delta^m(x) T^m(x)$ defined in Eq.~(\ref{eq:delta}) on different ensembles and { with different required precision}. Without loss of generality, we just show the first component $\Delta^1(x)$ { as illustration}. The shape of the distribution is similar on different ensembles and { for different precision}.}
\label{fig:DeltaDis}
\end{figure}

\subsection{Coulomb wall source}

As our focus is solely on the UV divergence { of} the quasi-PDF operator, we utilize the pion in the rest frame as the external state to enhance the signal-to-noise ratio. Additionally, we have employed { a Coulomb} wall source to further improve the signal. It is well-known that wall source { calculations include a gauge variation part}, but with a sufficient number of configurations, the gauge variant part will vanish. And one finds that applying Coulomb gauge fixing {  improves} the signal-to-noise ratio, but { does} not lead to an obvious systematic deviation. In this work, we aim to verify whether this holds true for non-local operator hadron matrix { elements}, and use the unpolarized quasi PDF as { illustration}.

\begin{figure}[tbph]
\centering
\includegraphics[width=8cm]{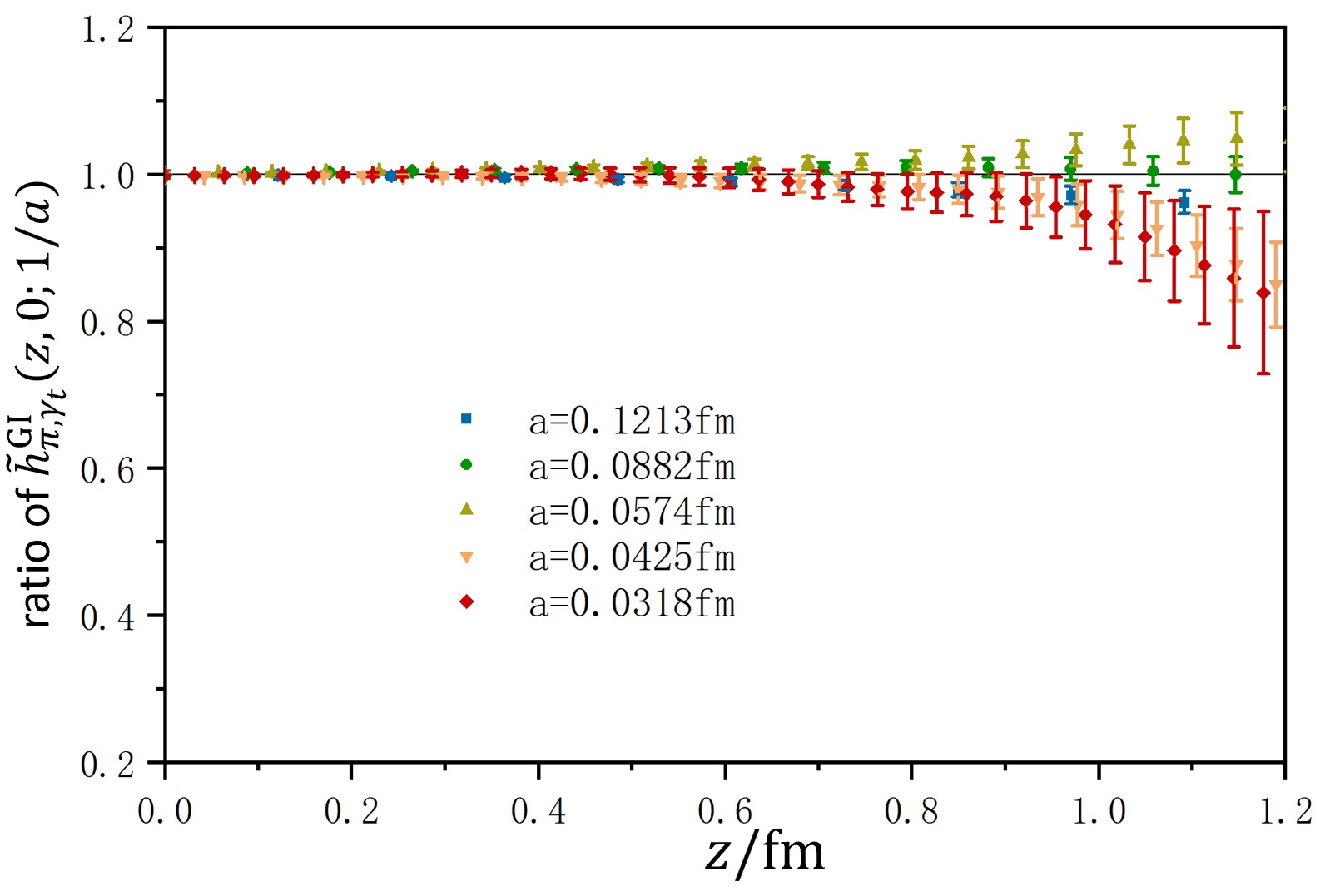}
\caption{The ratio of quasi PDF pion matrix element { in Coulomb gauge for $\delta^F=10^{-6}$ and $\delta^F=10^{-7}$}. They are calculated in the rest frame with { awall source}.}
\label{fig:CoulombWall}
\end{figure}

As depicted in Fig.~\ref{fig:CoulombWall}, the ratio remains { close to one for various required precisions}, as long as the Wilson lines are not excessively long. Consequently, we can conclude that the imprecise Coulomb gauge fixing in the wall source calculation in this study { does not change the central values}.

\end{widetext}

\end{document}